\documentclass[floats,floatfix,showpacs,amssymb,prd,twocolumn,superscriptaddress,nofootinbib,nolongbibliography,reprint]{revtex4-1}

\usepackage{amssymb,amsmath,verbatim,mathtools,needspace,enumitem,etoolbox,graphicx,physics,microtype,afterpage,xspace,tabularray,lmodern,multirow,bm}
\usepackage{textcomp, gensymb}
\usepackage{array}

\usepackage[normalem]{ulem}
\usepackage{placeins}
\usepackage{comment}
\usepackage[dvipsnames, usenames]{xcolor}
\definecolor{linkcolor}{rgb}{0.0,0.3,0.5}
\usepackage[unicode, colorlinks=true, linkcolor=linkcolor, citecolor=linkcolor, filecolor=linkcolor, urlcolor=linkcolor, linktocpage, breaklinks]{hyperref}
\usepackage[all]{hypcap}
\usepackage[T1]{fontenc}
\usepackage[utf8]{inputenc}
\usepackage{mathrsfs}
\usepackage[usenames,dvipsnames]{xcolor}
\hypersetup{colorlinks=true,citecolor=romared,linkcolor=romared,urlcolor=romared}

\definecolor{romared}{RGB}{142,0,28}

\newcommand{\be}{\begin{equation}}
\newcommand{\ee}{\end{equation}}

\def\be{\begin{equation}}
\def\ee{\end{equation}}
\newcommand{\beq}{\begin{eqnarray}}
\newcommand{\eeq}{\end{eqnarray}}

\usepackage{aas_macros}
\usepackage{makecell}
\usepackage{soul}
\usepackage[nolist,nohyperlinks]{acronym}
\usepackage{orcidlink}
\usepackage{lipsum}
\usepackage{cleveref}

\acrodef{LSC}[LSC]{LIGO Scientific Collaboration}
\acrodef{BH}{black hole}
\acrodef{NS}{neutron star}
\acrodef{PN}{Post-Newtonian}
\acrodef{BBH}{binary black-hole}
\acrodef{BNS}{binary neutron-star}
\acrodef{NSBH}{neutron-star black-hole}
\acrodef{NR}{numerical relativity}
\acrodef{GW}{gravitational wave}
\acrodef{PSD}{power spectral density}
\acrodef{aLIGO}{Advanced Laser interferometer Gravitational-Wave Observatory}
\acrodef{AZDHP}{aLIGO zero detuned high power density}
\acrodef{GR}{general relativity}
\acrodef{PE}{parameter estimation}
\acrodef{LAL}{LIGO algorithm library}
\acrodef{TPI}{tensor-product interpolant}
\acrodef{SVD}{singular value decomposition}
\acrodef{SNR}{signal-to-noise ratio}
\acrodef{ODE}{ordinary differential equation}
\acrodef{PDE}{partial differential equation}
\acrodef{ROM}{reduced order model}
\acrodef{QNM}{quasi-normal mode}
\acrodef{IMR}{inspiral-merger-ringdown}
\acrodef{LVK}{LIGO-Virgo-KAGRA}
\acrodef{SXS}{Simulating eXtreme Spacetimes}

\newcommand{\jhu}{\affiliation{William H. Miller III Department of Physics and Astronomy, Johns Hopkins University, 3400 North Charles Street, Baltimore, Maryland, 21218, USA}}
\newcommand{\ciera}{\affiliation{Center for Interdisciplinary Exploration and Research in Astrophysics (CIERA), 1800 Sherman Ave, Evanston, IL 60201, USA}}

\graphicspath{{../Plots/}}

\newcommand{\orcid}[1]{\href{https://orcid.org/#1}{\includegraphics[width=10pt]{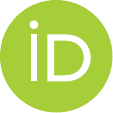}}}
\renewcommand{\vec}[1]{\boldsymbol{#1}}

\newcommand{\ben}{\begin{enumerate}}
\newcommand{\een}{\end{enumerate}}

\def\be{\begin{equation}}
\def\ee{\end{equation}}
\def\beq{\begin{eqnarray}}
\def\eeq{\end{eqnarray}}

\def\apjl{\rm{ApJ}}
\def\apjs{\rm{ApJS}}
\def\aap{\rm{A\&A}}

\graphicspath{{./Plots/}}

\allowdisplaybreaks

\begin{document}

\pagenumbering{arabic}

\title{Extracting linear and nonlinear quasinormal modes from black hole merger simulations}

\author{Mark Ho-Yeuk Cheung \orcid{0000-0002-7767-3428}}
\email{hcheung5@jhu.edu}
\jhu
\author{Emanuele Berti \orcid{0000-0003-0751-5130}}
\email{berti@jhu.edu}
\jhu
\author{Vishal Baibhav \orcid{0000-0002-2536-7752}}
\email{vishal.baibhav@northwestern.edu}
\ciera
\author{Roberto Cotesta \orcid{0000-0001-6568-6814}}
\email{rcotest1@jhu.edu}
\jhu
\date{\today}

\begin{abstract}
	In general relativity, when two black holes merge they produce a rotating (Kerr) black hole remnant. According to perturbation theory, the remnant emits ``ringdown'' radiation: a superposition of exponentials with characteristic complex frequencies that depend only on the remnant's mass and spin.
	While the goal of the black hole spectroscopy program is to measure the quasinormal mode frequencies, a knowledge of their amplitudes and phases is equally important to determine which modes are detectable, and possibly to perform additional consistency checks.
	Unlike the complex frequencies, the amplitudes and phases depend on the properties of the binary progenitors, such as the binary mass ratio and component spins.
	In this paper we develop a fitting algorithm designed to reliably identify the modes present in numerical simulations and to extract their amplitudes and phases.
	We apply the algorithm to over 500 binary black hole simulations from the public SXS numerical relativity simulation catalog, and we present fitting formulas for the resulting mode amplitudes and phases as functions of the properties of the progenitors.
	Crucially, our algorithm allows for the extraction of not only prograde fundamental modes and overtones, but also retrograde modes and second-order modes.
	We unveil interesting relations for the amplitude ratios of different modes. The fitting code and interactive versions of some of the plots are publicly available~\cite{jaxqualin}.
	The results presented in this paper can be updated as more and better simulations become available.
\end{abstract}

\maketitle

\section{Introduction}

The pioneering work on the linear perturbations of the Schwarzschild spacetime by Regge and Wheeler~\cite{Regge:1957td}, Zerilli~\cite{Zerilli:1970se,Zerilli:1970wzz} and Vishveshwara~\cite{Vishveshwara:1970zz} revealed that the response of a black hole (BH) to an incoming pulse of radiation is characterized by a superposition of damped exponentials with discrete frequencies and damping times, now commonly known as the ``ringdown'' by analogy with the dying tones of a vibrating bell. The damping occurs because BH spacetimes absorb gravitational waves (GWs) at the horizon and emit radiation at spatial infinity: the system is dissipative, hence the name ``quasinormal'' modes (QNMs), as opposed to the ``normal'' modes of self-adjoint physical systems~\cite{Nollert:1999ji,Kokkotas:1999bd,Berti:2009kk,Berti:2018vdi,Cardoso:2019rvt}.

The correspondence between BH spectra and atomic spectra was soon clear to early researchers who computed QNM spectra and clarified their physical interpretation~\cite{Press:1971wr,Davis:1971gg,Chandrasekhar:1975zza, Ferrari:1984zz,Mashhoon:1985cya,BLOME1984231,Schutz:1985km,Leaver:1985ax,Leaver:1986gd}.
The separation of angular variables results in multipolar components characterized by a discrete set of angular indices $(\ell,\,m)$. For fixed value of these indices, the spectrum contains a tower of discrete QNM frequencies $\omega_{\ell{,}m{,}n}$ that can be sorted by the magnitude of their imaginary part (the so-called ``overtones''). Typically $n=0$ denotes the fundamental mode, while increasing values of $n$ correspond to larger imaginary parts and shorter damping times.
Teukolsky proved that the perturbation equations for rotating (Kerr) BHs are also separable~\cite{Teukolsky:1972my,Teukolsky:1973ha,Press:1973zz,Teukolsky:1974yv}. Detweiler pointed out that the fundamental QNM frequency of a Kerr BH (the one with the smallest imaginary part and longest damping time) depends only on its mass and spin~\cite{Detweiler:1977gy}, so -- at least conceptually -- the relation can be inverted to identify the Kerr BH parameters from a knowledge of the frequency and damping time. In fact, {\em all} complex frequencies are fixed once we know the mass and spin of the remnant, so the observation of additional resonant frequencies can be used to identify Kerr BHs. This is the GW equivalent of atomic spectroscopy~\cite{Detweiler:1980gk}.

Within linearized BH perturbation theory, Leaver~\cite{Leaver:1986gd} proved that each multipolar component of the waveform at intermediate times -- after the ``prompt response'', and before the onset of power-law tails -- is described by a superposition of QNMs:
\begin{equation}\label{eq:model}
	h_{\ell m}(t) \equiv \sum_n A_{\ell{,}m{,}n} e^{-i\left[\omega_{\ell{,}m{,}n}(t - t_{\rm start}) + \phi_{\ell{,}m{,}n}\right]}\,.
\end{equation}
Here, $t_{\rm start}$ is an arbitrary starting time.
In linearized GR, the complex Kerr QNM frequencies $\omega_{\ell{,}m{,}n}$ depend only on the remnant BH mass $M_{\rm rem}$ and dimensionless spin $\chi_{\rm rem}$, but not on the nature of the perturbation, and are known to very high accuracy~\cite{Berti:2005ys,RDwebsites}.
On the contrary, the QNM amplitudes $A_{\ell{,}m{,}n}$ and phases $\phi_{\ell{,}m{,}n}$ depend on the astrophysical process causing the perturbation.

The merger of two comparable-mass BHs was identified early on as one of the most promising LIGO-Virgo sources~\cite{1987thyg.book..330T}, but predicting which QNMs are most excited as a result of the merger was essentially a matter of guesswork before the first numerical BH merger simulations.
This is important because practical attempts to implement the spectroscopy program in GW data analysis rely on the measurability of the frequency and damping time of the fundamental mode, and possibly of other modes~\cite{Echeverria:1989hg,Finn:1992wt,Dreyer:2003bv,Berti:2005ys}), but only the frequencies of modes with sufficiently large amplitude are measurable.
Green's function techniques imply that the QNM amplitudes $A_{\ell{,}m{,}n}$ can be factorized as a product of complex ``excitation factors'' $B_{\ell{,}m{,}n}$ that depend only on the remnant's mass and spin and complex-valued, initial-data dependent integrals $I_{\ell{,}m{,}n}$~\cite{Leaver:1986gd,Andersson:1995zk,Andersson:1996cm,Berti:2006wq,Dorband:2006gg,Zhang:2013ksa,Oshita:2021iyn,Lagos:2022otp}.
Before the first numerical BH merger simulations, heuristic arguments suggested that comparable-mass BH mergers may have ringdown signal-to-noise ratio (SNR) roughly comparable to the inspiral SNR~\cite{Flanagan:1997sx}, while other astrophysical processes would be much less efficient at exciting QNMs~\cite{Berti:2006hb}. However, early work trying to quantify the {\em measurability} of different multipoles and overtones had to rely on educated guesses~\cite{Berti:2005ys}.

Our understanding of ringdown excitation improved after the 2005 numerical relativity (NR) breakthrough~\cite{Pretorius:2005gq,Campanelli:2005dd,Baker:2005vv}. Fits of NR simulations showed that the radiation from a BBH merger is dominated by the $\ell{,}|m|=2,2$ spherical harmonic multipole, while higher multipoles are subdominant~\cite{Buonanno:2006ui,Berti:2007fi}.
Unbiased mass and spin estimates require the combination of different multipolar components~\cite{Berti:2007zu} or the inclusion of overtones~\cite{Baibhav:2017jhs,Giesler:2019uxc}.  Since multiple modes will always be excited to some extent, we must first understand which combination of modes will dominate the signal~\cite{Berti:2005ys}. Are we going to observe a combination of low-$n$ modes for different multipoles, or are higher overtones of the $\ell{,}m=2{,}2$ component dominant? Can we measure the frequencies and damping times accurately enough to resolve the modes? The answers to these questions depend on (i) the properties of the merger remnant progenitors, and (ii) the sensitivity of the detectors~\cite{Berti:2007zu,Berti:2007fi,Kamaretsos:2011um,Kamaretsos:2012bs,London:2014cma,Bhagwat:2016ntk,Baibhav:2017jhs,Thrane:2017lqn,London:2018gaq,Baibhav:2018rfk,Borhanian:2019kxt, Bhagwat:2019bwv,Baibhav:2020tma,Cook:2020otn,JimenezForteza:2020cve,Ota:2021ypb,Li:2021wgz,MaganaZertuche:2021syq}.

Initial attempts at fitting numerical simulations with a superposition of QNMs argued that it is possible to fit the numerical waveforms with a superposition of several overtones within linear perturbation theory~\cite{London:2014cma,Giesler:2019uxc,Forteza:2021wfq}, possibly including counterrotating modes and higher multipoles~\cite{Ota:2019bzl,Bhagwat:2019dtm,JimenezForteza:2020cve,Bustillo:2020buq,Okounkova:2020vwu,Mourier:2020mwa,Cook:2020otn,MaganaZertuche:2021syq,Dhani:2020nik,Dhani:2021vac,Finch:2021iip,Finch:2021qph,Ota:2021ypb,Jaramillo:2022oqn,Forteza:2022tgq}. More recent work (confirming early findings by London {\em et al.}~\cite{London:2014cma}) showed that nonlinearities are important: quadratic modes are present in NR merger simulations of quasicircular and head-on mergers~\cite{Ma:2022wpv,Mitman:2022qdl,Cheung:2022rbm,Khera:2023lnc}. This caused a recent surge of activity in the analytical modeling of nonlinear QNMs~\cite{Lagos:2022otp,Guerreiro:2023gdy,Kehagias:2023ctr,Kehagias:2023mcl,Redondo-Yuste:2023seq,Perrone:2023jzq,Bucciotti:2023ets} (see e.g.~\cite{Gleiser:1995gx,Campanelli:1998jv,Zlochower:2003yh,Ioka:2007ak,Nakano:2007cj,Brizuela:2009qd,Pazos:2010xf,Loutrel:2020wbw,Ripley:2020xby} for earlier work, and~\cite{Sberna:2021eui,MaganaZertuche:2021syq} for a discussion of other nonlinearities that could affect the ringdown).

In this paper we continue the ``agnostic'' BH spectroscopy program of Ref.~\cite{Baibhav:2023clw}. We develop a systematic fitting algorithm to extract the QNM content of NR waveforms. Our goal is, essentially, to build the most accurate ringdown model that avoids overfitting. We first identify which modes can be reliably found in NR data. Then we produce a catalog of their amplitudes, phases and (possibly) starting times. We obtain polynomial ``hyperfits'' of the resulting mode amplitudes as a function of the parameters of the merger progenitors, and we look for universal relations in the amplitude ratios of different modes.

The plan of the paper is as follows.
In Sec.~\ref{sec:QNMs} we review the classification of all possible QNMs that can be found from fits of a NR simulation.
In Sec.~\ref{sec:mode_extraction} we describe the algorithm underlying our QNM extraction procedure and show illustrative examples of its application to specific waveforms in the SXS catalog. This rather lengthy and technical section is the core of this paper. It is crucial to understand the assumptions underlying our QNM amplitude estimates, but it can be skipped by readers who want to focus on the physical implications of our work.
In Sec.~\ref{sec:hyperfit} we apply the algorithm to the dominant multipolar components of simulations from the whole SXS catalog, and we use the results of this large-scale fitting campaign to produce ``hyperfits'' of the QNM ampitudes for generic BH binary systems.
In Sec.~\ref{sec:amplratios} we study the relation between the amplitudes of various modes, including (i) the ratio between nonlinear modes and their ``parent'' linear modes, (ii) the ratio between the amplitudes of the first overtones and the fundamental modes, and (iii) the ratios between retrograde and prograde modes.
In Sec.~\ref{sec:tstart} we present some considerations on the thorny issue of defining a starting time for the ringdown for data analysis purposes.
In Sec.~\ref{sec:conclusions} we present some conclusions and directions for future research.

To improve readability, consistency checks and more technical material are relegated to several appendixes.
In Appendix~\ref{app:retro_definition} we clarify our definition of prograde and retrograde QNMs.
In Appendix~\ref{app:fitter} we complement the discussion of Sec.~\ref{sec:mode_extraction} with some technical details about our fitting routine.
In Appendix~\ref{app:toy} we develop a toy model to test the accuracy of the fitting algorithm.
In Appendixes~\ref{app:consistency} and \ref{app:procedure} we ``stress test'' our fitting routines using this toy model, and show that they do indeed give consistent results.
In Appendix~\ref{mixing} we study spherical-spheroidal mode mixing, and show that the fitting routines give results consistent with theoretical expectations.
The current SXS catalog does not, in general, make use of Cauchy characteristic extraction (CCE), but some simulations using CCE are publicly available.
In Appendix~\ref{app:CCE} we apply our fitting algorithm to the limited set of public CCE simulations, and show that the use of better wave extraction techniques gets rid of ``spurious'' QNMs.
In Appendix~\ref{app:stability} we investigate the effect of using different stability criteria to determine whether a mode is robustly present in the ringdown.
Finally, in Appendix~\ref{sec:outliers} we discuss some outliers in the large catalog of extracted QNM amplitudes and their possible origin.

\section{Quasinormal mode classification}
\label{sec:QNMs}

General relativity is a nonlinear theory of gravity.
The evolution of the full nonlinear equations for binary black holes (BBHs) poses theoretical challenges and is computationally expensive.
However, if our goal is to study the late-time postmerger behavior, the spacetime asymptotically approaches the Kerr solution, and we can linearize the Einstein equations on a Kerr metric background.

At late enough times, neglecting contributions from the back-scattering of GWs, the postmerger waveform can be modeled as a linear combination of QNMs with complex frequencies $\tilde{\omega}_k = \omega_{r,k} + i\omega_{i,k}$:
\begin{equation} \label{eq:QNM}
	h(t) = \sum_{k \in K} A_k e^{\omega_{i,k} t}e^{-i (\omega_{r,k} t + \phi_k)},
\end{equation}
where $A_k$, $\phi_k$, $\omega_{r,k}$ and $\omega_{i,k}$ are real.
In general, $\omega_{i,k} \leq 0$.
Each mode is labeled by a mode multi-index $k$, which has the form $k = \ell{,}m{,}n$ for linear modes, and the form $k = \ell_1{,}m_1{,}n_1 \! \times \! \ell_2{,}m_2{,}n_2$ for quadratic modes (we will neglect contributions beyond quadratic order).
The set of multi-indices $K$ includes all the modes $k$ that could be present in the waveform.
We will also define the mode $k = {\rm constant}$, where $\omega_{r, \rm constant} = \omega_{i, \rm constant} = 0$, denoting a ``pseudo-QNM'' contribution that acts as a complex constant in time, $A_{\rm constant} e^{i \phi_{\rm constant}}$. This is useful to model waveforms containing a constant component.

In this work, we are interested in recovering the list of all the modes $K$ that are present in NR waveforms, and the corresponding values of $A_k$ and $\phi_k$.
To do this, we will employ two types of fitting models:
(i) a ``frequency-agnostic'' model with $N_f$ free QNMs, where $A_k$, $\phi_k$, $\omega_{r,k}$ and $\omega_{i,k}$ of each mode are left as free parameters
\begin{equation}\label{eq:modelA}
	h(t) = \sum^{N_f}_{k} \underline{A_k} e^{\underline{\omega_{i,k}} t}e^{-i (\underline{\omega_{r,k}} t + \underline{\phi_k})};
\end{equation}
or (ii) a model with only $A_k$ and $\phi_k$ free, but with $\omega_{r,k}$ and $\omega_{i,k}$ fixed at the values corresponding to specific modes $k \in K_{\rm fix}$,
\begin{equation}\label{eq:modelB}
	h(t) = \sum_{k \in K_{\rm fix}} \underline{A_k} e^{\omega_{i,k} t}e^{-i (\omega_{r,k} t + \underline{\phi_k})}.
\end{equation}
The free parameters are underlined in Eqs.~\eqref{eq:modelA} and \eqref{eq:modelB}.
In model~\eqref{eq:modelB}, the fixed values of $\omega_{r,k}$ and $\omega_{i,k}$ for a particular mode $k$ can be computed through standard methods in BH perturbation theory given the remnant mass $M_{\rm rem}$ and dimensionless spin $\chi_{\rm rem}$.
We will call these fixed frequencies the Kerr QNMs frequencies.

While the QNM contribution to the strain decays exponentially over time, the values of $A_k$ and $\phi_k$, as defined in Eqs.~\eqref{eq:modelA} and \eqref{eq:modelB}, are constant.
Unless otherwise specified, we will always measure the time $t$ starting from $t_{\rm peak}$, the time corresponding to the peak strain for the $\ell{,}m = 2{,}2$ multipole of any given simulation.
This is equivalent to defining $t_{\rm peak} \equiv 0$ when using Eqs.~\eqref{eq:modelA} and \eqref{eq:modelB}, or to making a time shift by substituting $t \to t - t_{\rm peak}$.
By doing this, all of the $A_k$ and $\phi_k$ best-fit values are those ``extrapolated back to $t_{\rm peak}$'' regardless of the starting time $t_0$ ($\geq t_{\rm peak}$) of our fitting window.
In other words, we define the amplitude to be $A_k$ instead of $A_k e^{\omega_{i,k} (t_0 - t_{\rm peak})}$, and the phase to be $\phi_k$ instead of $\omega_{r, k} (t_0 - t_{\rm peak}) + \phi_k$.

The perturbation of a Kerr BH spacetime can be decomposed in terms of spin-weighted \textit{spheroidal} harmonics $S_{\ell m}$, which are labeled by polar and azimuthal numbers $\ell$ and $m$, similar to spherical harmonics $Y_{\ell m}$~\cite{Teukolsky:1973ha,Berti:2005gp}.
For a nonrotating (Schwarzschild) BH, the spin-weighted spheroidal harmonics $S_{\ell m}$ reduce to the spin-weighted spherical harmonics $Y_{\ell m}$, but the two are different in general.
In particular, $S_{\ell_1 m_1}$ is not orthogonal to $Y_{\ell_2 m_2}$ as long as $m_1 = m_2$, even if $\ell_1 \neq \ell_2$.

In NR, GWs are usually extracted using a spin-weighted spherical harmonic decomposition~\cite{Boyle:2019kee}.
While we still expect to see the $\ell{,}m{,}n$ Kerr modes in the $\ell{,}m$ harmonic because $S_{\ell_1 m_1}$ overlaps the most with $Y_{\ell_2 m_2}$ if $\ell_1{,}m_1 = \ell_2{,}m_2$, Kerr modes belonging to a different set of $\ell^\prime{,}m^\prime$ values can get mixed into the $\ell{,}m$ harmonic by various means.

In a given $\ell{,}m$ spherical harmonic, we expect to see a rich spectrum of different QNMs belonging to the following categories~\cite{Baibhav:2023clw}:
\begin{itemize}
	\item ``Natural'' linear modes: the $\ell{,}m{,}0$ fundamental mode and the $\ell{,}m{,}n$ overtones with $n \geq 1$.
	      These modes appear naturally in the $\ell{,}m$ harmonic, in the sense that their $\ell{,}m$ number are the same as the underlying harmonic.
	\item Spherical-spheroidal mixing modes: modes with indices $\ell^\prime{,}m{,}n$, $\ell^\prime \neq \ell$, mixed into the $\ell{,}m$ harmonic because $S_{\ell m}$ is not orthogonal to $Y_{\ell m}$~\cite{Berti:2005gp,Buonanno:2006ui,Kelly:2012nd,Berti:2014fga,Ma:2022wpv,Baibhav:2023clw}.
	\item Quadratic modes: These modes arise due to nonlinear coupling between two linear modes, and are labeled as $\ell_1{,}m_1{,}n_1 \! \times \! \ell_2{,}m_2{,}n_2$.
	      They are a direct consequence of the nonlinearity of Einstein's field equations, and their frequencies can be computed with second-order BH perturbation theory~\cite{Ioka:2007ak,Nakano:2007cj,London:2014cma}:
	      \begin{equation}
		      \omega_{\ell_1{,}m_1{,}n_1 \! \times \! \ell_2{,}m_2{,}n_2} = \omega_{\ell_1{,}m_1{,}n_1} + \omega_{\ell_2{,}m_2{,}n_2}.
	      \end{equation}
	      The amplitude of a quadratic mode in a certain $\ell{,}m$ harmonic depends on the Clebsch-Gordan coefficients~\cite{Lagos:2022otp}.
	      The $\ell_1{,}m_1{,}n_1 \! \times \! \ell_2{,}m_2{,}n_2$ Kerr mode is relevant in $m = m_1 + m_2$, most prominently in $\ell{,}m = \ell_1 + \ell_2, m_1 + m_2$~\cite{Ioka:2007ak,Nakano:2007cj,Pazos:2010xf,Lagos:2022otp}.
	\item Retrograde modes:
	      We define these to be the modes that counterrotate in space compared to the \textit{orbital angular momentum of the BBH}.
	      This is different from the more common definition in the literature (see e.g. Refs.~\cite{MaganaZertuche:2021syq,Ma:2022wpv}), which defines them to be those modes that are counterrotating with respect to the \textit{spin of the remnant BH}.
	      We clarify the rationale behind this definition in Fig.~\ref{fig:retro_definition} and Appendix~\ref{app:retro_definition}.
	      Each corotating mode has a retrograde counterpart that we will label with the prefix ``$r$'': for example, the $r2{,}2{,}0$ mode is the retrograde counterpart of the $2{,}2{,}0$ mode.
	      As they have a frequency $\omega_r$ of opposite sign compared to their prograde counterpart, they are counterrotating phasors in the complex plane~\cite{Dhani:2020nik,Dhani:2021vac}.
	      Note that modes with $m < 0$ might have a negative frequency $\omega_r$, but they might \textit{not} be retrograde in the sense of being counterrotating in space compared to the BBH orbital angular momentum.
	      To identify the rotational direction, we multiply $e^{-i \omega_r t}$ by $e^{i m \phi}$, where $\phi$ is the azimuthal coordinate.
	      The mode is only retrograde in our definition if ${\rm sgn}(m) \neq {\rm sgn}(\omega_r)$
	      (see Ref.~\cite{MaganaZertuche:2021syq} for a discussion of this point).
	\item Recoil modes: These are either linear modes with $\ell^\prime \neq \ell$ or quadratic modes with $\ell_1 + \ell_2 \neq \ell$.
	      They exist in the ``wrong'' $\ell{,}m$ harmonic because of imperfections in the choice of the Bondi-van der Burg-Metzner-Sachs (BMS) frame in NR simulations~\cite{Ma:2022wpv,Mitman:2022kwt}.
	      If the modes were extracted by CCE and the BMS frame was fixed appropriately, these modes would be (at least in principle) irrelevant~\cite{MaganaZertuche:2021syq}.
\end{itemize}
Note that the above categories are not mutually exclusive: for example, it is possible to find quadratic recoil modes.

\section{Quasinormal mode extraction procedure}
\label{sec:mode_extraction}

\begin{figure*}
	\centering
	\includegraphics[width=0.99\textwidth]{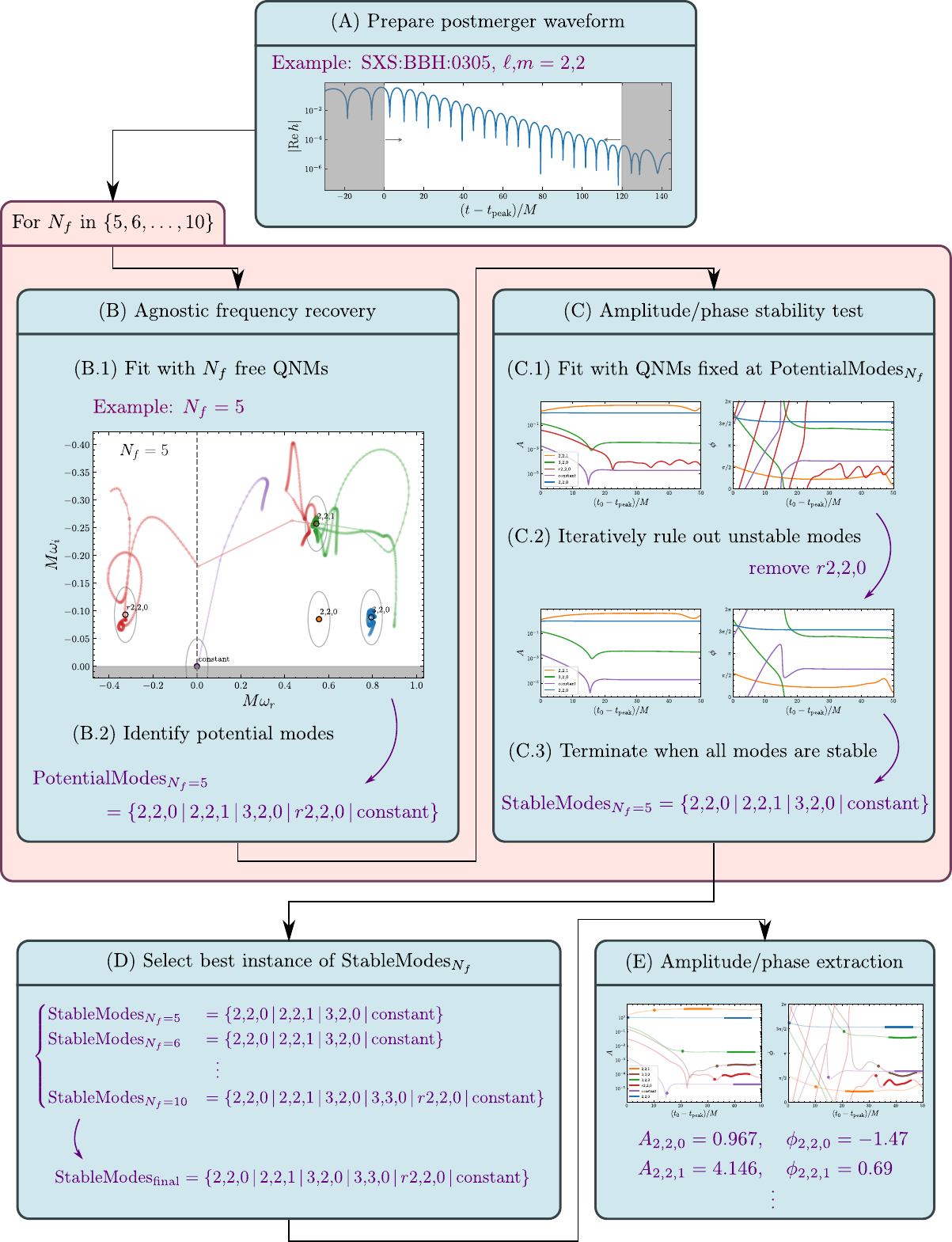}
	\caption{Schematic representation of the procedure we applied to extract QNMs from numerical simulations (see text in Sec.~\ref{sec:mode_extraction}).}
	\label{fig:flowchart}
\end{figure*}

The detection of QNMs is a nontrivial task, due to their exponentially decaying nature.
The difficulty arises both when detecting QNMs in real GW data and when trying to fit NR simulations, which are noise-free (modulo numerical noise).
On one hand, this has stirred some controversy on whether we have already detected multiple overtones in a GW event~\cite{Isi:2019aib,Capano:2021etf,Finch:2022ynt,Cotesta:2022pci,Ma:2023vvr,Siegel:2023lxl,Crisostomi:2023tle}. On the other hand, it led to a better understanding of the physical excitation of higher overtones and of our (in)ability to extract them from NR simulations~\cite{Giesler:2019uxc,Baibhav:2023clw,Nee:2023osy,Zhu:2023mzv}.
In this work we contribute to the ongoing discussion by pursuing the following goals:

\begin{itemize}
	\item[(1)] \textit{Mode finding.} We identify the largest set of Kerr modes that are likely to be present in the postmerger waveform.
	\item[(2)] \textit{Amplitude/phase extraction.} We extract the amplitude and phase of each of the Kerr QNMs thus found.
	\item[(3)] \textit{Ringdown starting time determination.} We determine the time at which each Kerr QNM starts to become stable.
\end{itemize}

For such a procedure to work, it is clear that we must set well-defined (and inevitably, somewhat arbitrary) criteria to decide whether a Kerr mode is present in the waveform, or to determine when the mode is ``stable.''
To do so in a quantifiable manner, we must specify some tolerance values for our QNM extraction procedure.

The three goals stated above are not mutually independent.
For example, missing a certain Kerr mode in the mode-finding stage can bias the amplitudes and phases of the other modes, because the missed Kerr mode is not included in our fitting model.

A schematic representation of the full procedure is illustrated in Fig.~\ref{fig:flowchart} for the SXS:BBH:0305 simulation previously considered in Refs.~\cite{Giesler:2019uxc,Baibhav:2023clw}.
The method is based on the following steps, which are explained in detail in the following subsections:
\begin{enumerate}
	\item[(A)] \textit{Prepare postmerger waveform.}
		We first select a gravitational waveform corresponding to a specific numerical simulation (e.g., SXS:BBH:0305) and a specific multipole (e.g. $\ell{,}m = 2{,}2$).
		We disregard the waveform data before $t_{\rm peak}$, defined to be the time at which $|h_{2{,}2}|$ attains its maximum value.
		This early segment of the waveform corresponds to the premerger signal.
	\item[(B)] \textit{Agnostic frequency recovery.} We identify potential modes in the waveform in a frequency-agnostic manner:
		\begin{enumerate}
			\item[(B.1)] We fit the waveform with $N_f$ free QNMs, with all four real parameters $A, \phi, \omega_r, \omega_i$ left as free parameters to be fitted, as in model~\eqref{eq:modelB}.
				For each $N_f$ we perform fits with a varying starting time $t_0$ to estimate the evolution of the QNM frequency content of the waveform with time.
			\item[(B.2)] We identify potential modes by checking whether the fitted values of $\omega_r$ and $\omega_i$ are consistent with the expected Kerr frequencies, as computed in BH perturbation theory.
		\end{enumerate}
	\item[(C)] \textit{Amplitude/phase stability test.} We assume that all the potential modes are present in the waveform and test their stability in time:
		\begin{enumerate}
			\item[(C.1)] We fit the waveform with a fitting model consisting of all the potential modes fixed at their corresponding $\omega_r$ and $\omega_i$, but with $A$ and $\phi$ as free parameters to be fitted.
			\item[(C.2)] We check whether the fitted amplitudes and phases of the modes are self-consistent (stable) across a specified time window.
				We iteratively remove the modes that fail the test.
			\item[(C.3)] We repeat the check until all the modes pass the test.
		\end{enumerate}
	\item[(D)] \textit{Select the best instance of stable modes.} We repeat steps (B) and (C) for a range of $N_f$ and select the instance where we extract the largest number of stable modes.
		Now we have an exhaustive list of the modes that are present in the postmerger waveform and have stable complex amplitudes.
	\item[(E)] \textit{Amplitude/phase extraction.} We extract the amplitude and phase of each mode within a time window over which the mode is present and its amplitude is stable.
\end{enumerate}

In some steps of the procedure we will define and use certain tolerance criteria to test whether a mode is present in the waveform.
For the reader's convenience, the quantities defining these criteria are listed in Table~\ref{table:definitions}.

\begin{table*}[t!]\label{table:definitions}
	\centering
	\begin{tblr}{
		colspec={ X Q[c,m] Q[c,m]},
		vlines,
		hlines,
		hline{1,2,Z} = {1.25pt,solid},
		vline{1,Z} = {1.25pt,solid},
				vspan=even,
				row{1} = {3em},
				row{2-Z} = {1.7em},
			}
		\SetCell{halign=c,valign=m}Description                                                                                                                                                                                                                                                                                                                                                                                                                                                                                     & Symbol                  & Value                           \\
		Sum of the masses of the BBH progenitors                                                                                                                                                                                                                                                                                                                                                                                                                                                                                   & $M$                     & for normalization               \\
		Mass of the remnant BH                                                                                                                                                                                                                                                                                                                                                                                                                                                                                                     & $M_{\rm rem}$           & simulation dependent            \\
		Magnitude of the dimensionless spin parameter of the remnant BH                                                                                                                                                                                                                                                                                                                                                                                                                                                            & $\chi_{\rm rem}$        & simulation dependent            \\
		Time of maximum strain $|h|$ of the $\ell{,}m=2{,}2$ multipole waveform                                                                                                                                                                                                                                                                                                                                                                                                                                                    & $t_{\rm peak}$          & simulation dependent            \\
		Truncation time at the end of the waveform                                                                                                                                                                                                                                                                                                                                                                                                                                                                                 & $t_{\rm end}$           & $t_{\rm peak} + 120 M$          \\
		Starting time of the fitting window                                                                                                                                                                                                                                                                                                                                                                                                                                                                                        & $t_0$                   & $\{0, 0.1, 0.2, \dots, 50 \} M$ \\
		Number of free modes used for agnostic frequency recovery                                                                                                                                                                                                                                                                                                                                                                                                                                                                  & $N_f$                   & $\{5, 6, 7, 8, 9, 10 \}$        \\
		\SetCell[r=2]{} If $|h_{{\rm peak},\ell{,}m}| > \gamma|h_{{\rm peak},2{,}2}|$, the $\ell{,}m$ component waveform will be fitted, and the $\ell{,}m{,}n$ modes will be included as potential recoil modes when fitting other component waveforms.        For the $\ell = m \leq 6$ waveforms of simulations with number higher than or equal to SXS:BBH:0305, we use $\gamma = \gamma_{\rm low}$. Otherwise, we use $\gamma = \gamma_{\rm hi}$.                                                                             & $\gamma_{\rm hi}$       & $0.02$                          \\
		                                                                                                                                                                                                                                                                                                                                                                                                                                                                                                                           & $\gamma_{\rm low}$      & $0.001$                         \\
		\SetCell[r=4]{} Agnostic frequency recovery: a mode passes the test if any recovered free QNM frequency lies within an ellipse of width $2\alpha_r$ and height $2\alpha_i$ centered at the expected frequency of the mode $p_{\rm ag}\%$ of the time within any continuous time window $(t_0, t_0 + \tau_{\rm ag})$                                                                                                                                                                                                        & $\alpha_r$              & $0.05$                          \\
		                                                                                                                                                                                                                                                                                                                                                                                                                                                                                                                           & $\alpha_i$              & $0.05$                          \\
		                                                                                                                                                                                                                                                                                                                                                                                                                                                                                                                           & $\tau_{\rm ag}$         & $10 M$                          \\
		                                                                                                                                                                                                                                                                                                                                                                                                                                                                                                                           & $p_{\rm ag} \%$         & $95 \%$                         \\
		\SetCell[r=6]{} Mode stability test:    $\Delta_{k, \rm stable}(t_0; \tau_{\rm stable}) = \sqrt{(\beta_{A}\delta A_k/A^\prime_k)^2 + (\beta_{\phi}\delta \phi_k/2\pi)^2}$, where $\delta A_k$ and $\delta \phi_k$ are the $p_{\rm stable}$ interpercentile range of $A_k$ and $\phi_k$ within the time window $(t_0, t_0 + \tau_{\rm stable})$, and $A^\prime_k = {\max}\left(A_k, A_{\rm tol} \right)$. A mode passes the test if $\Delta_{k, \rm stable}(t_0; \tau_{\rm stable}) < \epsilon_{\rm stable}$ for any $t_0$. & $\epsilon_{\rm stable}$ & {weak: $0.4$,                   \\ normal: $0.2$, \\ strong: $0.1$} \\
		                                                                                                                                                                                                                                                                                                                                                                                                                                                                                                                           & $\beta_A$               & 1.0                             \\
		                                                                                                                                                                                                                                                                                                                                                                                                                                                                                                                           & $\beta_\phi$            & 1.5                             \\
		                                                                                                                                                                                                                                                                                                                                                                                                                                                                                                                           & $A_{\rm tol} $          & {weak: $10^{-3}$,               \\ normal: $10^{-3}$, \\ strong: $10^{-7}$}       \\
		                                                                                                                                                                                                                                                                                                                                                                                                                                                                                                                           & $\tau_{\rm stable}$     & $10 M$                          \\
		                                                                                                                                                                                                                                                                                                                                                                                                                                                                                                                           & $p_{\rm stable} \%$     & $95 \%$                         \\
	\end{tblr}
	\caption{Definitions of some of the symbols used in this paper.}
\end{table*}

\subsection{Prepare postmerger waveform}\label{subsec:prepare}

In this work, we will consider publicly available waveforms from two catalogs: the SXS waveform catalog~\cite{Boyle:2019kee} and the Ext-CCE catalog. The SXS catalog contains NR simulations of BBH mergers extracted at null infinity through extrapolation, while the Ext-CCE catalog benefits from improvements such as center-of-mass corrections, improved Cauchy-characteristic evolution, the inclusion of gravitational memory using BMS balance laws, and improved wave extraction~\cite{Woodford:2019tlo,Moxon:2020gha,Mitman:2020bjf,Iozzo:2020jcu}.
The SXS waveforms are labeled as SXS:BBH:XXXX, and the Ext-CCE waveforms are labeled as SXS:BBH\_ExtCCE:XXXX, where XXXX is a four-digit number.
Each number corresponds to a specific numerical simulation of a BBH merger with different sets of intrinsic parameters (but, in general, simulations with the same label XXXX do not refer to simulations with the same intrinsic parameters in the SXS and Ext-CCE catalogs).
The waveforms available are decomposed in spin-weighted spherical harmonics, so each number corresponds to a set of waveforms decomposed in different multipolar components labeled by $\ell$ and $m$.

For a particular simulation, we will separately fit each multipolar component.
Once we have chosen the simulation and the multipolar indices, we determine the peak merger time $t_{\rm peak}$ where the absolute value of the strain of the dominant $\ell{,}m=2{,}2$ multipole has a maximum.
In what follows, we will be fitting the waveform within a time window that starts no earlier than $t_{\rm peak}$, so the waveform at $t < t_{\rm peak}$ will not be relevant for our analysis.
Certain harmonics in some simulations become noisy at late times, so we always trim the waveform at $t - t_{\rm peak} \gtrsim 120 M$.
We find that the late time cut-off of the fitting window does not significantly affect the results of our analysis, because the modes decay exponentially in time.

The SXS catalog lists waveform data for many multipolar components $\ell{,}m$ (up to $\ell \sim 8$).
However, in our analysis, we will only fit the multipolar component waveforms that have a sufficiently large peak strain: $|h_{{\rm peak},\ell{,}m}| > \gamma|h_{{\rm peak},2{,}2}|$, where $\gamma = 0.02$ or $0.001$.
This is because waveforms with a low strain are more affected by contaminations due to mode-mixing and numerical noise, and they are less relevant for data analysis anyway.
The quality of the higher $\ell{,}m$ components is typically better for more recent simulations.
Therefore, we use $\gamma = \gamma_{\rm low} =  0.001$ if the simulation number is SXS:BBH:XXXX $> 0305$ and $\ell = m \leq 6$.
Otherwise, we use $\gamma = \gamma_{\rm hi} = 0.02$.
We reduce the tolerance only for the $\ell = m \leq 6$ components because they are often more important, and we do not want to include too many other $\ell{,}m$ components due to computational costs.

For some of the next steps of the mode extraction procedure, we will have to compute the theoretically predicted QNM frequencies for specific values of $\ell{,}m{,}n$.
These are easily computed~\cite{Berti:2009kk,RDwebsites} from the mass $M_{\rm rem}$ and spin $\chi_{\rm rem}$ of the remnant BH, which are listed in the publicly available metadata file of each simulation.
We will use the \texttt{qnm} package~\cite{Stein2019} to compute the frequencies within our procedure.

\subsection{Agnostic frequency recovery} \label{subsec:agnostic}

In this subsection, we describe the procedure to identify potential modes in the waveform in a frequency-agnostic manner.

\subsubsection{Agnostic frequency fit}

\begin{figure*}
	\centering
	\includegraphics[width=0.99\textwidth]{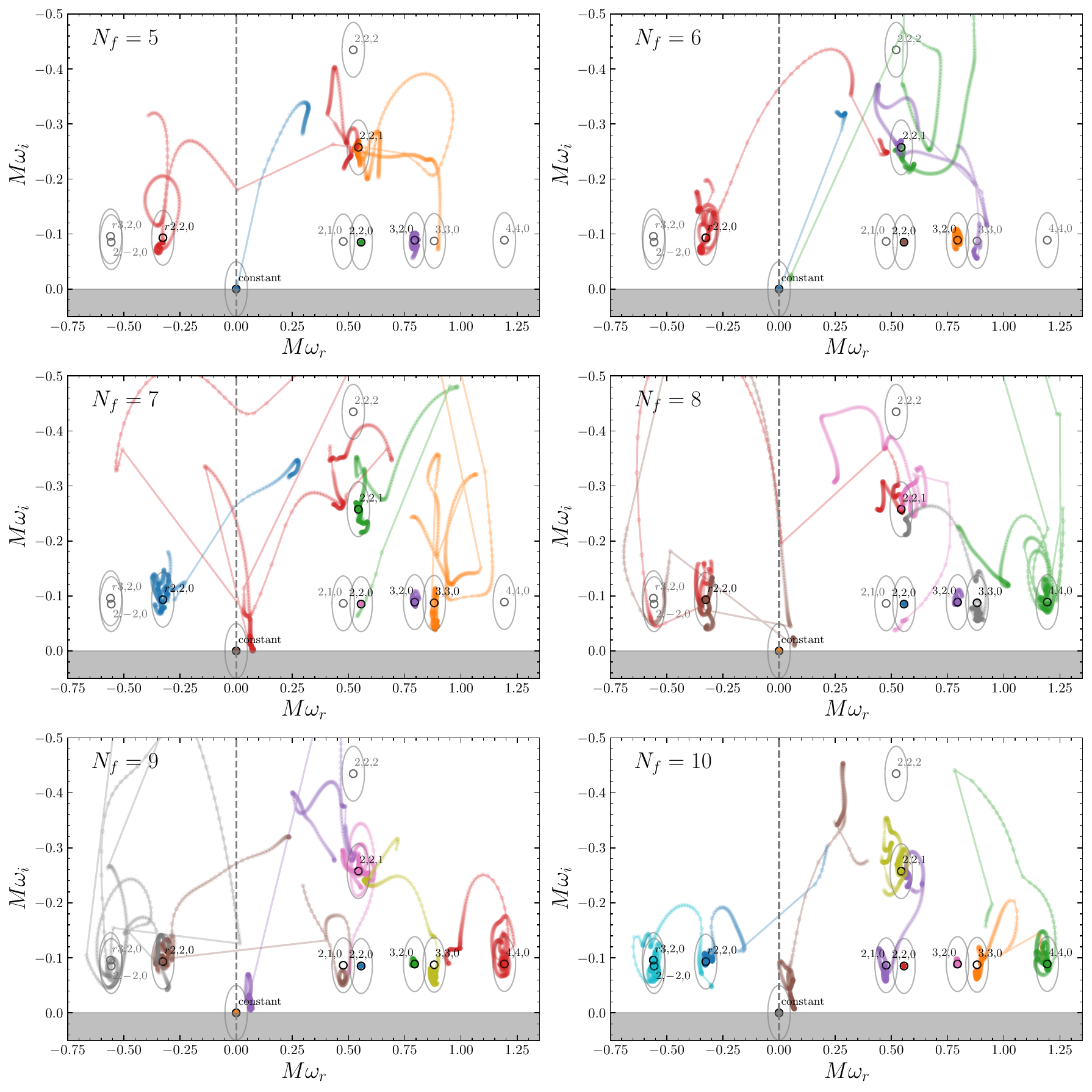}
	\caption{
		Agnostic fit results of the complex QNM frequencies for the $\ell{,}m = 2{,}2$ multipole of the SXS:BBH:0305 waveform.
		Each panel corresponds to a different number $N_f$ of free QNMs (labeled in the top-left corner) used in the fitting model~\eqref{eq:modelA}.
		Dots of a particular color correspond to the evolution of a particular free mode over time, i.e. when we vary the starting time of the fitting window from $(t_0 - t_{\rm peak}) / M = 0$ to $50$.
		Relevant Kerr modes are shown and labeled as gray circles, with those consistent with the agnostic fits shown in black.
		The gray ellipses around each circle are the tolerance zones of each mode.
		The gray vertical dashed line divides the complex plane into a positive-frequency region ($\omega_r > 0$) and a negative-frequency one ($\omega_r < 0$).
		We show the region $\omega_i > 0$ in gray as it corresponds to exponentially growing modes, which should be absent.
	}
	\label{fig:agnostic_fit}
\end{figure*}

We begin by fitting the waveform using $N_f$ free QNMs and all four real parameters $A, \phi, \omega_r$, and $\omega_i$ as free parameters to be fitted, as given by model~\eqref{eq:modelA}.
To capture the evolution of the free QNM frequencies over time, we vary the starting time of the fitting window from $(t_0 - t_{\rm peak})/M=0$ to $50$ with a time step of $\Delta t = 0.1 M$, while keeping the ending time of the window fixed at $t_{\rm end} - t_{\rm peak} = 120 M$.
We consider up to $N_f = 10$ free modes, which corresponds to $40$ real fitting parameters in model~\eqref{eq:modelA}.
To reduce the computational cost of the high-dimensional fit, we employ the \texttt{JaxFit} package~\cite{2022arXiv220812187H}, which implements a trust region method for nonlinear least-squares fitting.
The implementation is similar to that in the \texttt{curvefit} method of the \texttt{optimize} sub-package in \texttt{scipy}, but with auto-differentiation and hardware acceleration implemented with \texttt{Jax}~\cite{jax2018github}.
The technical details of the fitting algorithm and the relevant hyperparameters are described in Appendix~\ref{app:fitter}.
In Appendix~\ref{app:consistency}, we test the fitting algorithm and show that (with some caveats) it works well enough for our purposes.

In Fig.~\ref{fig:agnostic_fit} we show the results of the frequency-agnostic fit for $N_f \in {5, 6, \dots, 10}$ using the $\ell{,}m = 2{,}2$ multipole of the SXS:BBH:0305 waveform.
By visually inspecting the evolution of the frequencies, we can already identify some of the Kerr modes that have been captured by the agnostic fit.
Notably, the $2{,}2{,}0$, $2{,}2{,}1$, $3{,}2{,}0$, and constant modes are consistently approached by at least one free QNM for any value of $N_f$ we consider, strongly hinting at the presence of these Kerr modes in the waveform.
In the following subsection, we will employ more precise criteria based on the results of the agnostic fit to determine the presence of a Kerr mode.

By using as many as $N_f = 10$ free QNMs we could be overfitting the waveform, in the sense that some of the QNMs might be fitting away signal components unrelated to the ringdown (such as the prompt response, GW memory, back-scattering tails, or numerical noise).
Indeed, as shown in Appendix~\ref{app:consistency}, with $N_f \gtrsim 7$ the fit mismatch can be orders of magnitude {\em below} the estimated numerical noise floor.
At this step of the mode extraction procedure, our main goal is to exhaustively identify all of the potential QNMs that may be present in the waveform: we do our best not to miss any significant QNMs that have relatively higher amplitudes and are longer-lived, at the cost of possibly including spurious ones, so we \textit{choose} to overfit.

As shown in the $N_f = 10$ panel of Fig.~\ref{fig:agnostic_fit}, there could be as many as $9$ potential QNMs in the waveform, at least for the $\ell{,}m = 2{,}2$ multipole of the SXS:BBH:0305 waveform. Therefore using a value of $N_f$ as large as $10$ might be necessary, in general, if we want to do our best at extracting QNMs.
By inspecting the different panels, the $N_f = 10$ results are arguably the cleanest, with the free QNMs converging better to the known Kerr frequencies.
Also, while other nonQNM contaminations could be present in the waveform, this does not necessarily mean that they would prevent the extraction of QNMs, even if their amplitudes are comparable to the QNMs.
For example, Gaussian noise will not fatally impede a least-squares fitting algorithm, and the memory effect might be approximately mitigated by fitting away a constant at late times.
Moreover, if a free QNM is fitting away a nonQNM component, it should not linger around the same frequency in the complex plane over time;
conversely, if a free QNM lingers around the same Kerr QNM frequency over time, the component of the waveform it is fitting is then (by definition) a Kerr QNM.

Even if this stage of the fitting procedure were to pick up any spurious Kerr QNMs, at a later step we will use a more stringent criterion to rule them out.

\subsubsection{Potential mode identification}\label{subsec:potential_modes}

We now examine the evolution of free QNMs and check whether the results indicate the presence of certain Kerr QNMs in the waveform.

We first need to decide, a priori, what Kerr modes are relevant to our analysis, or what modes we expect to find.
For a strictly agnostic analysis, we could in principle deem all of the linear and quadratic modes with all possible combinations of indices ($\ell$, $m$ and $n$) to be relevant, regardless of the $\ell{,}m$ harmonic of the waveform we are considering.
However, we expect that most of these modes will be negligible, especially when they are recoil modes.
Therefore, while we deem all natural modes and spherical-spheroidal mixing modes relevant (including the retrograde modes within these categories), we only include recoil modes such that their natural-host component $\ell{,}m$ has a peak strain $|h_{{\rm peak},\ell{,}m}| > \gamma|h_{{\rm peak},2{,}2}|$. This is the same requirement used for deciding which waveform components to fit, as explained in Sec.~\ref{subsec:prepare}.
In this way we ensure that we include the recoil modes that could potentially be present in the multipolar component in question because they could have a significant amplitude in their natural-host component.

Given a list of relevant Kerr QNM frequencies, we can determine their proximity to the free QNMs to check whether some of them are consistent with the agnostic fit.
We do this by drawing an ellipse on the complex plane around each relevant mode frequency and checking whether any free QNM stays within the ellipse consistently (i.e., at least $p_{\rm ag}\%$ of the time, where ``$\rm ag$'' stands for ``agnostic'') in any continuous time window $t_0 \in (t, t+\tau_{\rm ag})$.
For example, if we choose $\tau_{\rm ag} = 10 M$ and $p_{\rm ag}\% = 95\%$, and a free QNM stays within the ellipse around the complex frequency $\tilde{\omega}_{2{,}2{,}0}$ for at least $95\%$ of the time between, say, $t_0 - t_{\rm peak} \in (16.7, 26.7) M$, then we say that the $2{,}2{,}0$ mode has passed the frequency-agnostic test.
We choose the 95\% criterion to avoid missing certain modes because of short fluctuations in the fitting results.

When we fit across different values of $t_0$, we use the results of the previous time step ($t_0 - \Delta t_0$) as an initial guess for the fitting algorithm. In Fig.~\ref{fig:agnostic_fit}, we label the corresponding frequencies (a ``stream'' of free QNM frequencies connected by iterative initial guesses) across different $t_0$ with the same color.
However, at a specific fit starting time $t_0$, as seen in model~\eqref{eq:modelA}, all free QNMs have the same form and contribute equally, meaning that we could very well shuffle the colors of different points corresponding to the same $t_0$:
when analyzing the results, at a specific $t_0$ we do not care about the color of each free mode (nor about the colors of the modes in a neighboring time step).
Therefore, in the proximity test for a relevant Kerr mode we do not require that the mode has to be approached by a free mode of the same color.
In other words, we do not care whether the mode ``switched color'' when it is within the ellipse.
In fact, this ``color switching'' behavior is frequently observed for overtones, as shown in Fig.~\ref{fig:agnostic_fit}, where the $2{,}2{,}1$ Kerr mode is often approached by modes with different colors.

We now have a list of potential Kerr modes, which will be subject to further screening in the next steps.

\subsection{Amplitude/phase stability test}\label{subsec:stability_test}

\begin{figure*}
	\centering
	\includegraphics[width=0.99\textwidth]{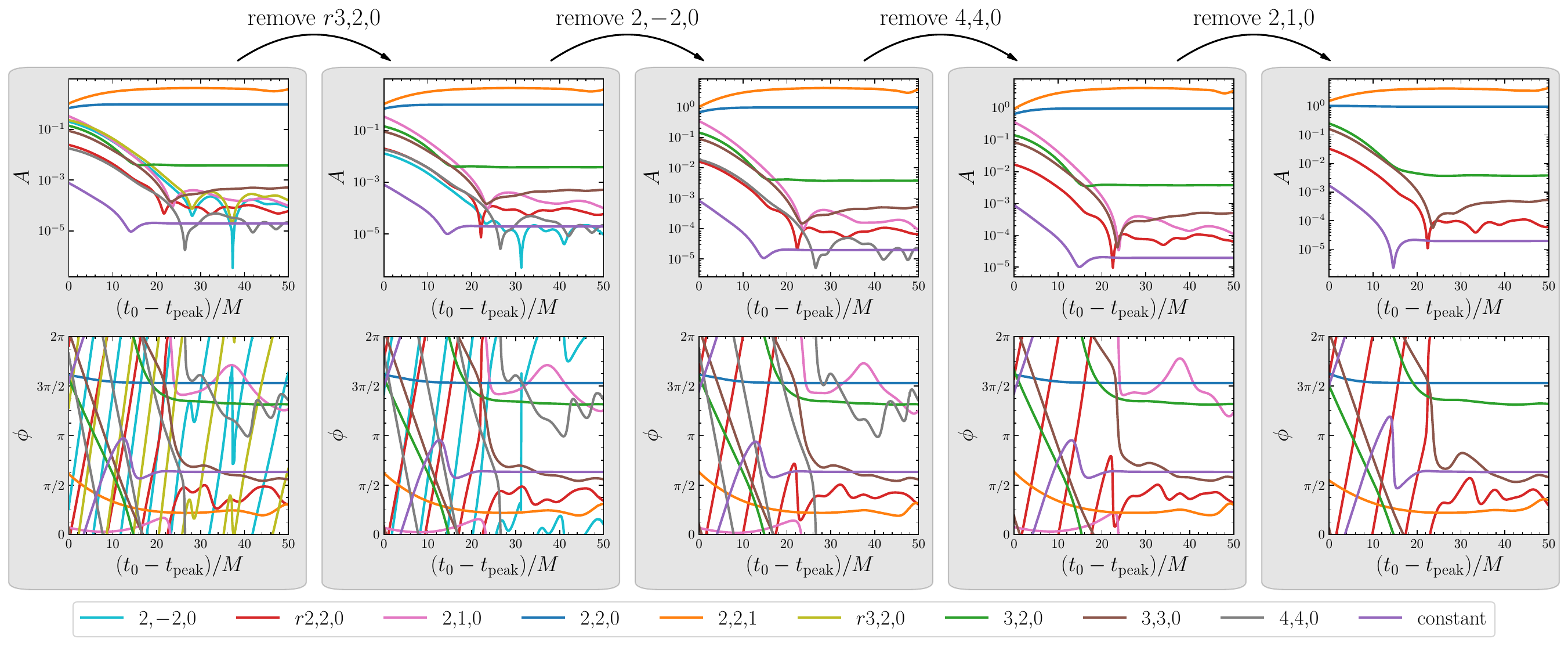}
	\caption{
		Amplitude (top row) and phase (bottom row) stability test for the $\ell{,}m = 2{,}2$ multipole of the SXS:BBH:0305 waveform.
		In the leftmost panel we fit the waveform with all of the potential modes identified in the frequency-agnostic fit (see legend at the bottom).
		Moving from the left to the right panels, we iteratively remove the Kerr mode whose amplitude and phase fluctuate the most over time, until (in the rightmost panel) all modes are stable within the tolerance specified.
	}
	\label{fig:flatness_check}
\end{figure*}

The frequency-agnostic test presented above checks whether a Kerr QNM \textit{frequency} $\tilde{\omega} = \omega_{r} + i \omega_{i}$ is consistently present, but a QNM also consists of two other parameters, $A$ and $\phi$.
A QNM is defined to be given by Eq.~\eqref{eq:QNM} with $A, \phi, \omega_r$ and $\omega_i$ \textit{all being constant in time.}
A nonQNM component could, in principle, mimic a QNM by giving the same fitted frequency, but with time-evolving values of $A$ and $\phi$.
Thus, an additional test is required to check whether $A$ and $\phi$ are stable (i.e., constant) in time.

At this point, there are two routes we can take.
The first is to examine the fitted $A$ and $\phi$ of the free QNMs in the frequency-agnostic fits and check for stability, and the second is to do a \textit{frequency-fixed} fit -- as in model~\eqref{eq:modelB} -- with frequencies chosen to match the potential Kerr modes identified by the frequency-agnostic test.
We will undertake the second route for two reasons.

The first reason is that the frequency-agnostic fits could contain more free modes than the potential Kerr modes.
The values of $A$ and $\phi$ for a free QNM that picked up a Kerr QNM might be affected by other free QNMs that were ``greedily'' overfitting nonQNM components.
One might wonder whether $\omega_r$ and $\omega_i$ of a free QNM approaching a Kerr mode could also be affected by the ``greedy'' QNMs.
In fact, as seen in Fig.~\ref{fig:agnostic_fit}, increasing $N_f$ does not cause $\omega_r$ nor $\omega_i$ to deviate from a potential Kerr mode: the $2{,}2{,}0, 2{,}2{,}1, 3{,}2{,}0, 2{,} \! - \! 2{,} 0$ and constant modes are always approached by a free QNM, even if some other free QNM is not picking up any Kerr mode (see e.g. the brown mode in the $N_f = 10$ panel).
This holds rather generally for all of the waveforms we examined.

The second reason is that in model~\eqref{eq:modelB} we fix $\omega_r$ and $\omega_i$, reducing the number of free parameters, and therefore making it easier to examine the behavior of $A$ and $\phi$ of a Kerr mode.

We fit the waveform with model~\eqref{eq:modelB}, fixing $\omega_{r,k}, \omega_{i,k}$ with $k \in K$, where $K$ is the set of potential Kerr modes, again over a wide range of starting times $t_0$.
We then test whether $A$ and $\phi$ are both stable (i.e., they fluctuate less than a certain tolerance) within a time window $t_0 \in (t, t+\tau_{\rm stable})$.
We quantify the fluctuation by a weighted quadrature sum of the fluctuations $\delta A$ and $\delta \phi$,
\begin{equation}\label{eq:delta_stable}
	\Delta_{k, \rm stable}(t; \tau_{\rm stable})   = \sqrt{\left(\frac{\beta_{A}\delta A_k}{A^\prime_k}\right)^2 + \left(\frac{\beta_{\phi}\delta \phi_k}{2 \pi}\right)^2},
\end{equation}
where $\delta A_k$ and $\delta \phi_k$ are defined to be the $p_{\rm stable}$-percentile range of $A_k$ and $\phi_k$ within the time window, and $A_k^\prime = {\max}\left(A_k, A_{\rm tol} \right)$.
If $\Delta_{k, \rm stable}(t) < \epsilon_{\rm stable}$ for some $t$, then we conclude that the potential mode is stable, and it is likely to be present in the waveform.
If a subset of modes $k \subseteq K$ is unstable, then we eliminate the mode with the highest $\min_t \Delta_{k, \rm stable}(t)$ and redo the frequency-fixed fits with the remaining modes.
The $A_k^\prime$ term encourages the algorithm to pick up subdominant modes with low amplitudes $A_k \lesssim A_{\rm tol}$. Effectively, it penalizes them less for a fluctuating relative amplitude, while still screening them based on the fluctuation of their phase.

We choose to eliminate only one mode at a time because a spurious Kerr mode could interfere with a robust Kerr mode and destabilize it, especially if the two modes have similar frequencies (e.g., $3{,}2{,}0$ and $3{,}3{,}0$).
By eliminating the most unstable mode first, which is likely the spurious mode, the robust mode will likely become stable in the next iteration and it will not be eliminated.
As the natural fundamental mode of any given multipolar component must be present in the waveform, we will never eliminate it with this test, even if its amplitude fluctuates more than the tolerance.
We iterate the procedure until all of the modes are stable.
At this point we have a list of modes that are likely to be present in the waveform.

\subsection{Finalizing stable modes} \label{subsec:finalize}

Steps (\hyperref[subsec:agnostic]{B}) and (\hyperref[subsec:stability_test]{C}) (agnostic frequency recovery and amplitude/phase stability tests) of the procedure are done in parallel for different values of $N_f$.
We choose to do multiple runs with different values of $N_f$ (instead of doing a single run with large $N_f$, say $N_f = 10$) to make sure that we will not miss certain modes because of overfitting the waveform with too many free QNMs.
Intuitively, there should be an optimal $N_f$ for a given waveform such that including fewer modes will cause underfitting, and including more causes overfitting.
Therefore, across all the results with different $N_f$, we select the instance that returns the longest list of potential modes to be our final result.
In practice, we find that we rarely find fewer modes when we increase $N_f$ (at least for $5 \leq N_f \leq 10$), meaning that overfitting the waveform seldom causes us to miss a mode.
Nevertheless, we still include the lower $N_f$ results to be safe, and because they are less costly in terms of computational resources.

There are alternatives to the above strategy.
For example, one could have chosen instead the union of all the stable modes returned by the stability test for all $N_f$.
However, the correct extraction of a certain mode could depend also on the extraction of the other modes.
For example, if for a certain $N_f$ we identify the Kerr modes $\{2{,}2{,}0 | 3{,}3{,}0\}$, and for a larger $N_f$ we find the set $\{2{,}2{,}0 | 2{,}2{,}1 | 3{,}2{,}0\}$, the union of the two will be $\{2{,}2{,}0 | 2{,}2{,}1 | 3{,}2{,}0 | 3{,}3{,}0\}$,
but it might very well be the case that the lower $N_f$ run was mistaking the $3{,}2{,}0$ mode as its neighbor $3{,}3{,}0$ because it missed the $2{,}2{,}1$ mode.
Our strategy is more conservative and avoids this problem.
Another alternative would be to do the frequency-agnostic fits for all different $N_f$, identify the potential modes in each run, take the union between all runs, and do one single stability test, iteratively removing unstable modes from the joint list of QNMs.
This raises concerns similar to those highlighted for the previous alternative strategy, so we do not use this method.

This step concludes the ``mode finding'' portion of the procedure.
In principle, the ringdown waveform could contain infinitely many QNMs with arbitrary mode numbers.
The main goal of our algorithm is to identify as many modes as possible, trying our best not to miss significant modes (i.e., those with a higher amplitude or those that are longer living) without including spurious modes that should have been undetectable.
Indeed, our algorithm could miss certain modes, especially those with a high overtone number $n$: see Appendix~\ref{app:procedure}.

\subsection{Amplitude/phase extraction}

\begin{figure}
	\centering
	\includegraphics[width=0.49\textwidth]{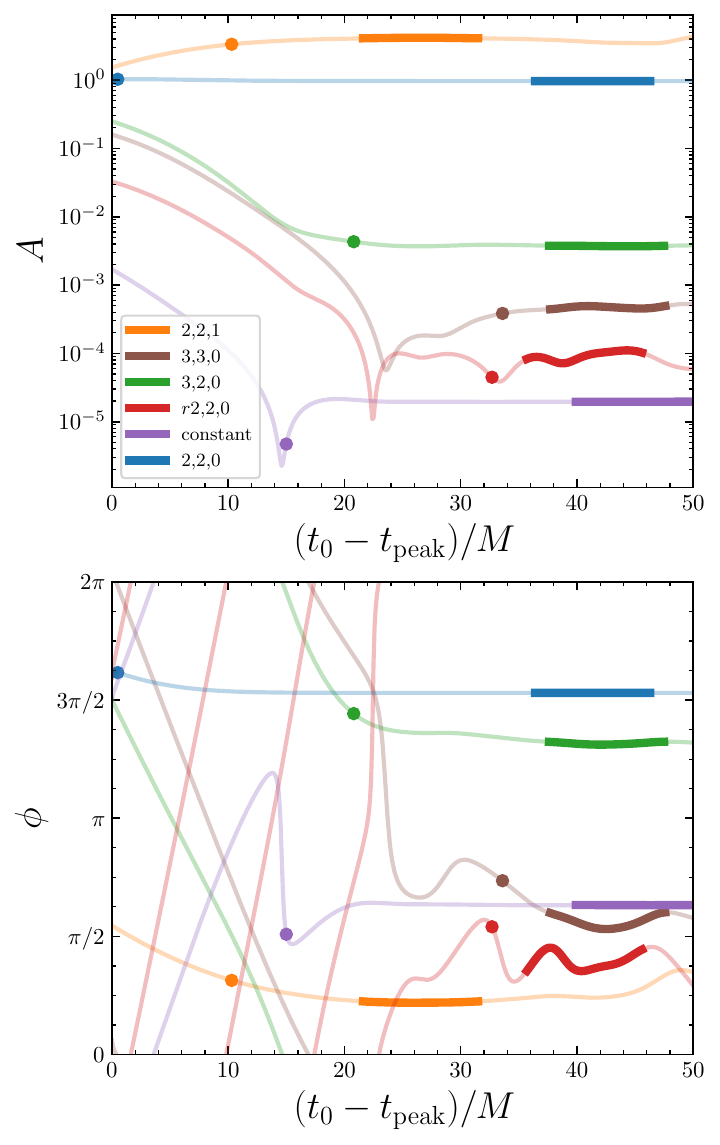}
	\caption{
		Amplitude (top) and phase (bottom) extraction for SXS:BBH:0305, $\ell{,}m = 2{,}2$.
		$A_k$ and $\phi_k$ are shown as bold lines within the $\tau_{\rm stable}$ time window when $\Delta_{k, \rm stable}$ is minimal, which is where the amplitudes and phases are extracted.
		We also label the time $t_{k, \rm start}$ with a solid circle.
	}
	\label{fig:flat_extraction}
\end{figure}

Given a set of Kerr modes $K$ that are likely to be present in the waveform, we would like to extract $A_k, \phi_k$ for all $k \in K$, so that we can specify completely the QNM content in the waveform.
This is a nontrivial task because $A_k$ and $\phi_k$ depend on the chosen fitting window.
As explained previously, the ending time $t_{\rm end}$ of the fitting window does not sensitively affect the fit. However, as can be seen in Fig.~\ref{fig:flatness_check}, the starting time $t_0$ does, so we should write the extracted amplitude and phase as $A_k(t_0)$ and $\phi_k(t_0)$.
If the fit starts too early ($t_0$ is too small), the results will be systematically biased because the GW response has not stabilized into constant-amplitude/phase QNMs due to a number of reasons, including the time-evolving background, delayed build-up of QNM amplitudes, prompt radiation, and nonlinear effects~\cite{Baibhav:2023clw,Zhu:2023mzv}.
If the fit starts too late, fast-decaying or small-amplitude modes can be missed because they dive beneath the noise floor.

Given these caveats, a simple way ahead is to extract $A_k$ and $\phi_k$ at times such that the mode $k$ is most stable.
We can find, for each mode, the time $t = t_{\rm best}$ that minimizes the quantity $\Delta_{k, \rm stable}(t; \tau_{\rm stable})$ defined in Eq.~\eqref{eq:delta_stable}.
Then, we can extract $A_k$ and $\phi_k$ as the median values within the time window $t_0 \in (t_{\rm best}, t_{\rm best} + \tau_{\rm stable})$.
We can also quantify the fluctuation of the mode, e.g. by the center $95$-percentile of $A_k$ and $\phi_k$ within the window (we do not use the maximum and minimum values within the window to quantify the fluctuation because we want to leave some margin for large sudden fluctuations in $A_k$ and $\phi_k$ due to nonconvergence, which rarely happens).
We show an example of this method in Fig.~\ref{fig:flat_extraction}.
Note that while we extract the amplitude and phases at times later than $t_{\rm peak}$, we always define their values to be those extrapolated back to $t_{\rm peak}$.

We now know which Kerr modes are likely to be present in the waveform, and with what amplitude and phase.
However, as we have shown explicitly, QNMs are only a valid description of the postmerger waveform (modulo memory effects, noise and backscattering) at late times.
To fully specify our model, we must identify an optimal starting time.
We can define the starting time of a particular mode to be the time at which it first becomes stable, i.e., the earliest time $t = t_{k, \rm start}$ for which $\Delta_{k, \rm stable} < \epsilon_{\rm stable}$.
This can be interpreted as the time at which we confidently believe the mode to be present in the waveform.
It can also be extracted from the frequency-fixed fits, which are labeled as filled circles in Fig.~\ref{fig:flat_extraction}.
Note in particular that $t_{k, \rm start} \neq t_{k, \rm best}$: the former is the time at which the mode \textit{first} becomes stable (marked by filled circles in Fig.~\ref{fig:flat_extraction}), while the latter is the starting time of the fitting window that gives the \textit{best} stability (the starting time of the bold lines in Fig.~\ref{fig:flat_extraction}).
We will return to a more detailed discussion of the definition of the starting time of the mode in Sec.~\ref{sec:tstart} below.

We stress once again that the results for a particular Kerr mode depend on which other modes are included in the fit.
Therefore, the amplitude/phases and starting times extracted from the fit refer to those obtained \textit{by fitting the whole waveform with a specific set of modes}.
For example, while the starting time of the $2{,}2{,}1$ mode in Fig.~\ref{fig:flat_extraction} is $t_0 - t_{\rm peak} \sim 10 M$, we should expect it to start even later if we were to use a fitting model including (say) only the $2{,}2{,}0$ and $2{,}2{,}1$ modes, as often done when analyzing real GW data, because missing other modes will hinder the stability of the fit.

\section{Hyperfits of the amplitudes}
\label{sec:hyperfit}

The simplicity of BH spectroscopy is largely due to the no-hair theorem.
As the remnant BH can be characterized fully by its final mass and dimensionless spin $M_{\rm rem}$ and $\chi_{\rm rem}$, its QNM frequencies do not depend on the initial conditions of the BBH merger, i.e., on the properties of the merger remnant progenitors.
However, the degree of excitation of different QNMs varies depending on the initial conditions. The amplitude and phase of each QNM are functions of the parameters of the progenitors, including their spins $\vec{\chi}_1$, $\vec{\chi}_2$ and the binary mass ratio $q = m_1 / m_2$. Here we choose the convention that $m_1 \geq m_2$ (so that $q \geq 1$).

Whether or not one cares about $A$ and $\phi$ of a QNM depends on their goal.
If one plans merely to test the no-hair theorem in a restrictive sense, then checking the consistency between the different QNM frequencies $\tilde{\omega}$ may suffice.
However, given a set of initial BBH parameters, $A$ and $\phi$ can be predicted in a well-defined manner within GR. More robust spectroscopy tests should also check the consistency of $A$ and $\phi$ with the predictions of GR~\cite{Forteza:2022tgq}.

To do this we must build a model of $A$ and $\phi$ as a function of the parameters of the binary progenitors. This is the goal of this section.
We will restrict our attention to nonprecessing waveforms, i.e., simulations where both progenitor BHs have spins either parallel or antiparallel to the direction of the orbital angular momentum.
These simulations are parameterized by $q, \chi_1$ and $\chi_2$, where $\chi_1 \equiv \chi_{1, z} = \pm |\vec{\chi_1}|$, where a positive (negative) sign means that $\vec{\chi_1}$ is parallel (antiparallel) to the orbital angular momentum of the binary, and similarly for $\chi_2$.

It will also be useful to use parameters that appear often in a post-Newtonian (PN) expansion of BBH dynamics:
\begin{align}
	\eta   & = \dfrac{q}{(1 + q)^2},             \\
	\chi_+ & = \dfrac{q \chi_1 + \chi_2}{1 + q}, \\
	\chi_- & = \dfrac{q \chi_1 - \chi_2}{1 + q}.
\end{align}
Note that $\chi_+$ is also called $\chi_{\rm eff}$ in the literature.

\begin{figure*}
	\centering
	\includegraphics[width=0.8\textwidth]{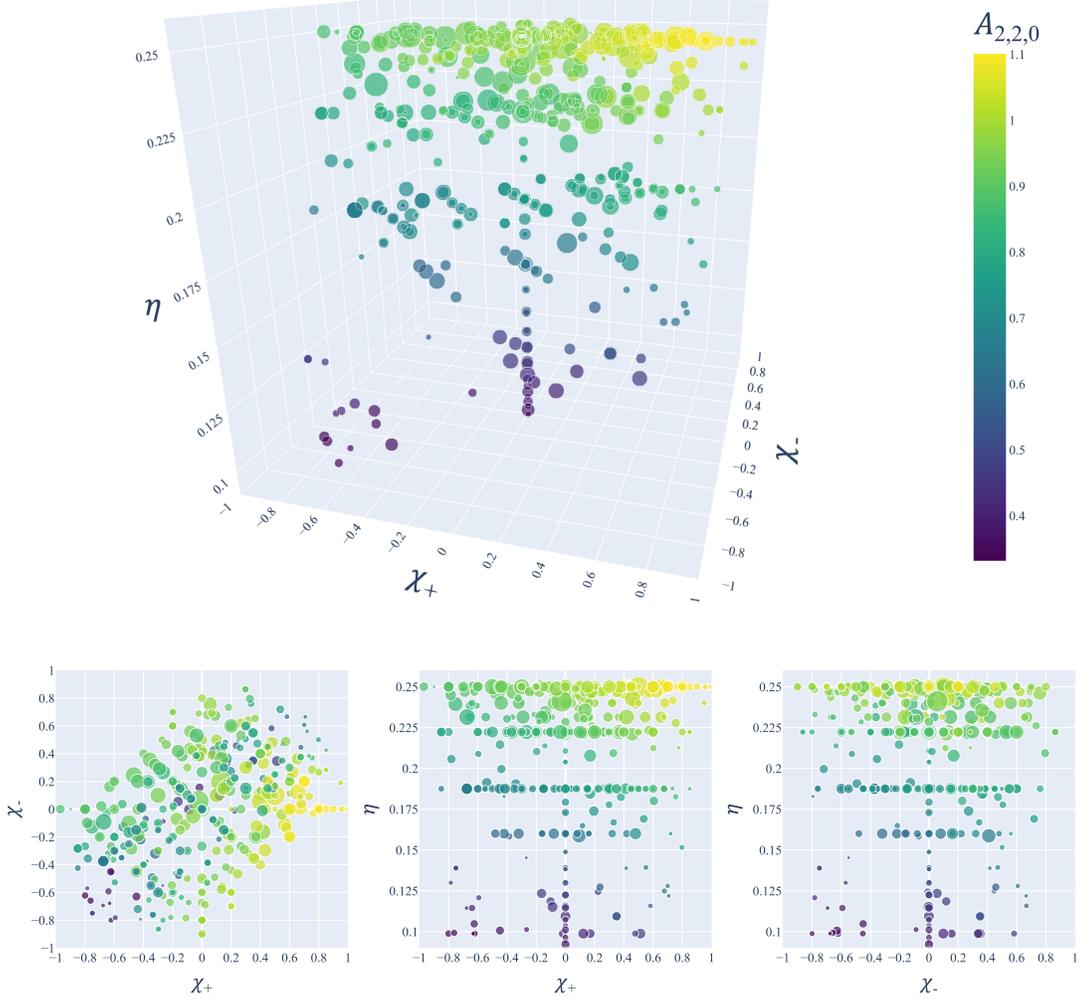}
	\caption{
	The amplitudes of the $2{,}2{,}0$ mode in the $\ell{,}m = 2{,}2$ harmonic waveform of all fitted nonprecessing SXS simulations.
	Top: the amplitudes visualized in the three-dimensional parameter space of $\eta$, $\chi_+$ and $\chi_-$.
	Each ball represents an SXS simulation at the corresponding location in parameter space, the color of the ball corresponds to the magnitude of $A_{2{,}2{,}0}$ extracted from that simulation.
	The size of the ball is proportional to the inverse of the fluctuation $\Delta_k$ of the extracted amplitude;
	in other words, the larger balls represent amplitudes we measured more confidently, hence they will be weighted heavier when performing the hyperfits.
	Bottom row: same as the top plot, but projected onto 2D parameter spaces, each panel being the space spanned by a selected pair of the three parameters $\eta$, $\chi_+$ and $\chi_-$.
	Interactive versions of this plot and similar plots for other modes can be found on the \texttt{jaxqualin} webpage~\cite{jaxqualin}.
	}
	\label{fig:3D_A_220}
\end{figure*}

For each simulation, we run the full mode-extraction algorithm presented in Sec.~\ref{sec:mode_extraction}.
This yields data points for the amplitudes of different modes over the whole parameter space $(\eta, \chi_+, \chi_-)$.
In Fig.~\ref{fig:3D_A_220} we visualize the amplitude of the $2{,}2{,}0$ mode in this three-dimensional parameter space.
Interactive versions of the plots for this and other QNMs can be found on the \texttt{jaxqualin} webpage~\cite{jaxqualin}.

Given the results of the fit at different points in parameter space, we can construct a phenomenological model of the amplitude as a function of the initial parameters of a BBH merger.
We will only construct these models for modes that reside in their natural $\ell{,}m$ multipole (e.g., for the $2{,}2{,}0$ mode in the $2{,}2$ multipole), but not for the mixing or recoil modes.
This is because the mixing and recoil modes are subdominant, making their extraction noisier.
In principle, one could map the amplitudes of a mode from their natural multipole to their mixing contribution in a different multipole by considering the spherical-spheroidal mixing coefficients (see Appendix~\ref{mixing}).

By inspection of Fig.~\ref{fig:3D_A_220}, we observe that the amplitude of the $2{,}2{,}0$ mode varies rather smoothly throughout the whole parameter domain.
This is true in general for all the modes that we will consider in this section.
In the literature, hyperfits of mode amplitudes are often performed by assuming a polynomial model in some combinations of the mass ratio ($\eta$ or $q$) and of the individual progenitor spins ($\chi_1$, $\chi_2$, $\chi_+$ or $\chi_-$)~\cite{London:2014cma,London:2018gaq,Baibhav:2017jhs}.
We will follow this approach, and we will adopt a strategy that tries to minimize overfitting due to the inclusion of too many terms in the polynomial.

When performing the hyperfits, we will consider the SXS simulations SXS:BBH:0209-0305 and SXS:BBH:1419-1509, for a total of 188 simulations.
The first set of simulations was performed to validate and improve waveform models~\cite{Chu:2015kft,Kumar:2016dhh} (we also included SXS:BBH:0305, the most studied waveform in BH spectroscopy studies), while the second set was used to construct the \texttt{NRHybSur3dq8}~\cite{Varma:2018mmi} and \texttt{NRHybSur3dq8\_CCE}~\cite{Yoo:2023spi} NR surrogate models.
These are high-quality simulations that cover the parameter space in question adequately.
For completeness, we will also include the data points or fitting loss of the other nonprecessing simulations in the catalogue, but those are not used to train nor validate the fitting model.
All of the fitting and validation are performed with the \texttt{scikit-learn} package~\cite{scikit-learn}.

\subsection{Amplitude adjustment}\label{subsec:amplitude_adjustment}

We start by using information from PN theory to remove the dominant dependence of the mode amplitude on $\eta$.
While the PN expansion is only valid significantly before the merger peak, the dominant dependence of the amplitude of the waveform is expected to be approximately the same for the ringdown amplitude, at least for the fundamental mode.
For example, the $2{,}2$ multipole waveform has a dominant amplitude dependence $\sim \eta$ in the PN expansion~\cite{Buonanno:2006ui,Berti:2007fi,Kidder:2007rt,Berti:2007nw,Borhanian:2019kxt}, and we find that the same is true for the amplitude of the $2{,}2{,}0$ ringdown mode in the $2{,}2$ multipole.
To remove the dominant PN dependence from the amplitudes before we perform the hyperfit, we transform to a new set of adjusted complex amplitudes~\cite{Borhanian:2019kxt}:
\begin{align}
	\tilde{A}_{\ell{,}m{,}n} =    & \dfrac{A_{\ell{,}m{,}n}}{\eta} , \label{eq:A_adjust}                 \\
	\tilde{\phi}_{\ell{,}m{,}n} = & 2 \phi_{\ell{,}m{,}n} - m \phi_{2{,}2{,}0} \,. \label{eq:phi_adjust}
\end{align}
Adjusting the amplitude reduces the complexity of our polynomial model, and we also adjust the phase with respect to $\phi_{2{,}2{,}0}$ as the coalescence phase of the BBH merger is simulation-dependent.
For most of the modes concerned, fitting $\tilde{A}_{\ell{,}m{,}n}$ with a polynomial in $\eta, \chi_+$ and $\chi_-$ would give us an adequate model of the amplitude.
However, for the fundamental modes with an odd $m$, $\tilde{A}_{\ell{,}m{,}n}$ almost vanishes as we approach the equal mass limit $\eta \to 1/4$.
A naive polynomial fit could struggle in this limit, or even return a negative amplitude, which is unphysical.
For example, for the $2{,}1{,}0$ mode, from PN theory we expect the amplitude dependence to be the sum of a term proportional to $\delta$ and of one proportional to $\chi_-$ when $\delta \to 0$, where $\delta = \sqrt{1 - 4 \eta}$.
The term proportional to $\sim \delta$ vanishes in the equal mass limit, but the small term proportional to $\sim \chi_-$ is still present.
Therefore, to ensure that the hyperfits perform as expected in this limit, we use an ansatz
\begin{equation}
	\tilde{A}_{\ell{,}m{,}n} = \delta \, \tilde{A}^\prime_{\ell{,}m{,}n} + b_{\ell{,}m{,}n} |\chi_-|,
\end{equation}
where $b_{\ell{,}m{,}n}$ is a constant determined by a linear fit of the amplitudes $\tilde{A}_{\ell{,}m{,}n}$ obtained from simulations at the equal mass limit $\delta = 0$, and we perform the polynomial hyperfit on $\tilde{A}^\prime_{\ell{,}m{,}n}$, but only with the data points for which $\eta < 0.245$.
We find that this procedure works well for the $2{,}1{,}0$, $3{,}3{,}0$ and $5{,}5{,}0$ modes and ensures that their hyperfits behave as expected as $\delta \to 0$, so in the following subsections we will be fitting $\tilde{A}^\prime_{\ell{,}m{,}n}$ instead of $\tilde{A}_{\ell{,}m{,}n}$ with polynomials for these modes.
However, for other modes with an odd $m$, i.e., the $r2{,}1{,}0$, $2{,}1{,}1$ and $3{,}3{,}1$ modes, we revert to fitting ${A}_{\ell{,}m{,}n}$, either because we do not find these modes in the equal-mass limit or because their amplitudes are not small in that limit.

\subsection{Fixing the degree of the polynomial model} \label{subsec:N_best}

\begin{figure}
	\centering
	\includegraphics[width=0.49\textwidth]{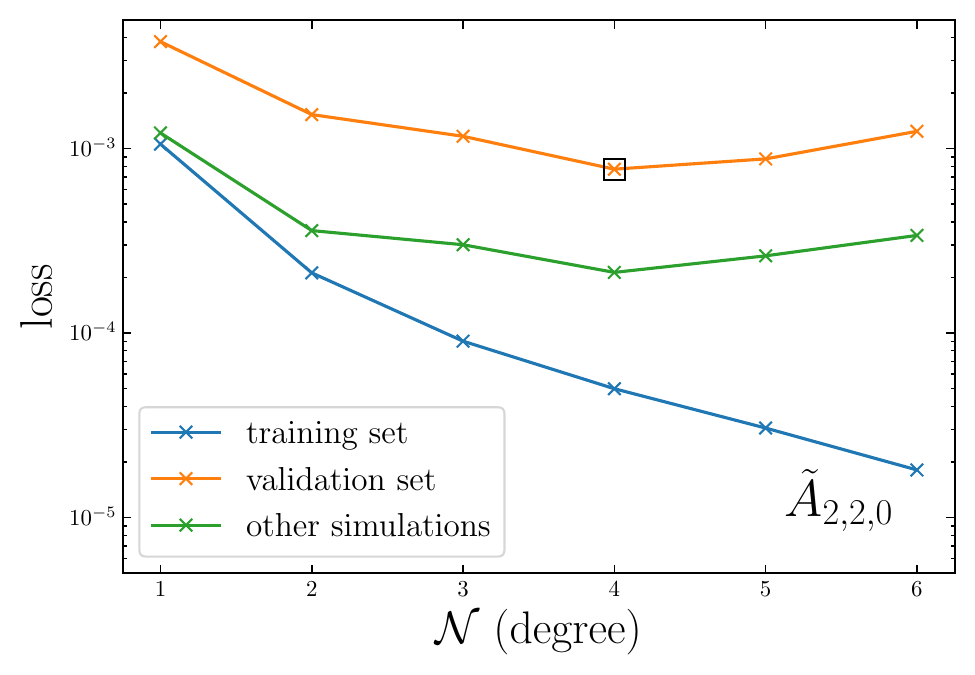}
	\caption{
	Fitting loss as a function of the degree of the multivariate polynomial model for the hyperfit of the amplitude $\tilde{A}_{2{,}2{,}0}$.
	The training and validation sets are obtained through a random $80\%/20\%$ split of the 188 amplitudes extracted from SXS:BBH:0209-0305,1419-1509.
	We also show the loss for all other nonprecessing and quasicircular simulations in the SXS catalog.
	The best degree $\mathcal{N}_{\rm best}$, determined by the validation set, is labeled by a black square.
	}
	\label{fig:loss_vs_degree_2.2.0}
\end{figure}

We now proceed to perform the hyperfits on the adjusted amplitudes $\tilde{A}_{\ell{,}m{,}n}$.
We choose to fit the amplitudes with a multi-variate polynomial in three variables ($\eta$, $\chi_+$ and $\chi_-$), but we must choose the degree of the polynomial.
To avoid underfitting or overfitting, we split the full set of data points randomly into a training set (consisting of $80\%$ of all simulations) and a validation set (the remaining $20\%$).
We fit the training set by a polynomial of three variables with a weighted least-squares loss for a varying degree $\mathcal{N}$, including all cross terms.
We use the inverse of the fluctuation ranges $\delta A$ and $\delta \phi$ as weights for the loss to penalize simulations that return less stable amplitude and phases.
Then, we compute the loss of the validation set when using the fitted polynomial model, and identify the value of $\mathcal{N}$ that minimizes the validation loss.
In Fig.~\ref{fig:loss_vs_degree_2.2.0}, we can see that the polynomial degree that gives the lowest validation loss is $\mathcal{N} = 4$: this is the optimal degree to use without overfitting the data.

For modes other than the $2{,}2{,}0$ mode shown in Fig.~\ref{fig:loss_vs_degree_2.2.0}, we will often also find $\mathcal{N} \sim 4$.
With $\mathcal{N} = 4$ and three variables, this polynomial is specified by 35 coefficients.
In the next subsection, we will show that the number of coefficients in the polynomial can be reduced further.

\begin{figure}[h!]
	\centering
	\includegraphics[width=0.49\textwidth]{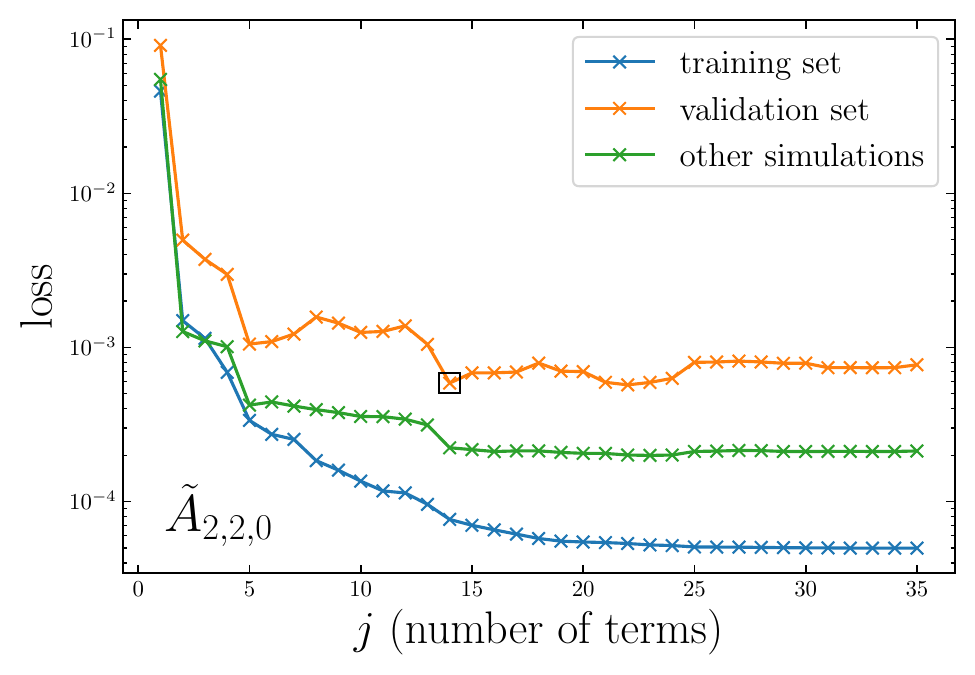}
	\caption{
	Fitting loss as a function of the number of terms we keep in the multivariate polynomial given the maximum degree $\mathcal{N}$ of the polynomial, for the case of $\tilde{A}_{2{,}2{,}0}$.
	In this case, we have already fixed $\mathcal{N} = 4$ (see Fig.~\ref{fig:loss_vs_degree_2.2.0}).
	In practice, we start by including all the terms of the multivariate polynomial of degree $\mathcal{N}$, and we remove terms iteratively, progressing from right to left in the figure.
	The best number of terms $j_{\rm best}$, determined by the validation set, is marked by a black square.
	}%
	\label{fig:loss_vs_n_terms_2.2.0}
\end{figure}

\subsection{Fixing the number of terms in the polynomial} \label{subsec:terms_remove}

A multivariate polynomial contains a significant number of cross terms even if the polynomial degree is moderate.
In our model, while a degree of $\mathcal{N}$ might be necessary to capture the dependence of the amplitude, some of the terms in the polynomial might not be required for the model to work well.
For this reason, we use an iterative scheme to remove unnecessary terms in the polynomial.

We begin with a polynomial of degree $\mathcal{N}$ containing all the $i$ possible terms.
For example, for $\mathcal{N} = 2$, there will be a set of $i = 10$ terms, including $1$, $\eta$, $\dots$, $\eta \chi_+$, $\eta \chi_-$, $\dots$, $\chi_-^2$, $\chi_+^2$.
We disregard one of the terms and fit the data with a polynomial that only includes the remaining $i - 1$ terms.
The training loss of the fit will be increased when we do so.
We repeat this step for each of the $i$ different terms, disregarding each term and keeping the other $i - 1$ terms, keeping track of the induced increase in the training loss.
In this way we can identify the term that leads to the smallest increase in the loss when it is neglected. This can be considered the term that contributes the least to the goodness of fit.
We remove this least-useful term from the set. At this point we are left with $i - 1$ terms, and we have completed the first iteration.
We use the same procedure to iteratively remove the least-useful term, until we are only left with one term.
This iterative procedure allows us to rank the importance of each term.

Given this ranking, we can try to determine the number of terms we have to include in order to adequately fit the data, i.e. the number $j \leq i$ for which the $j$ most important terms give a good enough model.
We do this, again, by inspecting the loss of the validation set.
As shown in Fig.~\ref{fig:loss_vs_n_terms_2.2.0}, for the $2{,}2{,}0$ mode with $\mathcal{N} = 4$ and $i = 35$, while the training loss decreases monotonically as we include more terms, the validation loss reaches a minimum and plateaus at around $j \sim 20$.
As a lower value of $j$ could already give a validation loss that is almost as good as the minimum value, we pick the best value $j_{\rm best}$ by searching for the lowest value of $j$ such that the loss is less than $1.5$ times the minimum validation loss.
For most modes, we find that $j_{\rm best} \lesssim 15$.
In this way we can compress the dimensions of the hyperparameter space of the model by a factor of $\gtrsim 2$.

We can now repeat the same procedure for the amplitudes and phases of different modes.
For some modes, we had to specify by hand a polynomial degree $\mathcal{N}$ that is larger than the ``best'' one found with the procedure in Sec.~\ref{subsec:N_best}, because the variation of the data contains maxima or minima within the fitting domain.
Explicitly, we used $\mathcal{N} = 4$ for $\tilde{A}_{2{,}1{,}1}$, $\mathcal{N} = 5$ for $\tilde{A}_{3{,}2{,}0}$ and $\tilde{A}_{4{,}4{,}0}$, and $\mathcal{N} = 6$ for $\tilde{\phi}_{4{,}4{,}0}$ and $\tilde{\phi}_{5{,}5{,}0}$.
Overfitting is avoided by the steps explained in this sub-section to remove unnecessary terms from the polynomial.

For the modes considered, we found the following best-fit functions:
\footnotetext{The error of the $\tilde{A}_{2{,}1{,}1}$ hyperfit model is significantly higher than that of $\tilde{A}_{2{,}1{,}0}$. 
When using the hyperfits to model the $\ell{,}m = 2{,}1$ ringdown waveform, including the $2{,}1{,}1$ mode could give a waveform with a lower match to NR compared to only using the $2{,}1{,}0$ mode.}
\addtocounter{footnote}{-1}
\begin{widetext}
	\vspace{-0.3cm}
	\begin{subequations}\label{eq:hyperfit_equations}
		\begin{align}
			\tilde{A}_{2{,}2{,}0} = \      & 4.004 + 1.349 \chi_+ + 0.333 \chi_- - 1.325 \eta^2 - 1.369 \eta \chi_- + 2.622 \chi_+ \chi_- - 32.74 \eta^2 \chi_+ + 4.313 \eta \chi_+^2 \nonumber                \\
			                               & - 25.18 \eta \chi_+ \chi_- + 83.37 \eta^3 \chi_+ - 13.39 \eta^2 \chi_+^2 + 58.01 \eta^2 \chi_+ \chi_- - 0.3837 \eta \chi_+^3 - 0.2075 \chi_+^4 \label{eq:A_2.2.0} \\
			\tilde{\phi}_{2{,}2{,}0} = \   & 0 \label{eq:phi_2.2.0}                                                                                                                                            \\
			\tilde{A}_{r2{,}2{,}0} = \     & 0.001657 - 0.07586 \chi_- + 0.1762 \eta \chi_+ + 1.358 \eta \chi_- + 0.1103 \chi_+^2 + 0.1361 \chi_-^2 - 0.03407 \eta^3 \nonumber                                 \\
			                               & - 2.147 \eta^2 \chi_+ - 7.814 \eta^2 \chi_- - 1.334 \eta \chi_+^2 - 1.295 \eta \chi_-^2 - 0.09387 \chi_+^3 - 0.01674 \chi_+ \chi_-^2 \nonumber                    \\
			                               & + 5.870 \eta^3 \chi_+ + 14.41 \eta^3 \chi_- + 3.323 \eta^2 \chi_+^2 + 2.948 \eta^2 \chi_-^2 + 0.1427 \eta \chi_+^3 - 0.03698 \chi_+^4 \label{eq:A_-2.2.0}         \\
			\tilde{\phi}_{r2{,}2{,}0} = \  & 13.14 - 31.89 \eta + 5.247 \chi_+ - 3.888 \chi_- + 12.24 \eta \chi_- + 2.571 \chi_+^2 \label{eq:phi_-2.2.0}                                                       \\
			\tilde{A}_{2{,}2{,}1} = \      & 15.46 - 407.0 \eta^2 + 55.43 \eta \chi_+ - 413.5 \eta \chi_- + 14.82 \chi_+^2 - 65.08 \chi_+ \chi_- + 17.99 \chi_-^2 + 1731 \eta^3 \nonumber                      \\
			                               & + 4245 \eta^2 \chi_- + 876.8 \eta \chi_+ \chi_- - 72.06 \eta \chi_-^2 + 11.46 \chi_+^3 + 101.2 \chi_+ \chi_-^2 - 2.499 \chi_-^3 \nonumber                         \\
			                               & - 10310 \eta^3 \chi_- - 2485 \eta^2 \chi_+ \chi_- - 400.0 \eta \chi_+ \chi_-^2 \label{eq:A_2.2.1}                                                                 \\
			\tilde{\phi}_{2{,}2{,}1} = \   & 3.918 + 30.68 \eta + 1.65 \chi_+ + 2.251 \chi_- - 196.8 \eta^2 - 15.94 \eta \chi_+ - 35.86 \eta \chi_- - 0.2809 \chi_+^2 \nonumber                                \\
			                               & - 2.797 \chi_+ \chi_- + 324.6 \eta^3 + 32.04 \eta^2 \chi_+ + 107 \eta^2 \chi_- + 11.19 \eta \chi_+ \chi_- - 0.2427 \chi_+^3 \label{eq:phi_2.2.1}                  \\
			\tilde{A}_{2{,}1{,}0} = \      & 0.9376 |\chi_-| + \delta (6.697 - 148.3 \eta - 1.035 \chi_- + 1603 \eta^2 - 0.960 \eta \chi_+ + 3.022 \chi_+ \chi_- - 4.270 \chi_-^2 \nonumber                    \\
			                               & - 7388 \eta^3 - 37.87 \eta^2 \chi_- - 15.85 \eta \chi_+ \chi_- + 12060 \eta^4 - 13.17 \eta \chi_+ \chi_-^2 + 11.61 \eta \chi_-^3 \nonumber                        \\
			                               & - 2.666 \chi_+^2 \chi_-^2 + 4.661 \chi_-^4) \label{eq:A_2.1.0}                                                                                                    \\
			\tilde{\phi}_{2{,}1{,}0} = \   & 4.282 + 2.075 \eta - 0.8584 \chi_+ - 5.040 \eta \chi_- - 1.626 \chi_+ \chi_- - 4.319 \eta^2 \chi_+ + 21.01 \eta^2 \chi_- - 2.270 \eta \chi_+^2 \nonumber          \\
			                               & + 5.414 \eta \chi_+ \chi_- \label{eq:phi_2.1.0}                                                                                                                   \\
			\tilde{A}_{r2{,}1{,}0} = \     & 0.08673 - 0.2838 \chi_+ - 0.08817 \chi_- - 10.79 \eta^2 + 2.238 \eta \chi_+ + 0.3544 \eta \chi_- + 0.1870 \chi_+ \chi_- + 0.2696 \chi_-^2 \nonumber               \\
			                               & + 71.95 \eta^3 - 4.639 \eta^2 \chi_+ - 2.673 \eta \chi_-^2 - 0.04592 \chi_+^2 \chi_- - 138.5 \eta^4 - 3.162 \eta^2 \chi_+ \chi_- \nonumber                        \\
			                               & + 6.734 \eta^2 \chi_-^2 \label{eq:A_-2.1.0}                                                                                                                       \\
			\tilde{\phi}_{r2{,}1{,}0} = \  & 9.273 - 21.85 \eta + 3.373 \chi_+ + 14.03 \eta^2 + 7.836 \eta \chi_+ - 3.304 \eta \chi_- + 3.543 \chi_+^2 + 3.424 \chi_+ \chi_- \nonumber                         \\
			                               & - 22.07 \eta^2 \chi_+ + 15.32 \eta^2 \chi_- - 11.25 \eta \chi_+ \chi_- + 4.089 \eta \chi_-^2 + 2.102 \chi_+^3 + 1.508 \chi_+ \chi_-^2 \label{eq:phi_-2.1.0}       \\
			\tilde{A}_{2{,}1{,}1}\text{\footnotemark} = \      & 13.85 - 48.23 \eta - 6.001 \chi_-^2 - 671.1 \eta^2 \chi_- + 45.03 \eta \chi_-^2 + 9.536 \chi_+ \chi_-^2 + 2648 \eta^3 \chi_- \label{eq:A_2.1.1}                   \\
			\tilde{\phi}_{2{,}1{,}1} = \   & 3.363 - 6.907 \eta - 5.204 \chi_+ + 26.98 \eta^2 + 41.15 \eta \chi_+ + 2.266 \chi_+^2 - 88.85 \eta^2 \chi_+ - 13.29 \eta \chi_+^2 \nonumber                       \\
			                               & - 0.8613 \eta \chi_-^2 - 1.767 \chi_+^3 - 0.2526 \chi_-^3 \label{eq:phi_2.1.1}                                                                                    \\
			\tilde{A}_{3{,}3{,}0} = \      & 0.2115 |\chi_-| + \delta (1.820 + 0.6007 \chi_+ + 0.4653 \chi_- + 16.49 \eta^2 + 0.9369 \chi_+ \chi_- - 0.2701 \chi_-^2 - 53.16 \eta^3 \nonumber                  \\
			                               & - 4.201 \eta^2 \chi_- + 2.180 \eta \chi_+^2 - 6.289 \eta \chi_+ \chi_-) \label{eq:A_3.3.0}                                                                        \\
			\tilde{\phi}_{3{,}3{,}0} = \   & 0.08988 + 1.049 \eta \chi_+ + 40.79 \eta^3 \label{eq:phi_3.3.0}                                                                                                   \\
			\tilde{A}_{3{,}3{,}1} = \      & 9.466 - 29.78 \eta - 35.38 \chi_+ + 404.6 \eta \chi_+ + 62.30 \eta \chi_- + 35.06 \chi_+^2 + 36.65 \chi_-^2 - 1021 \eta^2 \chi_+ \nonumber                        \\
			                               & - 264.3 \eta^2 \chi_- - 122.0 \eta \chi_+^2 - 155.2 \eta \chi_-^2 + 18.10 \chi_+^2 \chi_- \label{eq:A_3.3.1}                                                      \\
			\tilde{\phi}_{3{,}3{,}1} = \   & 4.984 - 1.686 \chi_- + 22.71 \eta \chi_- + 8.942 \eta^3 - 84.57 \eta^2 \chi_- - 6.581 \eta \chi_+ \chi_- \label{eq:phi_3.3.1}                                     \\
			\tilde{A}_{3{,}2{,}0} = \      & 0.7695 - 3.308 \eta - 1.446 \eta \chi_+ - 61.87 \eta^3 + 72.14 \eta^2 \chi_+ - 127.1 \eta^2 \chi_- - 2.769 \eta \chi_+ \chi_- \nonumber                           \\
			                               & + 0.3681 \eta \chi_-^2 - 0.5065 \chi_+ \chi_-^2 + 0.5483 \chi_-^3 + 293.4 \eta^4 - 527.6 \eta^3 \chi_+ + 1110 \eta^3 \chi_- \nonumber                             \\
			                               & + 11.14 \eta^2 \chi_+ \chi_- + 2.180 \eta \chi_+ \chi_-^2 - 2.023 \eta \chi_-^3 + 1014 \eta^4 \chi_+ - 2407 \eta^4 \chi_- \label{eq:A_3.2.0}                      \\
			\tilde{\phi}_{3{,}2{,}0} = \   & - 32.08 + 889.7 \eta - 81.88 \chi_+ + 93.05 \chi_- - 9292 \eta^2 + 1584 \eta \chi_+ - 1817 \eta \chi_- - 0.3888 \chi_-^2 \nonumber                                \\
			                               & + 40350 \eta^3 - 9588 \eta^2 \chi_+ + 10930 \eta^2 \chi_- - 6.121 \eta \chi_+^2 - 60250 \eta^4 + 18190 \eta^3 \chi_+ - 20600 \eta^3 \chi_- \label{eq:phi_3.2.0}   \\
			\tilde{A}_{4{,}4{,}0} = \      & 0.6505 + 2.978 \eta \chi_- + 0.4262 \chi_+ \chi_- + 106.1 \eta^3 + 67.45 \eta^2 \chi_+ - 12.08 \eta^2 \chi_- - 1.738 \eta \chi_+ \chi_- \nonumber                 \\
			                               & - 2041 \eta^4 - 614.2 \eta^3 \chi_+ + 5974 \eta^5 + 1387 \eta^4 \chi_+ \label{eq:A_4.4.0}                                                                         \\
			\tilde{\phi}_{4{,}4{,}0} = \   & 153.6 - 6463 \eta + 114700 \eta^2 - 1053000 \eta^3 + 5278000 \eta^4 + 478.4 \eta^3 \chi_+ - 13680000 \eta^5 \nonumber                                             \\
			                               & - 1960 \eta^4 \chi_+ + 65.40 \eta^4 \chi_- + 14320000 \eta^6 \label{eq:phi_4.4.0}                                                                                 \\
			\tilde{A}_{5{,}5{,}0} = \      & 0.04064 |\chi_-| + \delta (- 0.431 + 21.23 \eta + 0.2405 \chi_+ + 0.1427 \chi_- - 175.8 \eta^2 - 2.017 \eta \chi_+ + 0.03273 \chi_+^2 \nonumber                   \\
			                               & + 0.2473 \chi_-^2 + 414.9 \eta^3 - 1.526 \eta \chi_-^2 + 0.6688 \chi_+^3 - 1.876 \chi_+^2 \chi_- + 1.815 \chi_+ \chi_-^2 \nonumber                                \\
			                               & - 0.4803 \chi_-^3) \label{eq:A_5.5.0}                                                                                                                             \\
			\tilde{\phi}_{5{,}5{,}0} = \   & 6.400 - 296.5 \eta^3 - 56.71 \eta^2 \chi_- + 2.317 \eta \chi_+^2 - 4.757 \chi_-^3 - 353.7 \eta^2 \chi_-^2 - 41.22 \eta \chi_+ \chi_-^2 \nonumber                  \\
			                               & + 120.1 \eta \chi_-^3 + 4959 \eta^3 \chi_-^2 - 15780 \eta^4 \chi_-^2 + 131.9 \eta \chi_+ \chi_-^4 - 143.8 \eta \chi_-^5 \label{eq:phi_5.5.0}
		\end{align}
	\end{subequations}
\end{widetext}

From the hyperfits, the contribution of a particular QNM to the waveform strain at any time after $t_{\rm peak}$ can be obtained as
\begin{multline} \label{eq:individual_QNM}
	h_{\ell{,}m{,}n} \\ = \dfrac{M G}{d_L c^2} \eta \tilde{A}_{\ell{,}m{,}n} e^{-i [\omega_{\ell{,}m{,}n} (t - t_{\rm peak}) + (\tilde{\phi}_{\ell{,}m{,}n} + m \phi_{2{,}2{,}0})/2 ] }.
\end{multline}
Here $M$ is the total mass of the BBH progenitors, $d_L$ is the luminosity distance to the merger.
The contribution of the QNM in question to the detector strain can be obtained by projecting Eq.~\eqref{eq:individual_QNM} onto the antenna response function of the detector.
If needed, one can map the progenitor parameters to the remnant mass and spin with (e.g.) the \texttt{surfinBH} package~\cite{Varma:2018aht}.

All of the hyperfit functions are implemented in the \texttt{jaxqualin} package, with examples shown on the code webpage.
We omit modes that have a negligible amplitude in most of the simulations, as well as the overtones and quadratic modes whose amplitudes are too noisy.
In fact, as we will discuss below, it is easier to model the quadratic modes as a function of the amplitude of the linear modes that source them.
We have compared our amplitude hyperfits with Ref.~\cite{London:2018gaq} and confirmed that they follow the same trends, except at close-to-extremal values of $\chi_+$ or $\chi_-$, where there are fewer simulations available to train the hyperfit.

The hyperfit functions listed above should be used with caution.
First of all, one should proceed with care when using the fits outside of the range of the parameter space covered by the data used for training.
When doing the hyperfits for a given mode, we only used the simulations for which we identified the mode and extracted the amplitude and phase.
Therefore, in a region of parameter space where we did not find the mode, the hyperfit is an extrapolation.
Secondly, a high-dimensional multivariate polynomial might not be able to give a good fit for arbitrary data.
In fact, for some of the modes, the fits should only be taken as an order-of-magnitude estimate, and they can have large relative errors in regions where the mode amplitudes are small.
The error of the fits can be visualized on the \texttt{jaxqualin} webpage~\cite{jaxqualin}, and it should be carefully considered before using the fits.

If the fitting error is too large for the user's purposes, we recommend falling back to an interpolation scheme.
In the \texttt{jaxqualin} package, we provide an interpolation function along with the hyperfit functions.
The interpolator uses linear barycentric interpolation (as implemented in \texttt{scipy}~\cite{2020SciPy-NMeth}) to interpolate between data points, avoiding the use of any fitting model.
Using an interpolator bypasses the errors that inevitably arise from fitting, but it introduces fluctuations due to the noise of the data points.
The interpolator does not return a value if the requested point is outside of the convex hull of the data points.
This is somewhat restrictive, but it avoids returning spurious extrapolated results.

\section{Amplitude relationships between different modes}
\label{sec:amplratios}

We now turn to the interesting task of looking for correlations between the amplitudes of different modes. We will consider (A) the relation between quadratic modes and their ``parent'' linear modes, (B) the ratio between the amplitude of the first overtone and the amplitude of the fundamental mode for the dominant angular multipoles ($\ell{,}m=2{,}2$ and $\ell{,}m=3{,}3$), and (C) the amplitude ratio between retrograde and prograde modes, in this order.

\subsection{Quadratic modes}

\begin{figure*}
	\centering
	\includegraphics[width=0.99\textwidth]{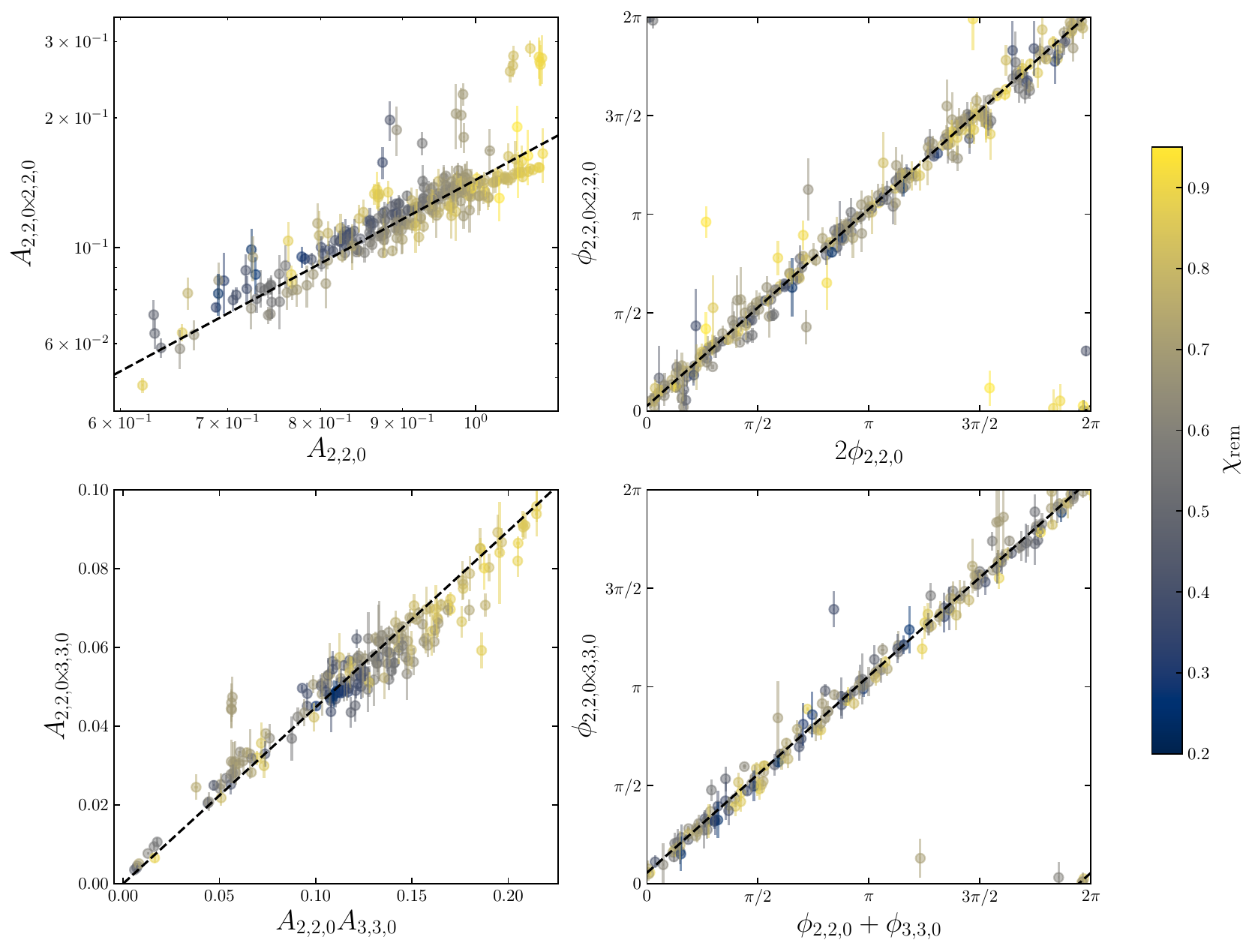}
	\caption{
	Dependence of the amplitude (left column) and phase (right column) of the $2{,}2{,}0 \! \times \! 2{,}2{,}0$ mode (top row) and the $2{,}2{,}0 \! \times \! 3{,}3{,}0$ mode (bottom row) on the linear modes that source them.
	The data points are colored according to the remnant spin $\chi_{\rm rem}$ of the corresponding SXS simulation.
	The error bars show the fluctuation of the amplitude within the $\tau_{\rm stable}$ window used for mode extraction.
	The black dashed lines are the best-fit lines when we assume the quadratic dependence of Eqs.~\eqref{eq:amp_quad} and \eqref{eq:phase_quad}, i.e., the line in the top-left panel has a slope of $2$, and the lines in the other panels have a slope of $1$.
	Note that the top-left panel is plotted on a logarithmic scale, while the bottom-left panel is plotted on a linear scale.
	}
	\label{fig:quad_dep}
\end{figure*}

The hyperfitting models presented above are phenomenological.
For quadratic modes, it has been shown, both theoretically and by examining NR simulations, that their amplitudes and phases depend quadratically on the linear modes that source them:
\begin{subequations}
	\begin{align}
		A_{\ell_1{,}m_1{,}n_1 \! \times \! \ell_2{,}m_2{,}n_2}    & = c_{\ell_1{,}m_1{,}n_1 \! \times \! \ell_2{,}m_2{,}n_2}(\chi_{\rm rem}) \nonumber \\ & \quad \times A_{\ell_1{,}m_1{,}n_1} A_{\ell_2{,}m_2{,}n_2},          \label{eq:amp_quad} \\
		\phi_{\ell_1{,}m_1{,}n_1 \! \times \! \ell_2{,}m_2{,}n_2} & = \phi_{\ell_1{,}m_1{,}n_1} + \phi_{\ell_2{,}m_2{,}n_2} \nonumber                  \\ & \quad + d_{\ell_1{,}m_1{,}n_1 \! \times \! \ell_2{,}m_2{,}n_2}(\chi_{\rm rem}), \label{eq:phase_quad}
	\end{align}
\end{subequations}
where all the amplitudes are assumed to be extracted from the ``natural'' multipolar components ($\ell, m = \ell_1 + \ell_2, m_1 + m_2$ for the quadratic mode).
The coefficients $c_{\ell_1{,}m_1{,}n_1 \! \times \! \ell_2{,}m_2{,}n_2}$ and $d_{\ell_1{,}m_1{,}n_1 \! \times \! \ell_2{,}m_2{,}n_2}$ can in principle be computed from the Teukolsky equation by plugging quadratic combinations of linear QNMs as source terms into the equation (although this has only been attempted in the nonrotating case and under several approximations: see e.g.~\cite{Ioka:2007ak,Nakano:2007cj,Kehagias:2023mcl,Perrone:2023jzq,Bucciotti:2023ets}). In general, these coefficients are expected to depend on $\chi_{\rm rem}$ and (to some extent) on the initial conditions~\cite{Redondo-Yuste:2023seq}, and therefore on $\eta, \chi_+$ and $\chi_-$.

With the fitting procedure in Sec.~\ref{sec:mode_extraction}, we have extracted both linear and quadratic modes.
Therefore, we can compare the amplitudes and phases between these modes for each simulation to check whether the quadratic relationship holds.
In Fig.~\ref{fig:quad_dep} we show how the $2{,}2{,}0 \! \times \! 2{,}2{,}0$ mode amplitude and phase depend on the corresponding quantities for the ``parent'' $2{,}2{,}0$ mode. We also illustrate the same relations for the $2{,}2{,}0 \! \times \! 3{,}3{,}0$ mode as a function of the properties of the $2{,}2{,}0$ and $3{,}3{,}0$ modes. These plots generalize and extend similar results presented in Ref.~\cite{Cheung:2022rbm}.
Interestingly, the dependence seems to be rather universal,
in the sense that the constants $c$ and $d$ are only mildly (if at all) dependent on $\chi_{\rm rem}$.

\begin{figure*}
	\centering
	\includegraphics[width=0.99\textwidth]{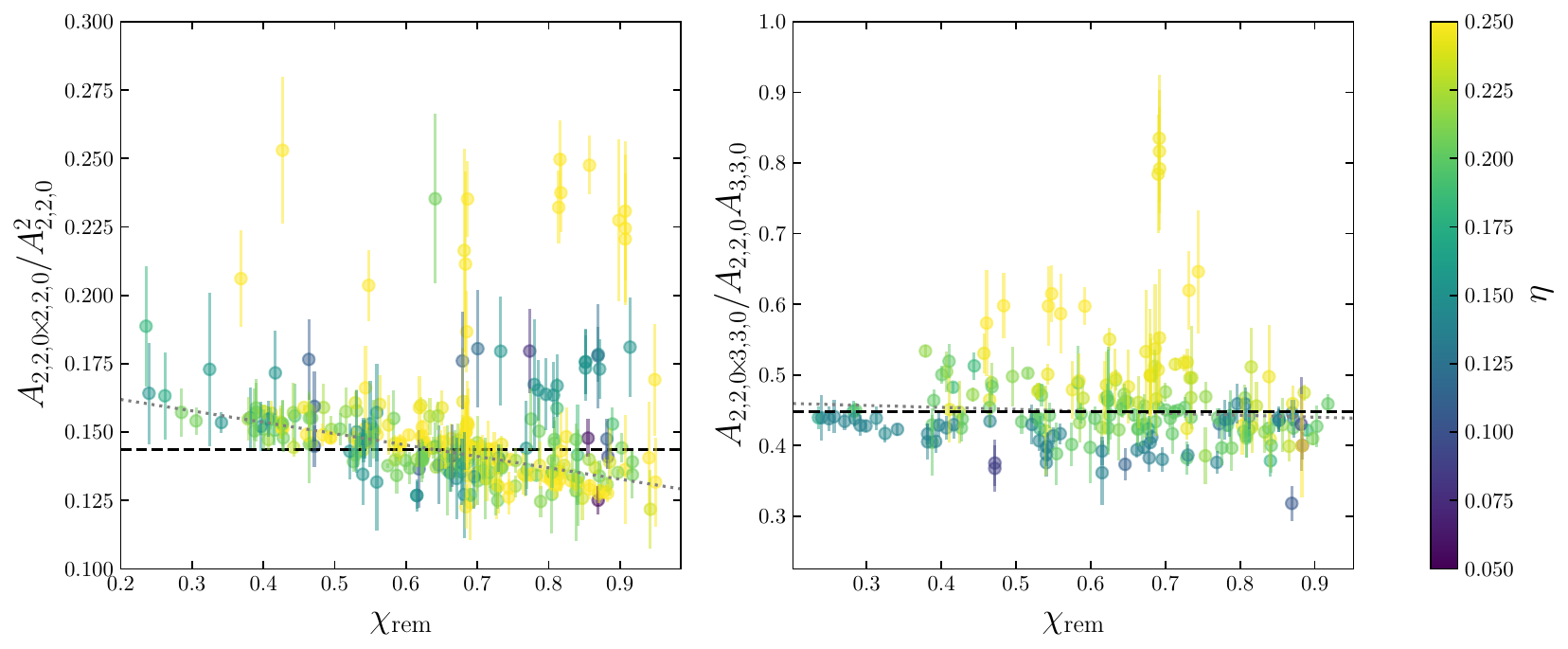}
	\caption{
	The ratios $A_{2{,}2{,}0 \! \times \! 2{,}2{,}0}/A_{2{,}2{,}0}^2$ (left) and $A_{2{,}2{,}0 \! \times \! 3{,}3{,}0}/A_{2{,}2{,}0}A_{3{,}3{,}0}$ (right) as a function of $\chi_{\rm rem}$.
	The ratio shown on the left is equal to the coefficient $c_{\ell_1{,}m_1{,}n_1 \! \times \! \ell_2{,}m_2{,}n_2}$ defined in Eq.~\eqref{eq:amp_quad}.
	The black dashed lines correspond to the lines shown in the left panels of Fig.~\ref{fig:quad_dep}, where $c_{\ell_1{,}m_1{,}n_1 \! \times \! \ell_2{,}m_2{,}n_2}$ is assumed to be a constant (independent of $\chi_{\rm rem}$).
	The gray dotted lines are the best-fit lines if we instead assume a linear relationship between the ratios and $\chi_{\rm rem}$.
	}
	\label{fig:quad_ratio}
\end{figure*}

In Fig.~\ref{fig:quad_ratio} we plot $c_{\ell_1{,}m_1{,}n_1 \! \times \! \ell_2{,}m_2{,}n_2}$ against $\chi_{\rm rem}$.
For both the $2{,}2{,}0 \! \times \! 2{,}2{,}0$ and $2{,}2{,}0 \! \times \! 3{,}3{,}0$ modes, the amplitude ratio follows an almost-constant linear trend with respect to $\chi_{\rm rem}$ for the majority of simulations. Only a small fraction of points deviate from this trend.
At least some of these outliers could result from biases, either because the mode extraction algorithm misses some modes, or because it picks up some spurious ones.
Note that $A_{3{,}3{,}0} \to 0$ as $\eta \to 1/4$, which could explain the bias in the ratio of the $2{,}2{,}0 \! \times \! 3{,}3{,}0$ mode in the comparable-mass limit.

Assuming that $c_{\ell_1{,}m_1{,}n_1 \! \times \! \ell_2{,}m_2{,}n_2}$ and $d_{\ell_1{,}m_1{,}n_1 \! \times \! \ell_2{,}m_2{,}n_2}$ do not depend on $\chi_{\rm rem}$, their best-fit values are
\begin{subequations}
	\begin{align}
		\bar{c}_{2{,}2{,}0 \! \times \! 2{,}2{,}0} = & 0.144, \label{eq:c_2.2.0x2.2.0} \\
		\bar{d}_{2{,}2{,}0 \! \times \! 2{,}2{,}0} = & 0.077, \label{eq:d_2.2.0x2.2.0} \\
		\bar{c}_{2{,}2{,}0 \! \times \! 3{,}3{,}0} = & 0.448, \label{eq:c_2.2.0x3.3.0} \\
		\bar{d}_{2{,}2{,}0 \! \times \! 3{,}3{,}0} = & 0.172,\label{eq:d_2.2.0x3.3.0}
	\end{align}
\end{subequations}
where the overbar means that we assume these coefficients to be constants.
If we allow the coefficient $c_{2{,}2{,}0 \! \times \! 2{,}2{,}0}$ to be dependent on $\chi_{\rm rem}$ and assume that the dependence is linear, the best fit is
\begin{equation}
	{c}_{2{,}2{,}0 \! \times \! 2{,}2{,}0} = -0.042 \chi_{\rm rem} + 0.170.
\end{equation}
For the other coefficients, we observe no significant dependence on $\chi_{\rm rem}$.

\subsection{Overtones}

\begin{figure*}
	\centering
	\includegraphics[width=0.99\textwidth]{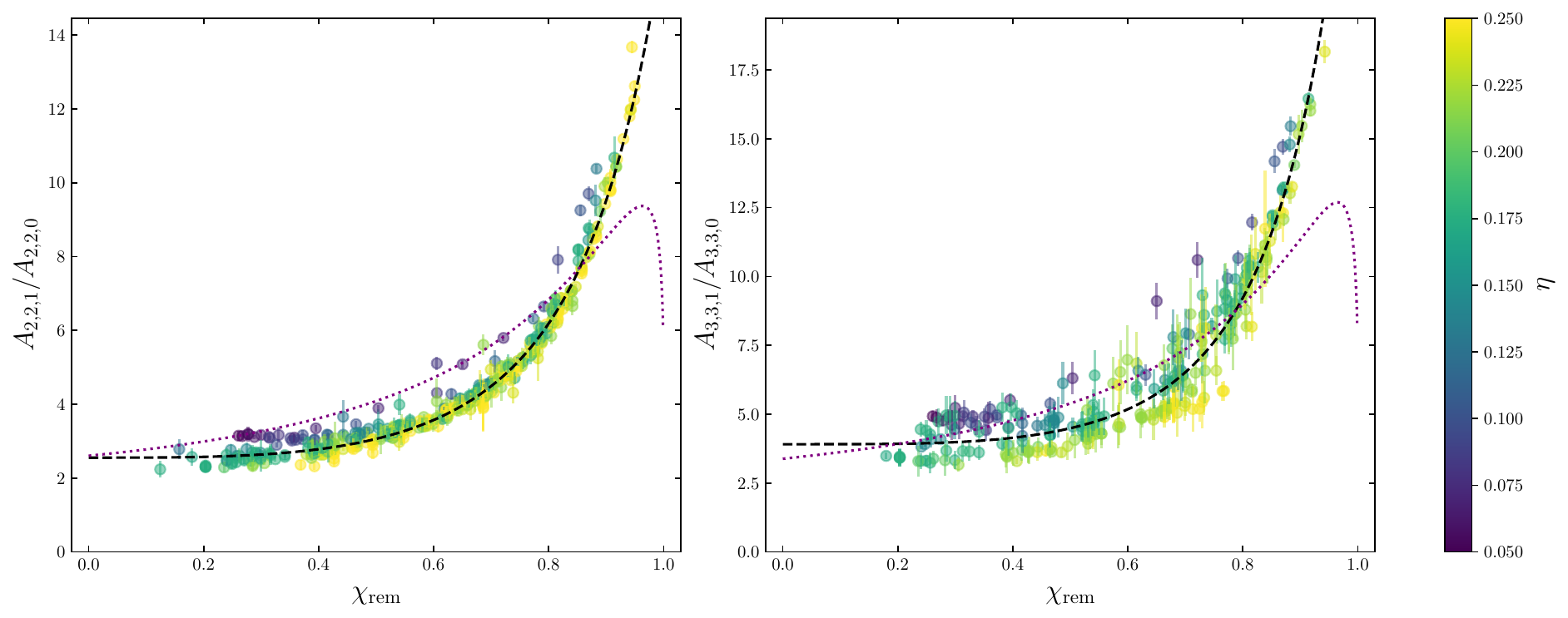}
	\caption{
	The ratios $A_{2{,}2{,}1}/A_{2{,}2{,}0}$ (left) and $A_{3{,}3{,}1}/A_{3{,}3{,}0}$ (right) as functions of $\chi_{\rm rem}$.
	The color of the circles corresponds to the value of $\eta$ for that particular simulation.
	The black dashed lines are the best-fit curves given in Eqs.~\eqref{eq:22_overtone_fit} and \eqref{eq:33_overtone_fit}.
	The purple dotted lines are the excitation factor ratios $|\omega_{2{,}2{,}0}^2 B_{2{,}2{,}1}/\omega_{2{,}2{,}1}^2 B_{2{,}2{,}0}|$ (left) and $|\omega_{3{,}3{,}0}^2 B_{3{,}3{,}1}/\omega_{3{,}3{,}1}^2 B_{3{,}3{,}0}|$ (right) adjusted by a factor of $1/\omega_{\ell{,}m{,}n}^2$ to convert from $\psi_4$ to $h$.
	}
	\label{fig:overtone_ratio}
\end{figure*}

There has been some discussion on whether the $2{,}2{,}1$ overtone has been observed in the first merger event GW150914~\cite{Isi:2020tac,Cotesta:2022pci,Finch:2022ynt,Ma:2023vvr}.
In fact, it is questionable whether the first overtone is useful at all for testing GR, even in the limit of infinite signal-to-noise ratio~\cite{Baibhav:2023clw,Nee:2023osy,Zhu:2023mzv}.  We contribute to this discussion by asking: is there some universal relation between the amplitude of the $2{,}2{,}1$ overtone and the amplitude of the $2{,}2{,}0$ fundamental mode?
In principle we could derive such a relation ``directly'' from the fitting formulas~\eqref{eq:A_2.2.0} and \eqref{eq:A_2.2.1}. By looking directly at the amplitudes, we find that their ratios is rather universal.
As shown in the left panel of Fig.~\ref{fig:overtone_ratio}, the amplitude ratio $A_{2{,}2{,}1}/A_{2{,}2{,}0}$ has a strong dependence on $\chi_{\rm rem}$, but only a very mild dependence on the (symmetric) mass ratio.
The ratio $A_{3{,}3{,}1}/A_{3{,}3{,}0}$ follows a similar trend, although there is a larger spread in the data.
An exponential in a power of $\chi_{\rm rem}$ seems to fit quite well most of the data.  The best fit curves shown in the figure are given by
\begin{subequations}\label{eq:overtone_fits}
	\begin{align}
		A_{2{,}2{,}1}/A_{2{,}2{,}0} = & 6.519^{(\chi_{\rm rem}^{3.366} + 0.500)}, \label{eq:22_overtone_fit} \\
		A_{3{,}3{,}1}/A_{3{,}3{,}0} = & 7.679^{(\chi_{\rm rem}^{3.877} + 0.669)}. \label{eq:33_overtone_fit}
	\end{align}
\end{subequations}
Note that this relation might not hold in the extrapolation region, i.e., for $\chi_{\rm rem} \lesssim 0.1$ or $\chi_{\rm rem} \gtrsim 0.95$.

\begin{figure*}
	\centering
	\includegraphics[width=0.99\textwidth]{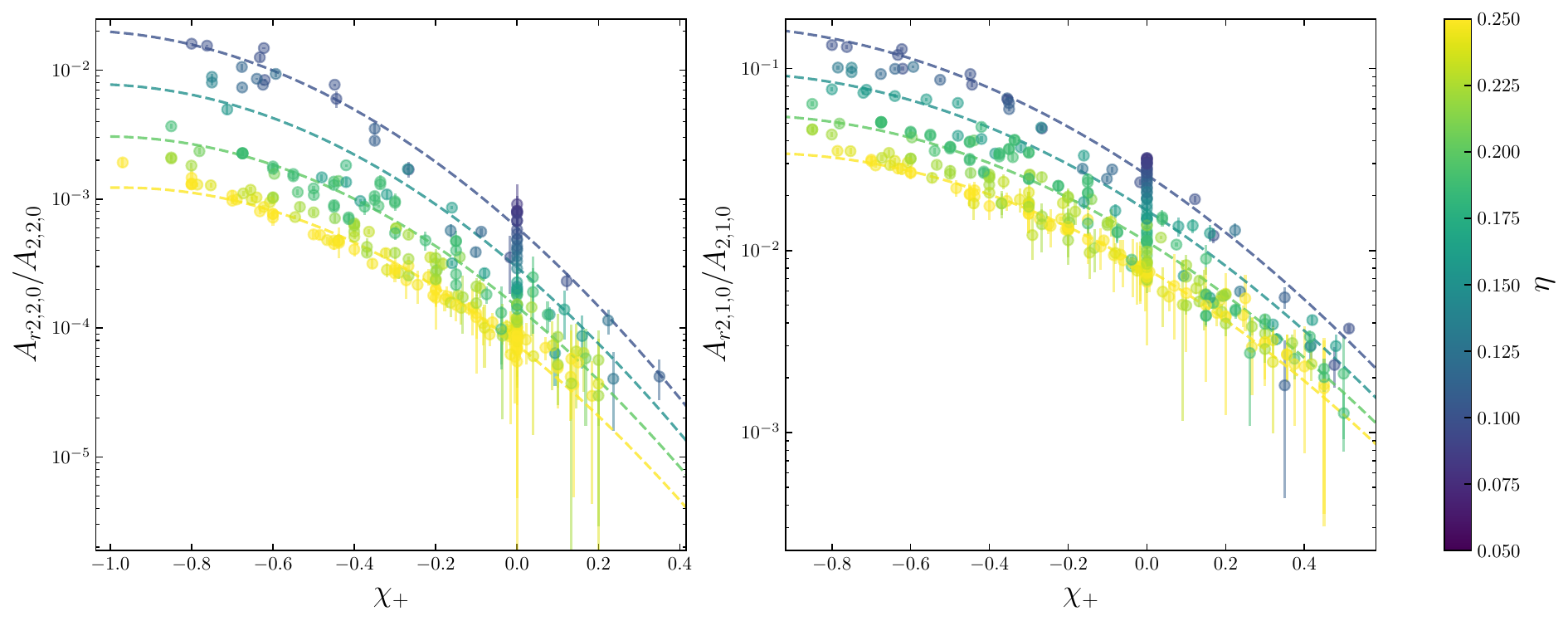}
	\caption{
	The ratios $A_{r2{,}2{,}0}/A_{2{,}2{,}0}$ (left) and $A_{r2{,}1{,}0}/A_{2{,}1{,}0}$ (right) as functions of $\chi_+$.
	The color of the circles corresponds to the value of $\eta$ for that particular simulation.
	The parabolic fitting surfaces of Eqs.~\eqref{eq:retro_ratio_220} and \eqref{eq:retro_ratio_210} are shown at discrete values of $\eta = 0.1, 0.15, 0.2$ and $0.25$ as dashed lines of the corresponding color.
	}
	\label{fig:retrograde_ratio}
\end{figure*}

In principle, it should be possible to compute the ratios in Eqs.~\eqref{eq:22_overtone_fit} and \eqref{eq:33_overtone_fit} given the initial conditions for the perturbations of the remnant Kerr BH.
In practice this is a difficult problem, as there is a large amount of arbitrariness in the choice of the initial time slice, gauges, etceteras.
The ratios we extract from the numerical data are equivalent to the ratio between the so-called ``excitation coefficients'' $C_{\ell{,}m{,}n}$ of the two modes, which can be factorized as
\begin{equation} \label{eq:EF}
	C_{\ell{,}m{,}n} = B_{\ell{,}m{,}n} I_{\ell{,}m{,}n}\,.
\end{equation}
Here the ``excitation factors'' $B_{\ell{,}m{,}n} \equiv B_{\ell{,}m{,}n}(\chi_{\rm rem})$ are functions only of the remnant BH spin, but independent of the initial conditions, while the integral $I_{\ell{,}m{,}n}$ depends on both the initial conditions and on the source exciting the perturbations~\cite{Leaver:1986gd,Berti:2006wq,Zhang:2013ksa,Oshita:2021iyn}.
Determining appropriate initial conditions for a comparable-mass BBH merger is highly nontrivial, and (to the best of our knowledge) the coefficients $C_{\ell{,}m{,}n}$ have never been computed ``from first principles'' in this context.
However, recently there are claims that the ratio between overtone amplitudes is closely approximated by the ratio between the corresponding excitation factors $B_{\ell{,}m{,}n}$. This would imply that the integral $I_{\ell{,}m{,}n}$ does not strongly depend on $n$ for fixed valus of the multipolar indices $\ell$ and $m$.
In Fig.~\ref{fig:overtone_ratio} we check the accuracy of this approximation by plotting the absolute value of the ratios between the excitation factors $|\omega_{\ell{,}m{,}0}^2 B_{\ell{,}m{,}1}/\omega_{\ell{,}m{,}1}^2 B_{\ell{,}m{,}0}|$ (the numerical values were computed in Refs.~\cite{Berti:2006wq,Zhang:2013ksa} and are publicly available online~\cite{RDwebsites}).
The numbers $B_{\ell{,}m{,}n}$ are defined as the excitation factor of QNMs in the $\psi_4$ scalar, so a factor of $1/\omega_{\ell{,}m{,}n}^2$ is included to convert from the ratio of amplitudes in $\psi_4$ to that of the waveform strain $h$.
The $A_{\ell{,}m{,}n}$ and $B_{\ell{,}m{,}n}$ ratios follow similar trends for low values of the dimensionless remnant spin $\chi_{\rm rem}$, but there are significant deviations between the two at higher values of $\chi_{\rm rem}$.
This implies that the ratios $A_{\ell{,}m{,}1}/A_{\ell{,}m{,}0}$ will, in general, depend on the initial data and on the source exciting the perturbations, although the dependence may be mild in certain favorable situations.
This is an interesting problem that deserves further study.

\subsection{Retrograde modes}

We now focus on ``retrograde'' modes -- i.e., modes that counterrotate with respect to the orbit of the BBH: see Appendix~\ref{app:retro_definition} for a more precise definition.
Out of all possible retrograde modes, we find the $r2{,}2{,}0$ and $r2{,}1{,}0$ modes in a significant fraction of the simulations.
Work on extreme mass-ratio inspirals~\cite{Bernuzzi:2010ty,Barausse:2011kb,Harms:2014dqa} and comparable-mass binaries~\cite{London:2014cma,Dhani:2020nik,Dhani:2021vac,MaganaZertuche:2021syq} has shown that retrograde modes are most excited when (i) the progenitor spins $\vec{\chi_1}$ and $\vec{\chi_2}$ are antiparallel to the orbital angular momentum of the binary, and (ii) for binaries with more asymmetric mass ratio ($\eta \to 0$), for fixed spin magnitudes.

Both trends are visible in Fig.~\ref{fig:retrograde_ratio}, where the ratios between the retrograde and prograde modes $A_{r2{,}2{,}0}/A_{2{,}2{,}0}$ and $A_{r2{,}1{,}0}/A_{2{,}1{,}0}$ both increase as $\eta \to 0$ and as $\chi_+$ becomes more negative.
The logarithm of the ratios is well fitted by a quadratic polynomial in $\eta$ and $\chi_+$:
\begin{subequations}
	\begin{align}
		 & \log_{10}\left(A_{r2{,}2{,}0}/ A_{2{,}2{,}0}\right) \nonumber                      \\
		 & \quad = 1.168 \eta^2 -   1.281\chi_+^2 +  1.983 \eta \chi_+ - 6.480\eta  \nonumber \\ & \quad \quad -2.997 \chi_+ -2.584, \label{eq:retro_ratio_220} \\
		 & \log_{10}\left(A_{r2{,}1{,}0}/ A_{2{,}1{,}0}\right) \nonumber                      \\
		 & \quad = 4.500 \eta^2 - 0.6437  \chi_+^2 + 1.148 \eta \chi_+ -5.016 \eta \nonumber \\ & \quad \quad -1.568 \chi_+ -1.129. \label{eq:retro_ratio_210}
	\end{align}
\end{subequations}
These parabolic surfaces are visualized as dashed lines of different colors in Fig.~\ref{fig:retrograde_ratio}.

\section{Ringdown starting time}
\label{sec:tstart}

When extracting QNMs from data, it is important to determine the optimal time to do so.
If the extraction is performed too early, then the waveform would still be contaminated by the nonlinearities at the merger, or (even if we disregard those nonlinearities) by the prompt response of the initial data within linear theory.
If performed too late, the exponentially decaying QNMs might be significantly contaminated by noise and other subdominant contributions, such as tails.
These considerations apply to both the extraction of QNMs from waveforms computed in NR and perturbation theory~\cite{Baibhav:2023clw,Nee:2023osy,Zhu:2023mzv} and to ringdown data analysis of BBH mergers detected by GW detectors.
In this section we discuss possible ways to define such an optimal mode extraction time, with an emphasis on applications to the analysis of real data.

\subsection{An overoptimistic definition}\label{subsec:time_optimistic}

Our goal so far has been to extract the amplitude and phase of a mode to the best accuracy from NR simulations.
The natural procedure that we employed is to extract them within a time window where they are the most stable (i.e., where their amplitude varies the least).

However, when analyzing real data it is important to maximize the signal-to-noise ratio (SNR), so (ideally) we would like to extract the modes as early as possible without biasing the results.
As the numerical noise in the NR simulations is negligible when compared to the instrumental noise of GW detectors, we can treat the mode extraction results from NR as ``ground truths'' for finding the optimal extraction time in real detections.

For ringdown signals detected in real data, we can define the optimal starting time for extracting each QNM $k$ to be the time $t_{k, \rm start}$ at which the variation of the mode amplitude and phase ($\Delta_{\rm NR} A_k$ and $\Delta_{\rm NR} \phi_k$, as extracted from NR) is smaller than their statistical measurement errors ($\Delta_{\rm stat} A_k$ and $\Delta_{\rm stat} \phi_k$, as extracted from real data, e.g. at the $2\sigma$ level).
Assuming that the statistical error is sensible ($\Delta_{\rm stat} A_k < A_k$ and $\Delta_{\rm stat} \phi_k < 2\pi$ at the very least), this can be seen as the earliest time at which the mode extraction from real data will not be significantly biased compared to the statistical error.
The quantities $\Delta_{\rm stat} A_k$ and $\Delta_{\rm stat} \phi_k$ are well-defined, and they can be read off from the posterior of the analysis.
The quantities $\Delta_{\rm NR} A_k$ and $\Delta_{\rm NR} \phi_k$ depend on the length of the time window $\tau_{\rm NR}$ over which we measure the spread of $A$ and $\phi$ (compare the $\tau_{\rm stable}$ window used in Sec.~\ref{sec:mode_extraction}).

Clearly, this definition of $t_{k, \rm start}$ depends on the level of instrumental noise in a GW detector.
However, we can still develop some intuition on $t_{k, \rm start}$ by imposing that $\Delta_{\rm NR} A_k$ and $\Delta_{\rm NR} \phi_k$ should vary less than a sensible range.
For example, we can reuse our criteria in Sec.~\ref{subsec:stability_test}, i.e. the quantities $\epsilon_{\rm stable}, \beta_A, \beta_\phi, A_{\rm tol}, \tau_{\rm stable}$ and $p_{\rm stable}\%$, using the ``normal'' stability condition listed in Table~\ref{table:definitions}, and define $t_{k, \rm start}$ to be the earliest time $t_0$ at which $\Delta_{k, \rm stable}(t_0; \tau_{\rm stable}) < \epsilon_{\rm stable}$.
For example, $t_{k, \rm start}$ is labeled as a solid circles for each mode in Fig.~\ref{fig:flat_extraction}.
While there is some arbitrariness in the choice of the parameters, a value of $\epsilon_{\rm stable}$ of order $O(0.1)$ is in line with the sensitivity of current GW detectors, while $\tau_{\rm stable} \sim 10 M$ is comparable to half a period of the dominant $2{,}2{,}0$ mode (the natural time scale of the ringdown).

Such a definition of $t_{k, \rm start}$ is not only somewhat arbitrary, but it also has a crucial shortcoming.
Because we define a different $t_{k, \rm start}$ for each mode $k$, there is no guarantee that at the $t_{k, \rm start}$ appropriate for a particular mode, the other modes will also be stable.
This can be seen from Fig.~\ref{fig:flat_extraction}. If we focus, for example, on $t_{2{,}2{,}0, \rm start}$, then the amplitudes and phases of all other modes are unstable.
As discussed in Ref.~\cite{Baibhav:2023clw}, by including multiple QNMs in the fitting model we can overfit the merger part of the signal, and thus drive the amplitude of the dominant QNMs (in this case, the $2{,}2{,}0$ mode) to their ``correct'' late-time value at earlier and earlier times.
In other words, while from Fig.~\ref{fig:flat_extraction} we may be tempted to conclude that $t_{2{,}2{,}0, \rm start} - t_{\rm peak}\sim 0 M$, this could be an artifact of overfitting the merger with all the other modes, as can be deduced from their unstable amplitudes.
This implies that the $t_{k, \rm start}$ defined in this way could be overoptimistic (i.e., it tends to be ``too early'').

Such overoptimism would only be exacerbated by using a model containing fewer modes than those used for extracting $t_{k, \rm start}$ in this work (e.g., less than the $6$ modes listed in Fig.~\ref{fig:flat_extraction}).
The goal of the algorithm specified in Sec.~\ref{sec:mode_extraction} is to find {\em all} identifiable modes within NR, and to include them all in the fitting model when extracting the amplitudes and phases of the modes.
At current detector sensitivity, we could only hope to measure a few ($\sim 2$) QNMs in real data, so it would not be sensible to always include so many modes in the fitting model.
When that is the case, we should use a {\em larger} $t_{k, \rm start}$ than those shown in this work, because the negligence of subdominant modes could affect the stability of the dominant ones.

\subsection{An overconservative definition}

On the other hand, an overconservative way to define the optimal starting time would be to take the {\em largest} $t_{k, \rm start}$ out of all of the modes found in the waveform (e.g., $t_{3{,}3{,}0, \rm start}$ in Fig.~\ref{fig:flat_extraction}), and define that to be the starting time of the whole QNM model \textit{containing all of the modes}.
This is, indeed, the earliest time at which {\em all of the modes} in the model have a stable amplitude.

This requirement is overconservative: for example, the subdominant mixing and recoil modes will only be found at late times, when other contaminations have died down. At those times the dominant modes (including the fundamental mode and some of the overtones) should have long stabilized.

\subsection{A recommended strategy}

In practice, the ``true'' optimal starting time should be somewhere in between these overoptimistic and overconservative definitions.
A recommended strategy is to first decide which modes should be included in the model, fit only with those modes, and then use the overconservative strategy to decide the optimal starting time for the \textit{full model} (including all QNMs, not just one particular QNM).

A sketch of such a procedure is as follows:
\begin{enumerate}
	\item Prepare the waveform to be fitted. \label{optimal:waveform}
	\item Decide the set of modes $K = \{k\}$ to use and the tolerance for stability, e.g., the values of $\epsilon_{\rm stable}$ and $\tau_{\rm stable}$.
	      The tolerance parameters could be determined from the statistical error obtained when performing data analysis.
	\item Perform a fit with all modes $k \in K$, and find $t_{k, \rm start}$ for each of them.
	\item The optimal starting time $t_{\rm start}$ is the largest $t_{k, \rm start}$ among all modes $k \in K$.
\end{enumerate}
For example, if one is using the $\ell{,}m = 2{,}2$ multipole of a BBH simulation and chooses $K = \{2{,}2{,}0\}$ or $\{2{,}2{,}0 | 2{,}2{,}1\}$, one would find $t_{\rm start} \gtrsim 10 M$ in both cases.

If we want to quantify the optimal time for performing data analysis on real signals in detector noise, it is important to make sure that the waveform prepared in Step~\ref{optimal:waveform} should be as close as possible to the one expected in the detector.
For example, the actual detected waveform should contain all of the multipolar components of the waveform projected onto the detectors via the antenna pattern functions, which depend on the sky position of the source and on the orientation of the detectors.
This is significantly more complicated than dealing with the individual multipolar components considered in this work.

We stress once again that the $t_{k, \rm start}$ values shown in our figures uses the \textit{overoptimistic} definition of Sec.~\ref{subsec:time_optimistic} and, {\em they should not be applied to analyze real GW data.}
This is because the merger is generally overfitted by the subdominant modes,
because the the number of modes used in this work will usually be different from the number of modes included in real data analysis, and because we only considered individual multipoles rather than the (sensibly more complicated) sum of multipoles that is relevant for real detections.
The steps laid out in this subsection suggest a better strategy for determining $t_{\rm start}$ in real data.

Another possibility for determining $t_{k, \rm start}$ is to take the amplitude and phases extracted for different modes in Sec.~\ref{sec:mode_extraction} (or the hyperfits in Sec.~\ref{sec:hyperfit}) as the ground truth, and to perform an array of data analysis runs for different values of the starting time $t_0$ to find the earliest time for which all the QNM frequencies, amplitude and phases are consistent with the ``true'' values.
This could be a viable strategy, but it requires performing multiple data analysis runs (as in Refs.~\cite{Cotesta:2022pci,Isi:2022mhy,Ma:2023vvr}) or marginalizing over the ringdown starting time (as in Ref.~\cite{Finch:2022ynt}).

\section{Conclusions}
\label{sec:conclusions}

In this work, we have presented a strategy for extracting QNMs from BBH merger simulations.
The core procedure is sketched in Fig.~\ref{fig:flowchart} and detailed in Sec.~\ref{sec:mode_extraction}, with tolerance criteria summarized in Table~\ref{table:definitions}.
The strategy is implemented in the publicly available \texttt{jaxqualin} package~\cite{jaxqualin}, and the tolerance parameters can be tuned for the user's purposes.

By fitting an exhaustive set of nonprecessing quasicircular BBH waveforms in the SXS catalog, we have constructed hyperfitting models for the amplitudes and phases of different QNMs.
We have verified the expected scaling between quadratic modes and the linear modes that source them, confirming and extending previous results~\cite{London:2014cma,Mitman:2022qdl,Cheung:2022rbm}.
We have also found quasiuniversal relations between different modes, including the amplitude ratio between the first overtone and the fundamental mode shown in Fig.~\ref{fig:overtone_ratio}, and the amplitude ratio between the retrograde and prograde modes shown in Fig.~\ref{fig:retrograde_ratio}. To the best of our knowledge, these results are new.
These quasiuniversal relations warrant further investigation, both in waveform modeling and in BH perturbation theory.
The apparent discrepancy between the remnant spin dependence of nonlinear modes extracted from BBH simulations (see Fig.~\ref{fig:quad_ratio}) and the dependence found for Gaussian scattering at second order in BH perturbation theory~\cite{Redondo-Yuste:2023seq} deserves further investigation.

Finally, we have presented some practical proposals to define a convenient ringdown starting time for data analysis purposes.
These proposals should be tested and refined by analyzing ringdown signals in real data.

It is trivial to apply the \texttt{jaxqualin} code to other NR waveforms or simulation catalogs, such as those using CCE, or precessing BBH simulations.
We expect that higher-quality simulations including center-of-mass corrections, improved Cauchy-characteristic evolution, the inclusion of gravitational memory using BMS balance laws, and improved wave extraction methods would be very useful to remove spurious QNMs, to improve the accuracy of our hyperfitting models, and to shed further light on the relation between the amplitudes and phases of different modes.

\begin{figure*}[t!]
	\centering
	\includegraphics[width=0.99\textwidth]{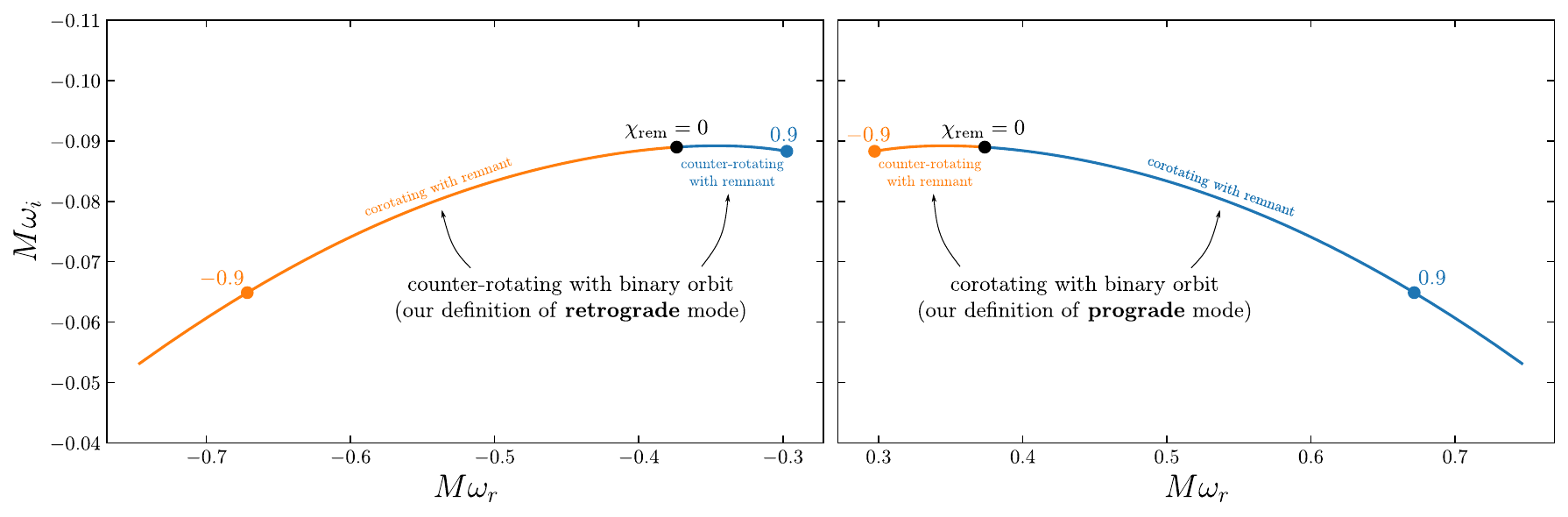}
	\caption{
	Our definition of prograde and retrograde modes.
	For illustration, we show the two QNM frequency solutions for $\ell{,}m = 2{,}2$.
	We show the negative-$\omega_r$ region of the complex plane in the left panel, and the positive-$\omega_r$ region in the right panel.
	The blue (orange) branches correspond to the case when the remnant BH spin is aligned (antialigned) with the BBH orbit.
	Each branch goes from $|\chi_{\rm rem}| = 0$ up to $0.95$, with the points at $|\chi_{\rm rem}| = 0.9$ labeled by circles.
	In our definition, we call both branches in the right panel the prograde mode ($2{,}2{,}0$), while those in the left panel are the retrograde mode ($r2{,}2{,}0$).
	This definition generalizes to other $\ell{,}m$ multipoles, but for $m < 0$ the plot should be mirrored (i.e., $\omega_r \to - \omega_r$).
	With this definition, we have implicitly assumed that the waveform is extracted in a frame such that the $z$-axis is coaligned with orbital angular momentum of the BBH.
	}
	\label{fig:retro_definition}
\end{figure*}

\acknowledgments
We thank Costantino Pacilio, Swetha Bhagwat, Francesco Nobili and Davide Gerosa for pointing out that the hyperfit functions should not go to zero in the equal mass limit when $\chi_- \neq 0$.
We thank Alessandra Buonanno, Giada Caneva Santoro, Vitor Cardoso, Gregorio Carullo, Adrian Chung, Marina de Amicis, Jonathan Gair, Leda Gao, Thomas Helfer, Lionel London, Sizheng Ma, James Marsden, Andrea Maselli, Keefe Mitman, Peter James Nee, Lorenzo Pierini, Jaime Redondo-Yuste, Laura Sberna, Alexandre Toubiana, Vijay Varma, Sebastian V\"olkel and Nicol\'as Yunes for discussions.
M.H.Y.C. is a Croucher Scholar supported by the Croucher Foundation.
M.H.Y.C., E.B. and R.C. are supported by NSF Grants No. AST-2006538, PHY-2207502, PHY-090003 and PHY-20043; by NASA Grants No.~20-LPS20-0011 and 21-ATP21-0010; and by the John Templeton Foundation Grant 62840.
M.H.Y.C., E.B. and R.C. acknowledge support from the ITA-USA Science and Technology Cooperation program, supported by the Ministry of Foreign Affairs of Italy (MAECI), and from the Indo-US Science and Technology Forum through the Indo-US Centre for Gravitational-Physics and Astronomy, grant IUSSTF/JC-142/2019.
The authors also acknowledge the Texas Advanced Computing Center (TACC) at The University of Texas at Austin for providing resources that have contributed to the research results reported within this paper~\cite{10.1145/3311790.3396656}.
We made use of the following \texttt{python} packages to produce the results of this paper: \texttt{adjustText}~\cite{ilya_flyamer_2023_7648985}, \texttt{Jax}~\cite{jax2018github}, \texttt{JaxFit}~\cite{2022arXiv220812187H}, \texttt{matplotlib}~\cite{Hunter:2007}, \texttt{numpy}~\cite{harris2020array}, \texttt{pandas}~\cite{reback2020pandas,mckinney-proc-scipy-2010}, \texttt{qnm}~\cite{Stein2019}, \texttt{scikit-learn}~\cite{scikit-learn}, \texttt{scipy}~\cite{2020SciPy-NMeth}, \texttt{scri}~\cite{mike_boyle_2020_4041972} and \texttt{sxs}~\cite{boyle_michael_2023_8021607}.

\appendix

\section{Definition of prograde and retrograde modes}
\label{app:retro_definition}

Given a BH remnant with spin $\chi_{\rm rem}$, there are two solutions for the QNM frequencies with given values of $\ell$ and $m$: one is corotating with the BH remnant, and the other is counterrotating.
In the literature, these are usually called ``prograde'' and ``retrograde'' modes, respectively.
Here we employ a different definition that is more useful for our work.
We define the modes corotating (counterrotating) \textit{with the BBH orbit} to be prograde (retrograde), respectively.
This definition is clarified in Fig.~\ref{fig:retro_definition}.
We implicity assume that the waveform is extracted in a frame aligned with the BBH angular momentum, so prograde modes with positive frequency are corotating with the BBH orbit.
We also assume that $\chi_{\rm rem}<0$ corresponds to the remnant BH spin being antialigned with the orbital angular momentum.

We use this unconventional definition because in BBH simulations, the orbital direction is easy to control, while the remnant spin direction depends nontrivially on the initial conditions.
By using the present definition, when we vary the initial conditions (e.g. $q, \chi_1, \chi_2$) smoothly, the amplitude and phase of the mode will also vary smoothly, \textit{even when the spin direction of the remnant flips.}
A similar behavior can be observed for the QNM frequency of a mode: for example, both the prograde mode ($2{,}2{,}0$, right panel) and the retrograde mode ($r2{,}2{,}0$, left panel) shown in Fig.~\ref{fig:retro_definition} evolve smoothly on the same line from negative to positive $\chi_{\rm rem}$.
If we were to use the conventional definition, the QNM frequencies, amplitude and phase would undergo nonsmooth transitions when the remnant spin flips.

To summarize, our definition follows the usual convention for remnants with spins aligned with the BBH orbital direction (which is the case for the large majority of the nonprecessing simulations in the SXS catalog).
However, when the remnant spin is flipped, our definition is the opposite of the usual convention.

\section{Technical details of the fitting algorithm}
\label{app:fitter}

Throughout this paper we use nonlinear least squares fitting to extract QNMs from the ringdown.
In this appendix we present the technical details of the fitting algorithm.

For both the frequency-agnostic and fixed-frequency fits, we made use of an implementation of the trust-region reflective method in the \texttt{JaxFit} package~\cite{2022arXiv220812187H}, optimized with \texttt{Jax}~\cite{jax2018github}.
Although a trust-region method supports the imposition of bounds, we find that the fitting algorithm finds lower minima more effectively when we do not specify any bounds.
Thus, we use the boundless implementation of trust-region in \texttt{JaxFit}.
We use a least-squares loss function, implemented as the square of the norm of the complex residue due to the complex nature of our waveform.

For most of the postmerger waveforms examined in this work, we fit the waveform at different starting times $t_0$, and we use values of $t_0 \in (0, 50) M$ in steps of $\Delta t = 0.1 M$ unless otherwise specified.
We do the fits in series, starting at $t_0 = 0$, and then using the results of the current time step as the initial guess of the next time step.
At $t_0 = 0$, if we are doing a fixed-frequency fit, an arbitrary guess suffices because there are fewer free parameters, as long as the guess is not too many orders of magnitude away from the best-fit value, so we initialize $A_k$ at $|h_{\rm peak}|$ and $\phi_k$ at $1$ for all modes.
As for the frequency-agnostic fits, as it could be harder for the fit to locate a global minimum, or at least a good enough minimum, we use $10$ initial guess points and select the results that give the lowest loss.
The initial guess for $A$ of each mode is drawn from a log-uniform distribution $A \in (0.1, 10) A_{\rm peak}$, while the other parameters are drawn from uniform distributions $\phi \in (0, 2 \pi)$, $\omega_r \in (-2, 2)$, $\omega_i \in (-1, 0)$.

For the trust-region algorithm, we set the maximum number of iterations to $200{,}000$.
When there are many modes in our fitting model, the fit will not converge within the maximum number of iterations for some time steps.
If that is the case, we skip that time step and move on to the next one, using the results from the last converged fit as an initial guess.
When we are identifying the potential modes from the frequency-agnostic fits, or when we are determining the stability of a mode with the fixed-frequency fits, we take a conservative approach by deeming the skipped time-steps as data points that do not meet the criteria we imposed.

\section{Toy waveforms}
\label{app:toy}

We now introduce two classes of toy waveforms that will be used in Appendix~\ref{app:consistency} to test our fitting algorithm, and in Appendix \ref{app:procedure} to test our QNM extraction procedure.

In the literature, there are two points of view regarding the QNM content of waveforms.
The first one argues on physical grounds that the QNMs are unclean (because they have an evolving amplitude/phases and because they are contaminated by other physics, including nonlinearities) close to the merger peak, and they can only be cleanly identified at later times ($\sim 10 M$ after the peak). The second argues phenomenologically that the postmerger can be specified by a combination of clean overtones starting at the peak.
Our goal is to construct two toy waveforms that follow the above hypotheses (a more systematic study will be performed in Appendix~\ref{app:procedure}).

The first class of toy waveforms consists of QNMs that ``turn on slowly'' following a sigmoid function, approaching a stable QNM at late times:
\begin{align}
	\tilde{A}_k(t)    & =  A_k\left(1- \dfrac{A_{k, \rm red}}{2}\left[1 - \tanh\left(\dfrac{t - t_{A_k}}{\sigma_{A_k}}\right)\right]\right), \\
	\tilde{\phi}_k(t) & = \phi_k - \dfrac{\phi_{k, \rm red}}{2}\left[1- \tanh\left(\dfrac{t - t_{\phi_k}}{\sigma_{\phi_k}}\right)\right],    \\
	h(t)              & = \sum_{k \in K} \tilde{A}_k(t) e^{\omega_{i,k} t}e^{-i (\omega_{r,k} t + \tilde{\phi}_k(t))} + n(t),
\end{align}
where $A_{k, \rm red}$ governs the degree of suppression of the mode amplitude at earlier times, while $t_{A_k}$ and $\sigma_{\phi_k}$ govern the time scale of the evolution, and similarly for $\phi$.
We also add a Gaussian white noise component $n(t)$ to the toy waveform.
We use Kerr QNMs of mode index $k$, so $\omega_{r, k}$ and $\omega_{i,k}$ are given by their Kerr values.
To avoid fine-tuning, we draw $A_k$ and $\phi_k$ randomly from uniform distributions, except for the natural fundamental mode (i.e., the $2{,}2{,}0$ mode if we are constructing a toy waveform for the $\ell{,}m = 2{,}2$ multipole, which is our main focus in these appendixes, although the results should generalize to other multipoles) because only the relative amplitude/phases between modes matter.
We always fix $A_{\rm fund}$ and $\phi_{\rm fund}$, the amplitude and phase of the natural fundamental mode, to be $1$ and $0$ respectively.

To push the limits of our algorithm, we include $10$ Kerr modes in our toy waveform, namely $K = \{2{,}2{,}0|\linebreak[0] 2{,}2{,}1|\linebreak[0]3{,}2{,}0 |\linebreak[0] 4{,}2{,}0 |\linebreak[0] 3{,}3{,}0 |\linebreak[0]2{,}1{,}0 |\linebreak[0]4{,}4{,}0 |\linebreak[0] r2{,} 2{,} 0 |\linebreak[0]r3{,}2{,}0 |\linebreak[0] {\rm constant}\}$.
We also impose a loose hierarchy among the modes by drawing the first overtone amplitude from a uniform distribution $(3, 5)$, the spheroidal-mixing modes from $(1 \times 10^{-3}, 1 \times 10^{-2})$, the recoil modes from $(1 \times 10^{-4}, 1 \times 10^{-2})$, mirror modes from $(1 \times 10^{-5}, 1 \times 10^{-3})$, and the constant ``mode'' from $(1 \times 10^{-6}, 5 \times 10^{-5})$.
The phases are all drawn uniformly from $(0, 2\pi)$.
We use random realizations of stationary Gaussian noise with standard deviation $\sigma_n = 5 \times 10^{-7}$.
We choose not to inject the estimated resolution error of an SXS simulation because it will likely contain components that resemble Kerr QNMs.
The resolution error is estimated by computing the difference between the simulations of the highest and second-highest resolutions. The two waveforms will both contain the same Kerr QNMs, albeit with slightly different amplitude/phase/frequency, and their difference could still resemble a Kerr mode with lower amplitude.

The second class of toy waveforms corresponds to the second hypothesis, for which the waveform contains clean overtones as early as the peak with little contamination.
Explicitly, these toy waveforms contain QNMs with strictly constant amplitudes and phases:
\begin{equation}
	h(t) = \sum_{k \in K} A_k e^{\omega_{i,k} t}e^{-i (\omega_{r,k} t + \phi_k)} + n(t).
\end{equation}
For simplicity, we include only the modes $K = \{2{,}2{,}0|\linebreak[0]2{,}2{,}1 |\linebreak[0] 2{,}2{,}2 |\linebreak[0] 2{,}2{,}3 |\linebreak[0]2{,}2{,}4 |\linebreak[0] 3{,}2{,}0\}$ in the waveform.
The amplitude and phase of the $2{,}2{,}0$ mode are fixed at $1$ and $0$, respectively.
The amplitude of the $2{,}2{,}1$ mode is drawn uniformly from $(3, 5)$, $2{,}2{,}2$ from $(8, 12)$, $2{,}2{,}3$ from $(12, 24)$, $2{,}2{,}4$ from $(20, 30)$, and $3{,}2{,}0$ from $(1 \times 10^{-3}, 1 \times 10^{-2})$.
Similar to the previous toy waveform, we inject Gaussian noise $n(t)$ with standard deviation $\sigma_n = 5 \times 10^{-7}$.

\section{Consistency test of the fitting algorithm}
\label{app:consistency}

In the main text we fit the postmerger waveform with up to $N_f = 10$ free QNMs using model~\eqref{eq:modelA}, corresponding to $40$ free parameters in total.
In general it will be difficult for the algorithm to locate the global minimum of the least-squares cost function in such a high-dimensional space.
Nonetheless, we will see that we do not have to insist on finding the global minimum best fit, because a \textit{local} minimum could already serve our purposes.
As long as we have fitted all of the physical (QNM) content of the waveform by diving below the noise floor, we do not have to find a better minimum, because by doing so we are only fitting the noise better.
Moreover, even if we only found a local minimum, we will show that a free QNM will approach a Kerr mode as long as it stays long enough above the noise floor.

We first test the behavior of the fitting algorithm with the two waveforms specified in Appendix~\ref{app:toy}.
Recall that, for the toy waveforms of the first class, the waveform approaches a combination of constant amplitude/phase QNMs at late times (modulo Gaussian noise), while those of the second class are always pure QNMs.
To test the ability of the algorithm to find an adequate least-squares minimum, we perform the same fit with different initial guesses.
For each free QNM $k$ in the fitting model~\eqref{eq:modelA}, we draw the initial guess of $A_k$ from a log-uniform distribution in $(0.1, 10)h_{\rm peak}$, where $h_{\rm peak}$ is the amplitude of the waveform at $t_{\rm peak}$.
We also draw the initial guesses for $\phi_k$ from $(0, 2\pi)$, $\omega_r$ from $(-2, 2)$, and $\omega_i$ from $(-1, 0)$.
Note that we do \textit{not} put bounds on $\phi$ in the trust region algorithm for nonlinear least squares fitting, but we take $\phi_k \! \mod 2\pi$ as the extracted phase value.
Nonetheless, an initial guess drawn between in $(0, 2\pi)$ covers adequately the relevant space of values that $\phi_k$ can take.

\begin{figure}
	\centering
	\includegraphics[width=0.49\textwidth]{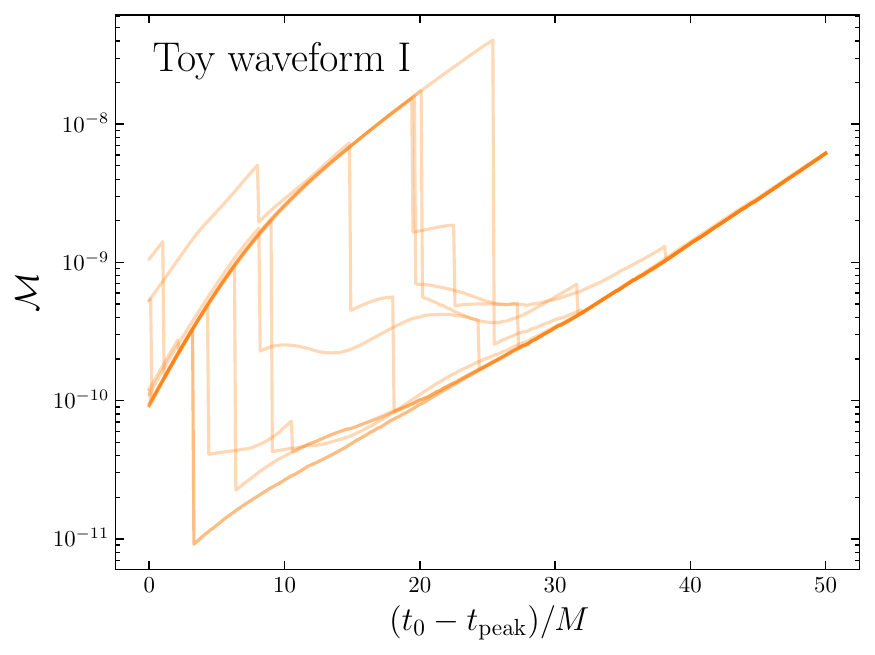}
	\includegraphics[width=0.49\textwidth]{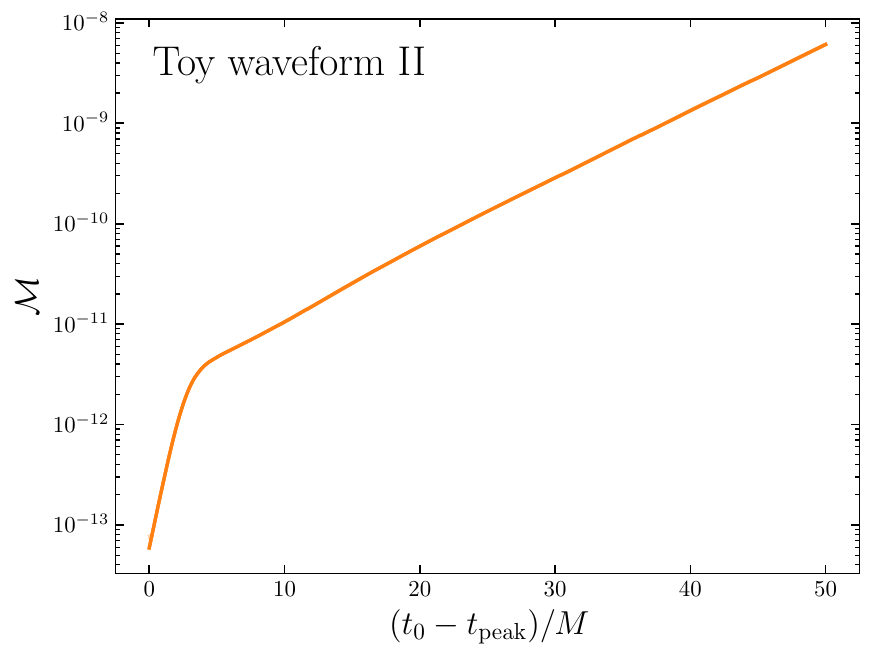}
	\caption{
		Fit mismatch versus fit starting time for different initial guesses on a realization of the toy waveform of the first kind (top panel) and of the second kind (bottom panel).
		Each orange curve corresponds to a series of fits where the initial guess of the fitting algorithm at $t_0 - t_{\rm peak} = 0 M$ is drawn randomly (a different realization for a different curve), and subsequent time steps make use iteratively of results from the previous time step as an initial guess.
	}
	\label{fig:toy_convergence}
\end{figure}

\begin{figure*}
	\centering
	\includegraphics[width=0.99\textwidth]{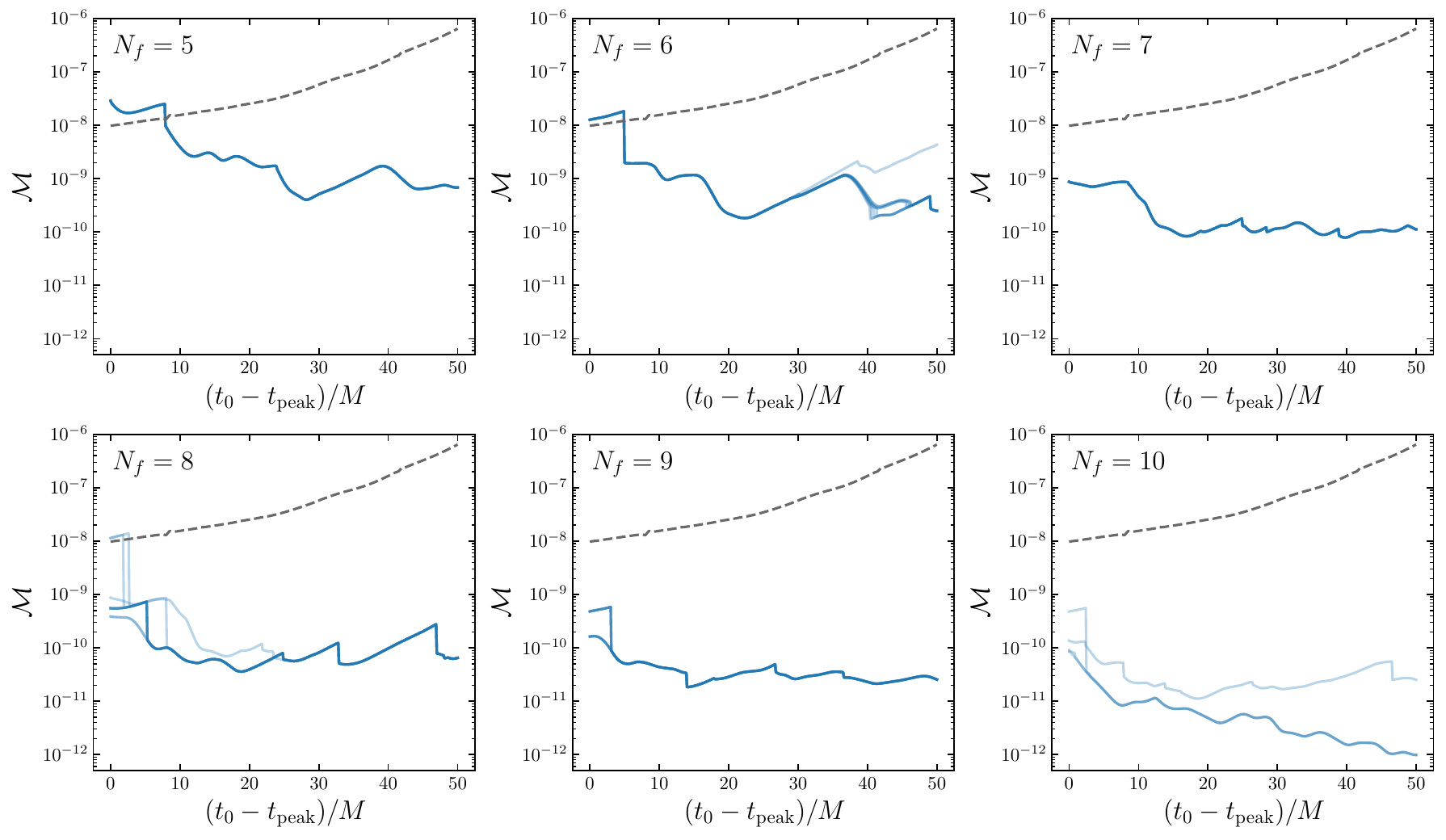}
	\caption{
		Fit mismatch for different initial guess realizations for SXS:BBH:0305, $\ell{,}m = 2{,}2$.
		We use model~\eqref{eq:modelA} for these fits.
		Different panels correspond to fits using a different number of free QNMs $N_f$.
		Each blue curve corresponds to a series of fits where the initial guess of the fit algorithm at $t_0 - t_{\rm peak} = 0 M$ is drawn randomly, and subsequent time steps make use of the previous time step results as initial guesses.
		The dashed line indicates the mismatch induced by the resolution error of the simulation, which is estimated by the (phase-minimized) difference between the waveforms extracted from the best and second-best resolution simulations.
		At any given $t_0$ in any panel, either most of the blue curves are under the resolution-induced mismatch, or all of them converge to the same values.
	}
	\label{fig:0305_convergence}
\end{figure*}

In Fig.~\ref{fig:toy_convergence} we plot the fit mismatch versus $t_0$ for $10$ runs with different initial guesses on the same waveform.
We do this separately for a realization of the toy waveform of the first class (top panel) and of the second class (bottom panel).
Recall that for a particular time step $t_0$, we use the best-fit value from the previous time step $t_0 - \Delta t_0$ as the initial guess, so the initial guess that we specify is only used for the first time step ($t_0 - t_{\rm peak} = 0 M$).

For the toy waveform of the first class, no matter which initial guess we choose for $t_0 - t_{\rm peak} = 0 M$, the fits always converge to the same mismatch at times $t_0 - t_{\rm peak} \gtrsim 40$, when the waveform approaches a combination of clean QNMs.
For the toy waveform of the second class the fit mismatch is always the same, regardless of the fit starting time $t_0 - t_{\rm peak}$ or the initial guess realization.
This test suggests that \textit{if a waveform were a combination of clean QNMs (as long as we fit it with a number $N_f$ of free QNMs greater than its actual QNM content), the goodness of fit would be minimally affected by the initial guess}.

The disagreement between the fits with different initial guesses when the waveform is not a clean combination of QNMs might look concerning.
However, we know that at early times the waveform contains nonQNM contributions anyway, and this is the reason for the disagreement in the first place.
At early times, if we were to insist in looking for the global minimum, we could have used a large number of initial guesses that exhaustively cover the relevant parameter space and select the instance that gives the lowest mismatch. Such an attempt would not help us extract more of the \textit{QNM content of the waveform} as long as the original (local) minimum already has a mismatch lower than that induced by nonQNM components, such as numerical noise.
Nonetheless, as mentioned in Appendix~\ref{app:fitter}, we use $10$ initial guesses for each waveform and select the best instance. In our experience this strikes a balance between the computational resources required and the ability of the fit to find an adequate minimum.
In fact, as seen in the top panel of Fig.~\ref{fig:toy_convergence}, the fit mismatch is always below $10^{-7}$ -- well below the noise level, regardless of the initial guess we use.

Next, we perform the same test on an NR waveform, namely the $\ell{,}m = 2{,}2$ multipole of SXS:BBH:0305.
The results are shown in Fig.~\ref{fig:0305_convergence}.
For $N_f \in \{5, 6, \dots, 10\}$, a majority of the $10$ initial guesses give similar mismatch curves.
More importantly, they either converge to the same mismatch (e.g. for $N_f = 5$ and $6$, close to $t_0 - t_{\rm peak} \sim 0 M$), or they dive beneath the resolution-error induced mismatch for $N_f \geq 7$.

\begin{figure*}
	\centering
	\includegraphics[width=0.99\textwidth]{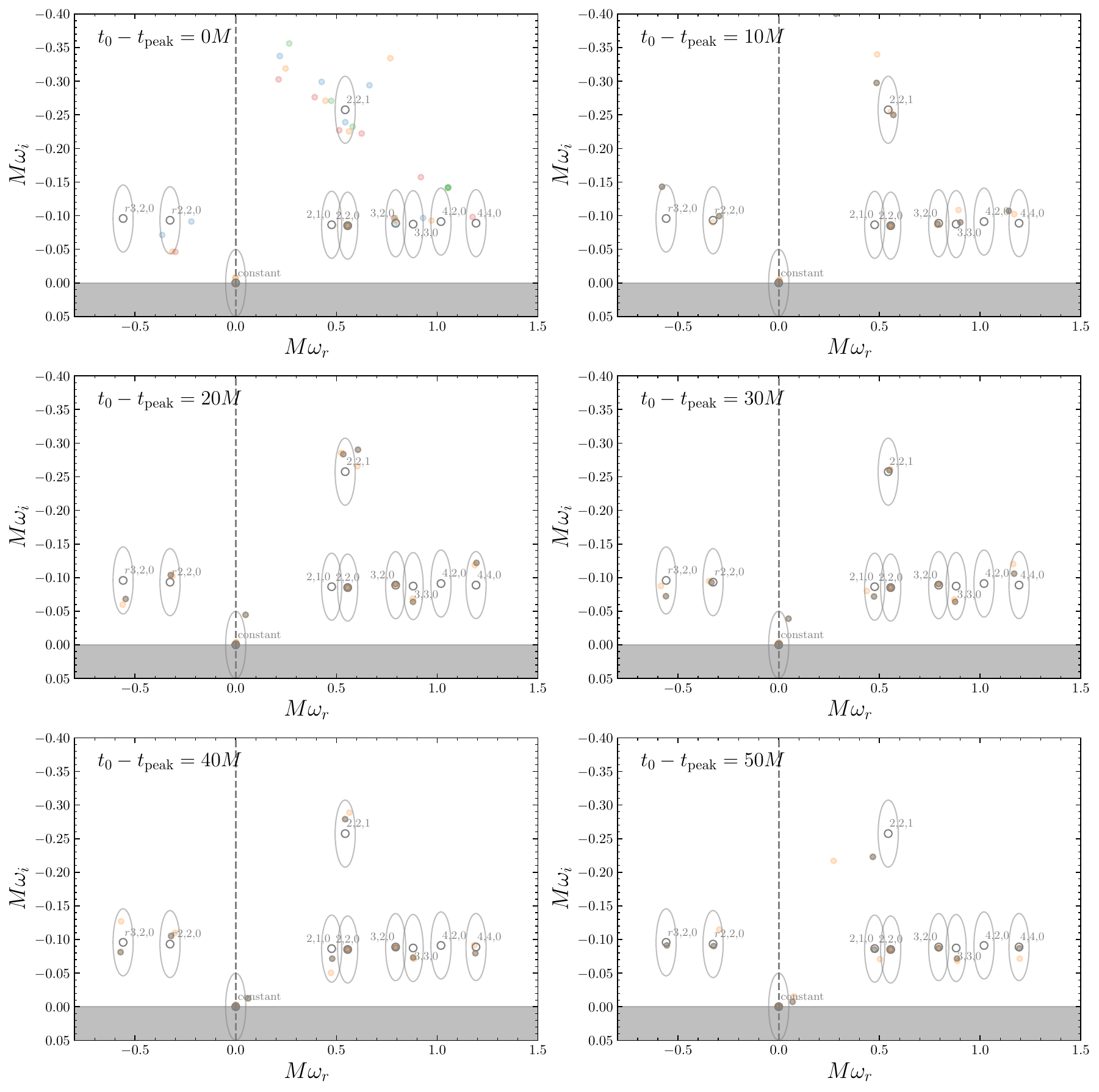}
	\caption{
		Fitted QNM frequencies at different $t_0$ with different initial guesses, for the $\ell{,}m = 2{,}2$ multipole of SXS:BBH:0305.
		Different panels correspond to different $t_0$, and different colored circles in the same panel correspond to the recovered $\tilde{\omega_k} = \omega_{k, r} + i \omega_{k, i}$ for different initial guesses.
		We make the circles transparent to better visualize the locations where different points overlap, which appears with a deep gray color.
		For each initial guess, we fit the waveform with $N_f = 10$ free QNMs.
		Notice how at late times ($t_0 - t_{\rm peak} \gtrsim 40$), different initial guesses give mostly overlapping points.
		The same qualitative behavior can be seen in other waveforms in general:
		as long as a Kerr mode is stable above the noise long enough at a given time, free QNMs will converge around the mode, regardless of the initial guess.
	}
	\label{fig:0305_lm_22_omega_vs_t}
\end{figure*}

Although the mismatch converges for different initial guesses when the data matches the model, there is no guarantee that the recovered $\omega_{k,r}$ and $\omega_{k,i}$ are also minimally affected by the initial guess.
As shown in Fig.~\ref{fig:0305_lm_22_omega_vs_t}, for SXS:BBH:0305 with $\ell{,}m = 2{,}2$, the values of $\omega_{k,r}$ and $\omega_{k,i}$ obtained with different initial guesses could differ initially at $t_0 - t_{\rm peak} = 0 M$, but most of them converge to the values corresponding to the initial guess that gives the lowest mismatch at late times. An exception are the points around the $2{,}2{,}1$ mode, which converge at around $t_0 - t_{\rm peak} = 30 M$ only, the time when the $2{,}2{,}1$ mode can be cleanly identified above the noise without a significantly varying amplitude or phase.
The two toy waveforms examined earlier, as well as other $\ell{,}m$ multipoles of SXS:BBH:0305, show the same qualitative behavior as Fig.~\ref{fig:0305_lm_22_omega_vs_t}: the consistency between free QNMs of different initial guesses is better around a Kerr mode when it is stably above the noise.

\begin{figure*}
	\centering
	\includegraphics[width=0.99\textwidth]{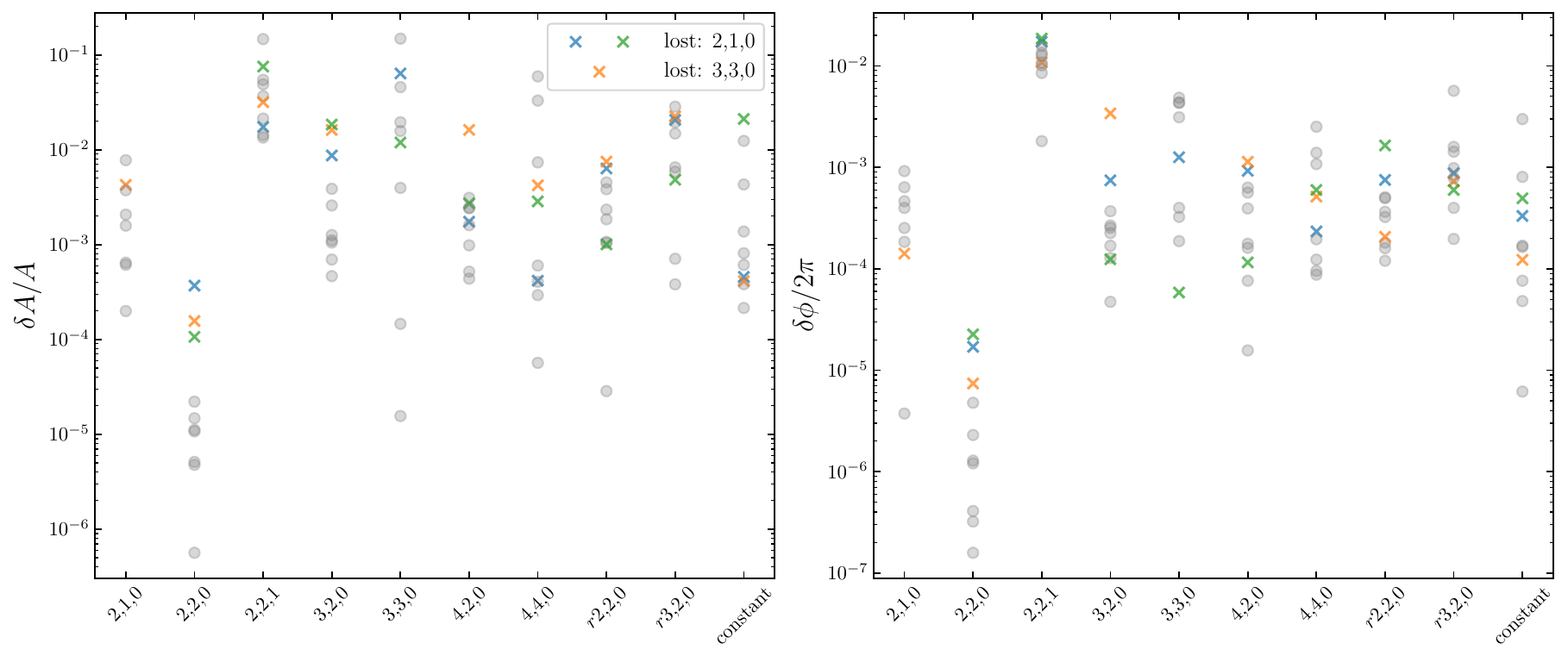}
	\caption{
	Mode extraction results for a constructed toy ringdown waveform containing the modes $K = \{2{,}2{,}0| 2{,}2{,}1|3{,}2{,}0 | 4{,}2{,}0 | 3{,}3{,}0 | 2{,}1{,}0 | 4{,}4{,}0 | r2{,} 2{,} 0 | r3{,} 2{,}0 | {\rm constant}\}$, for which the remnant BH spin is set to be $\chi_{\rm rem} = 0.7$.
	We use ten realizations of the waveform, each constructed by drawing the amplitudes and phases of different modes randomly, while loosely respecting a hierarchy as prescribed in Appendix.~\ref{app:toy}.
	Each ``column'' of points represents the relative extraction error of the amplitude (left panel) and phase (right panel), defined as $\delta A / A$ and $\delta \phi / 2 \pi$, for different realizations of the waveform.
	Gray circles are the points corresponding to realizations for which all of the injected modes are identified by our procedure without any spurious modes, while colored crosses correspond to realizations where some modes are missed by the algorithm.
	Crosses of the same color correspond to the same realization, and the mode (or modes) that are missed for each realization are labeled in the legend.
	}
	\label{fig:eff_a_0.7_delta}
\end{figure*}

However, if certain Kerr modes exist only briefly above the noise floor (e.g., a mode with very low amplitude or one with an overtone number $n \geq 2$), they might not be picked up by the free QNMs at all,
especially if $N_f$ is smaller than the number of Kerr modes actually present in the waveform.
Some free QNMs might try to pick up these modes, but the precise value of the free QNM could depend on the initial guess -- i.e., at a particular $t_0$ different initial-guess runs would give different frequencies around the brief Kerr mode, forming a small spread.
In our procedure, we introduce a margin for error by requiring only that the free QNM approaches a Kerr mode within an ellipse in the complex plane. This allows for a small bias due to an initial guess that does not return the global minimum.
If the bias is larger than this margin of error, the Kerr mode that the free QNM is trying to pick up has too low an amplitude (or decays too quickly) anyways, so it is not too harmful to lose the mode.
On the other hand, if $N_f$ is larger than the number of Kerr modes that are consistently above the noise floor, some free QNMs will be (over)fitting the nonQNM components of the waveform, and their frequency evolution could depend significantly on the initial guess.
As explained in the main text, this is not a big concern, because even if the overfitting causes us to pick up spurious modes, they will be removed in a later step of the mode-extraction procedure (i.e., they will fail the mode stability tests).

With the above caveats in mind, the results in this appendix indicate that while there is no guarantee that we can find the global least-squares minimum,
\textit{as long as a Kerr mode stably exists above the noise floor for a long enough time, the mode will most likely be picked up, no matter the initial guess.}
Using $\sim 10$ different realizations of the initial guess and picking the best instance (the one that gives the lowest mismatch) is adequate for extracting all of the QNM components of a waveform.

\section{Test of the QNM extraction procedure}
\label{app:procedure}

In this appendix we test the full QNM extraction procedure on the toy waveforms constructed in Appendix~\ref{app:toy}.
We will stress test our procedure by considering toy waveforms with $10$ or more modes, including higher overtones.
We will also consider waveforms containing modes with frequencies that are close to each other, which will be the case when $\chi_{\rm rem}$ is small.

\subsection{When $n \geq 2$ overtones are absent}

\begin{figure*}
	\centering
	\includegraphics[width=0.99\textwidth]{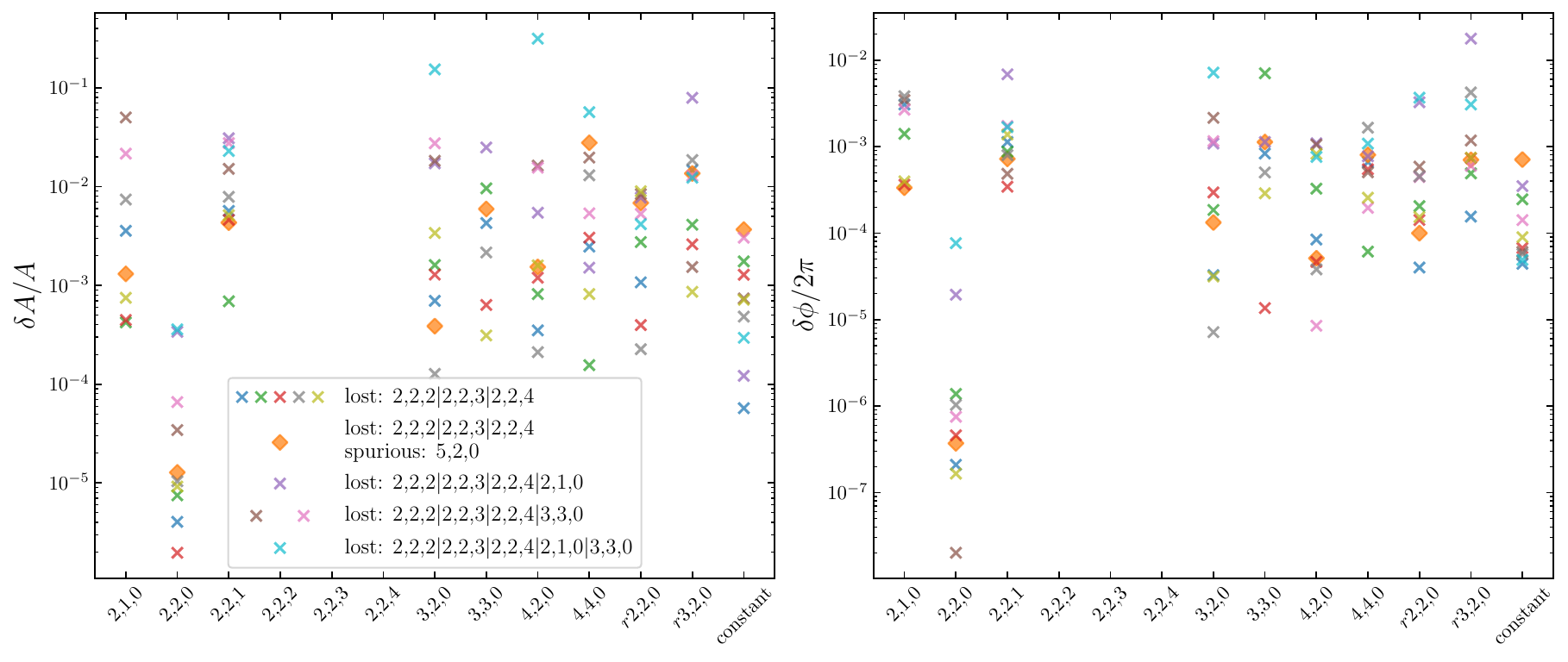}
	\includegraphics[width=0.99\textwidth]{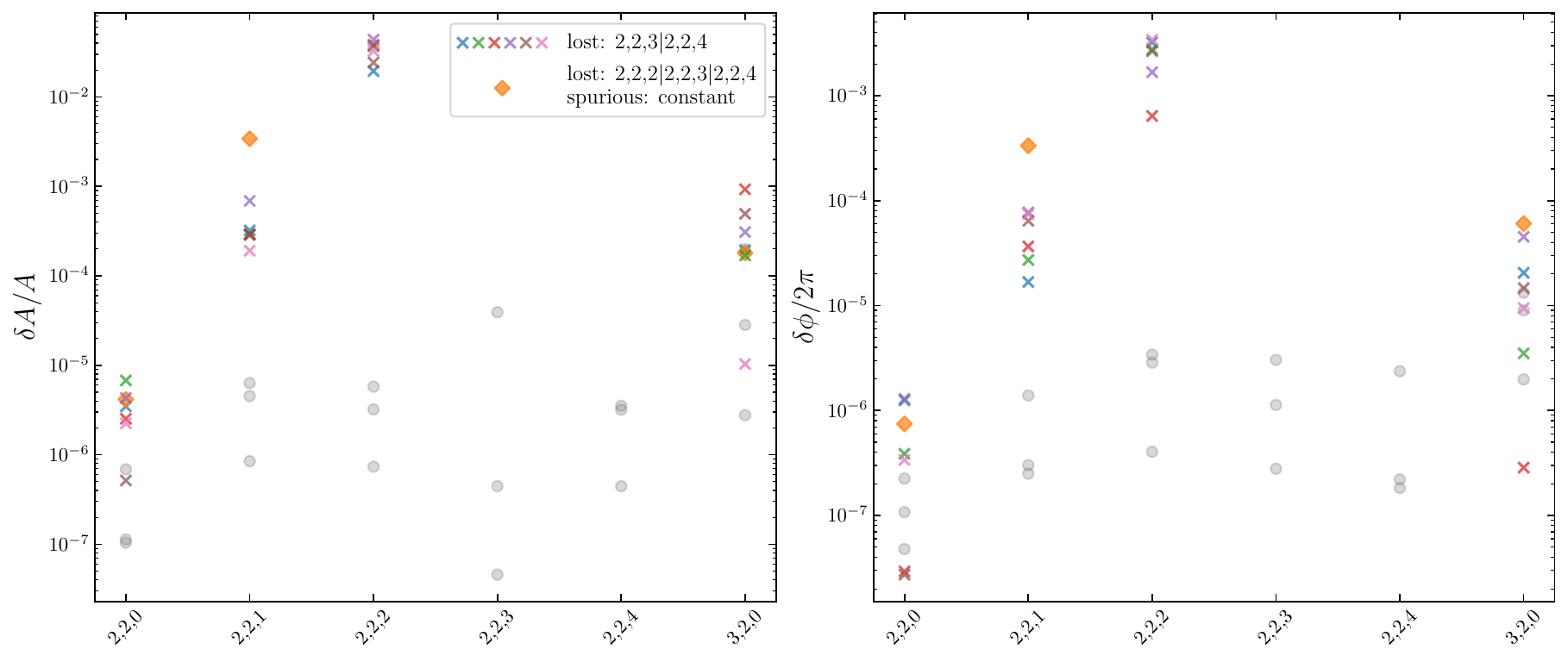}
	\caption{
	Same as Fig.~\ref{fig:eff_a_0.7_delta}, but for toy waveforms including $4$ overtones $\{2{,}2{,}1|2{,}2{,}2|2{,}2{,}3|2{,}2{,}4 \}$ and keeping all the other modes used in Fig.~\ref{fig:eff_a_0.7_delta} (top), and including only the $2{,}2{,}0$ and $3{,}2{,}0$ mode in addition to the overtones (bottom).
	We label realizations where modes are missed as crosses and those with both missed modes and spurious modes as diamonds.
	Points shown as gray circles correspond to realizations with no missed modes or spurious modes.
	}
	\label{fig:eff_a_0.7_delta_overtone}
\end{figure*}

\begin{figure*}
	\centering
	\includegraphics[width=0.99\textwidth]{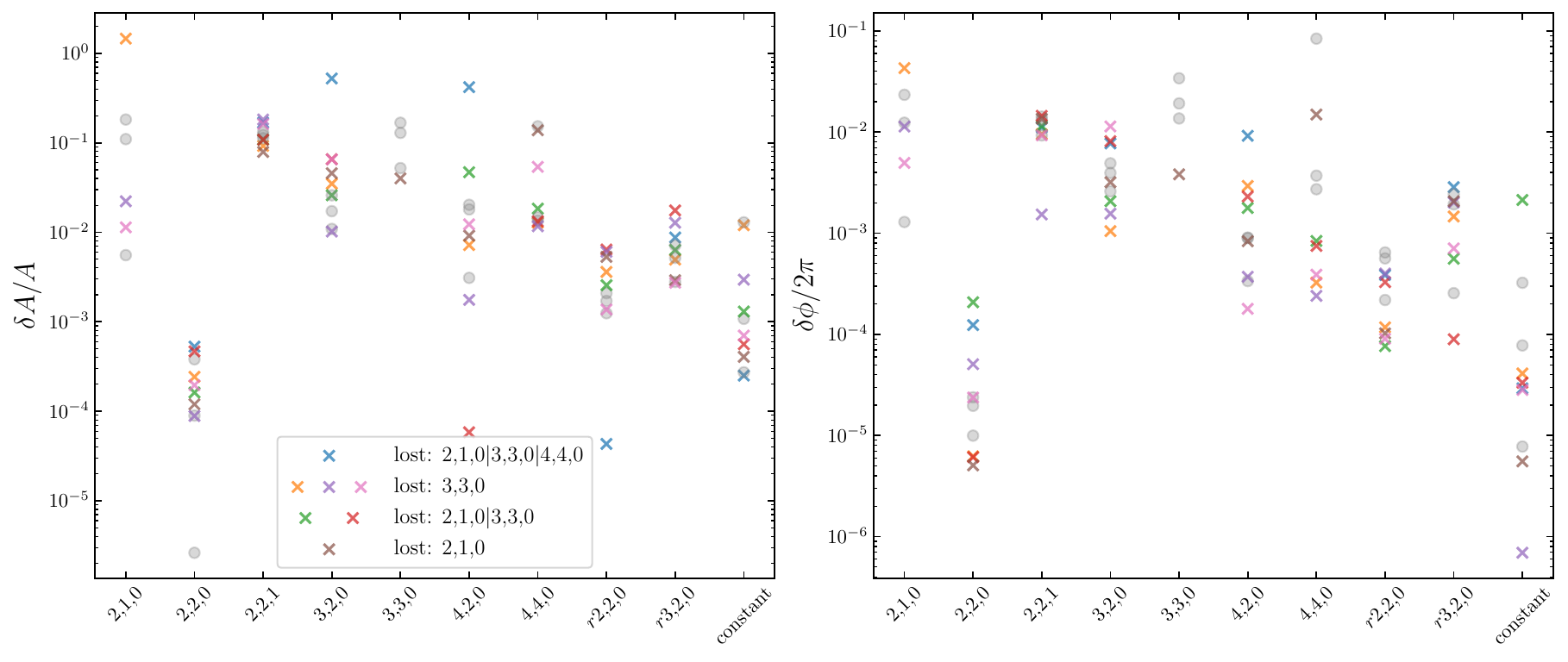}
	\includegraphics[width=0.99\textwidth]{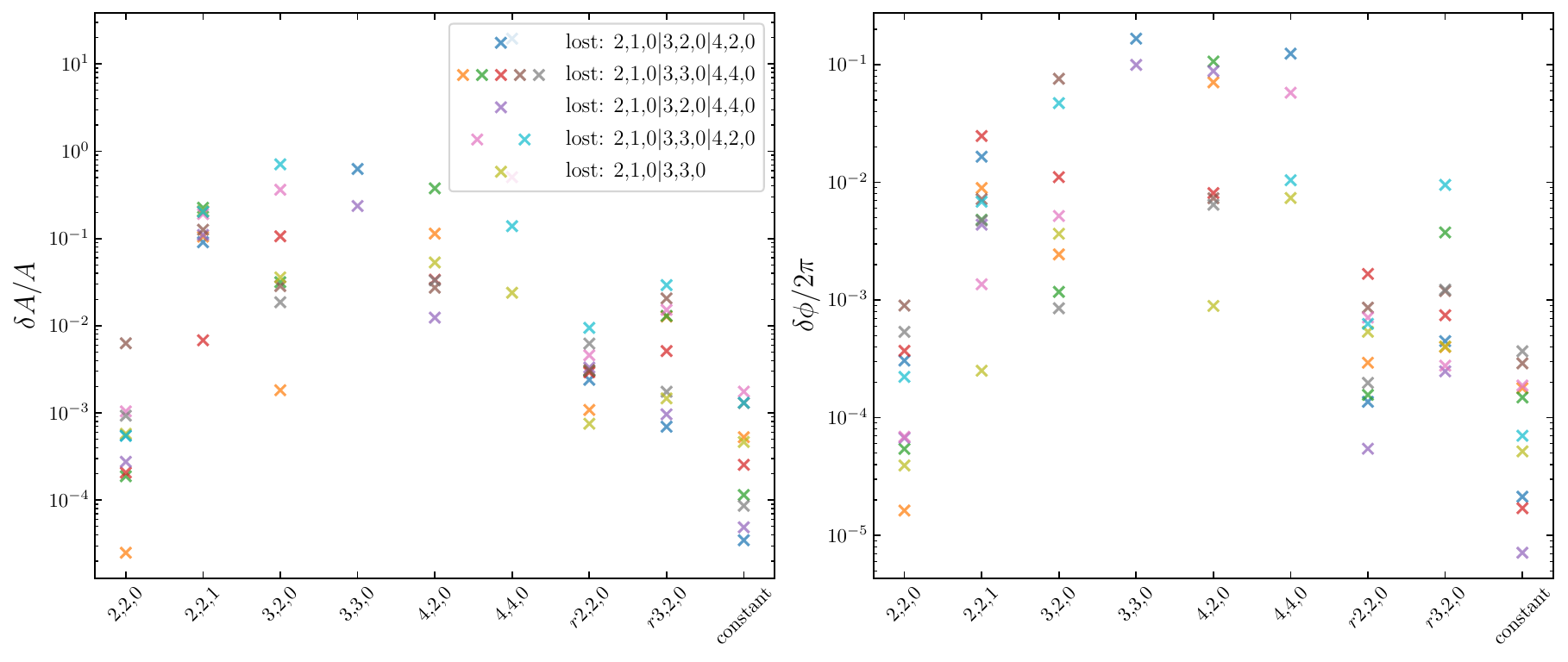}
	\caption{
		Same as Fig.~\ref{fig:eff_a_0.7_delta}, but for the case $\chi_{\rm rem} = 0.1$ (top) and $0.01$ (bottom).
	}
	\label{fig:eff_a_0.1_0.01_delta}
\end{figure*}

Let us first apply our procedure to a toy waveform containing fundamental modes but without overtones.
The toy waveform is constructed by randomizing the amplitudes and phases of the modes, while maintaining a loose hierarchy of the amplitudes between them.
We ``distort'' the modes close to the merger peak and include a Gaussian noise floor to test the robustness of our procedure to contributions different from the ringdown, that could be present close to the peak and at late times.
We choose a typical remnant spin of $\chi_{\rm rem} = 0.7$, and use the QNM frequencies corresponding to this value.

We apply the full QNM extraction procedure, as described in the main text and schematically represented in Fig.~\ref{fig:flowchart}, and test whether the procedure correctly recovers the ``injected'' modes and their corresponding amplitudes.
We perform this test on $10$ realizations of the toy waveform, each time picking the amplitudes randomly while respecting the amplitude hierarchy. We summarize the results in Fig.~\ref{fig:eff_a_0.7_delta}.
As shown in the legend of Fig.~\ref{fig:eff_a_0.7_delta}, for $3$ out of $10$ of the realizations, the procedure misses one of the injected modes (either the $2{,}1{,}0$ or the $3{,}3{,}0$ mode).
This is expected, as these two modes are the recoil modes, which we assigned to the bottom of the amplitude hierarchy in Appendix~\ref{app:toy}, meaning that their amplitudes are the lowest and close to the noise floor.
For the other $7$ simulations, all of the injected modes are identified, and no spurious modes are detected.
Crucially, no matter if all the modes are detected or not, $\delta A / A$ and $\delta \phi / 2 \pi$ are within $10\%$ most of the time, except for isolated cases involving overtones and recoil modes.
However, note that missing a mode in the procedure often leads to a higher error in the extraction of another mode (in this case, the $2{,}2{,}0$ mode).

\subsection{When $n \geq 2$ overtones are present}

Next, we test whether our procedure is capable of extracting overtones with mode number $n \geq 2$.
Similar to the previous subsection, we construct toy waveforms by including different modes with randomly drawn amplitudes and phases.
However, in contrast to the previous case, we keep the mode amplitudes and phases constant close to the merger peak, meaning that the QNMs are ``clean'' throughout the whole ringdown waveform.
We do this because it has been hypothesized in the literature that overtones exist cleanly close to the merger peak, and we want to test whether our procedure will work well if that is the case.
Recent work has found that, if such a hypothesis were true, the overtone amplitudes will also respect a hierarchy, where the amplitude increase with $n$ for $n \lesssim 5$~\cite{Giesler:2019uxc}.
Therefore, we impose such a hierarchy when drawing the amplitudes of the overtones, as specified at the end of Appendix~\ref{app:toy}.

\begin{figure*}
	\centering
	\includegraphics[width=0.99\textwidth]{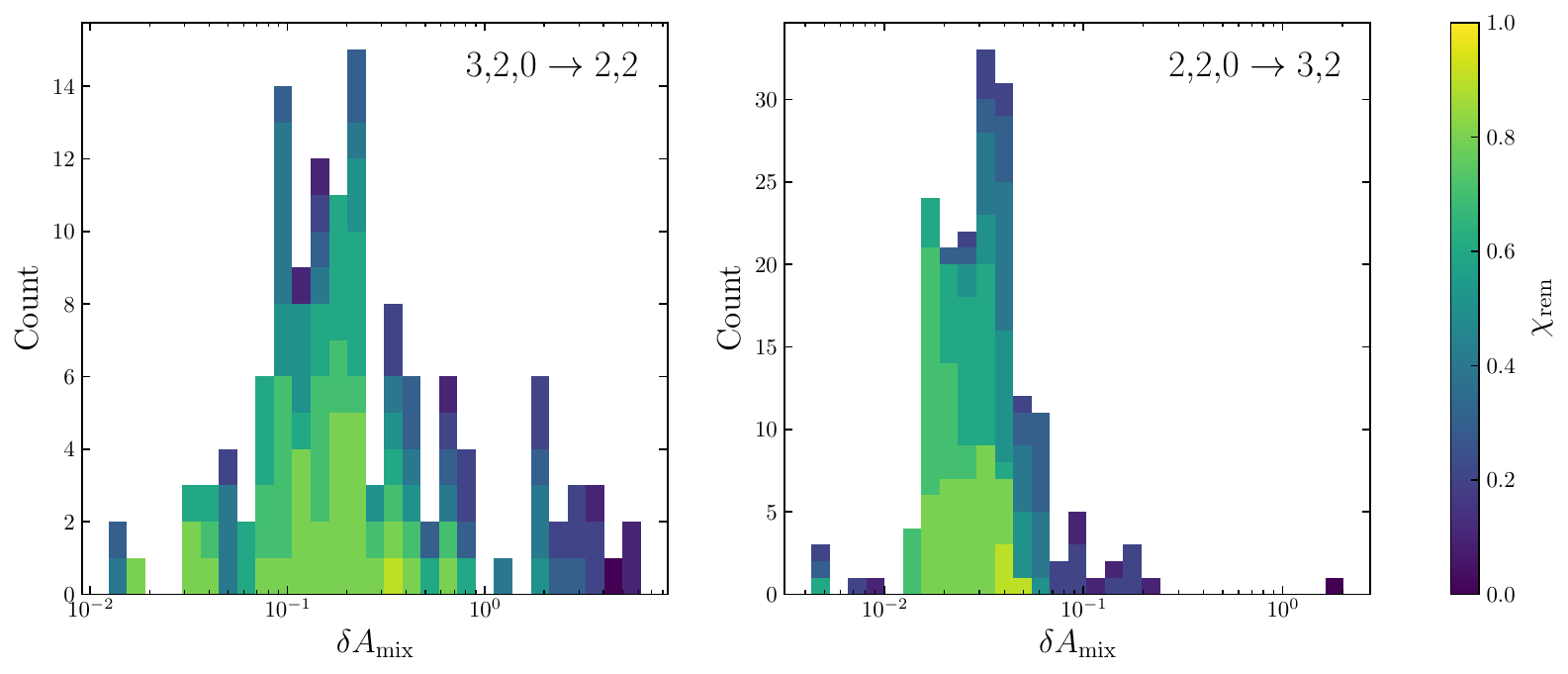}
	\caption{
	The relative difference $\delta A_{\rm mix} / A$, as defined in Eq.~\eqref{eq:delta_A_mix}, between the mixing amplitudes computed in theory and those extracted by our procedure. We consider the $3{,}2{,}0$ mode mixing into the $\ell{,}m = 2{,}2$ multipole (left panel) and the $2{,}2{,}0$ mode mixing into the $\ell{,}m = 3{,}2$ multipole (right panel).
	This plot refers to the SXS simulations SXS:BBH:0209-0305,1419-1509.
	Items in each bin correspond to individual SXS simulations, and they are colored by the value of their remnant spin $\chi_{\rm rem}$.
	}
	\label{fig:22_32_mixing}
\end{figure*}

The top row of Fig.~\ref{fig:eff_a_0.7_delta_overtone} shows the mode extraction results for the case when we include the $n = 2, 3, 4$ overtones, in addition to the $10$ modes included in the previous section.
We find that our procedure always prefers to pick up modes that are less quickly decaying, instead of the $n \geq 2$ overtones~\cite{Zhu:2023mzv}.
To test whether it is inherently impossible to find these overtones with our procedure, we do another test, this time including only the $3{,}2{,}0$ mode in addition to the $2{,}2{,}n$ fundamental mode and overtones, with $n \leq 4$.
In this case, we can find the $n = 2$ overtone for $9$ out of $10$ realizations, but we still cannot identify the $n \geq 3$ overtones.
Note that the detectability of the modes depends on their amplitude, so if the modes do not follow the hierarchy that we prescribed, the results could be different.

The results in this subsection indicate that our procedure is not suitable for extracting overtones with $n \geq 3$, while it can in principle extract the $n = 2$ mode if few other modes are present.
This is expected, because higher overtones decay faster.
Other methods might be necessary for extracting these overtones. However it is hard to imagine that these higher overtones may be useful in GW data analysis when they cannot even be extracted with a least-squares fit in a minimal amount of noise, as we show here.

\subsection{When $\chi_{\rm rem} \lesssim 0.1$}

In the limit $\chi_{\rm rem} \to 0$, modes labeled by the same $\ell$ and $n$ but by a different azimuthal index $m$ will converge to the same (Schwarzschild) frequency, making it difficult to distinguish between them when fitting the waveform.
This is a problem because of the presence of recoil modes, which can have the same $\ell$ but a different $m$.

In Fig.~\ref{fig:eff_a_0.1_0.01_delta}, we show the mode extraction results for $\chi_{\rm rem} = 0.1$ and $0.01$.
When $\chi_{\rm rem} = 0.1$, at least one of the recoil modes is missed in most of the cases, while for $\chi_{\rm rem} = 0.01$ we always miss either some recoil modes or spherical-spheroidal mixing modes.
Comparing with Fig.~\ref{fig:eff_a_0.7_delta}, we can also see that the amplitudes and phases of the other modes are not recovered as accurately when the remnant spin is low.
For this reason, we do not include the simulations that give $\chi_{\rm rem} \leq 0.1$ when constructing the hyperfitting model in Sec.~\ref{sec:hyperfit} of the main text.

\section{Verification of spherical-spheroidal mixing mode amplitudes}
\label{mixing}

Spherical-spheroidal mode mixing implies that within a waveform of multipolar component $\ell{,}m$ we can find QNMs with the same $m$ but different $\ell$~\cite{Berti:2005gp,Buonanno:2006ui,Kelly:2012nd,Berti:2014fga,Ma:2022wpv,Baibhav:2023clw}.
The degree of mixing can be characterized by the spherical-spheroidal mixing coefficients $\mu_{m{,}\ell{,}\ell^\prime{,}n}$, defined by
\begin{equation}
	\int S^*_{\ell^\prime{,}m^\prime{,}n} Y_{\ell{,}m} d\Omega = \mu_{m{,}\ell{,}\ell^\prime{,}n}(\chi_{\rm rem}) \delta_{m{,}m^\prime}, \label{eq:mixing_coeff}
\end{equation}
where $S_{\ell^\prime{,}m^\prime{,}n}$ is the spheroidal harmonic eigenfunction for the $\ell{,}m$ multipole and $n$th overtone with spin weight $s = -2$.
Thus, the complex amplitude $A_{\ell{,}m{,}n | \ell{,}m}$of the $\ell{,}m{,}n$ QNM in its natural multipole $\ell{,}m$ is related to the amplitude $A_{\ell{,}m{,}n|\ell^\prime{,}m}$ in the $\ell^\prime{,}m$ multipole by
\begin{equation}
	A_{\ell{,}m{,}n|\ell^\prime{,}m} = \dfrac{\mu_{m{,}\ell{,}\ell^\prime{,}n}}{\mu_{m{,}\ell{,}\ell{,}n}} A_{\ell{,}m{,}n | \ell{,}m}.
\end{equation}

In our mode extraction procedure, we do not assume any relation between the amplitudes of a QNM in its ``natural'' multipole and its amplitude in other multipoles where it may appear because of spherical-spheroidal mixing.
Therefore, we can use the known values for $\mu_{m{,}\ell{,}\ell^\prime{,}n}$ to verify whether the amplitudes of the mixing modes we extract are indeed consistent with the theory.

\begin{figure*}[h]
	\centering
	\includegraphics[width=0.99\textwidth]{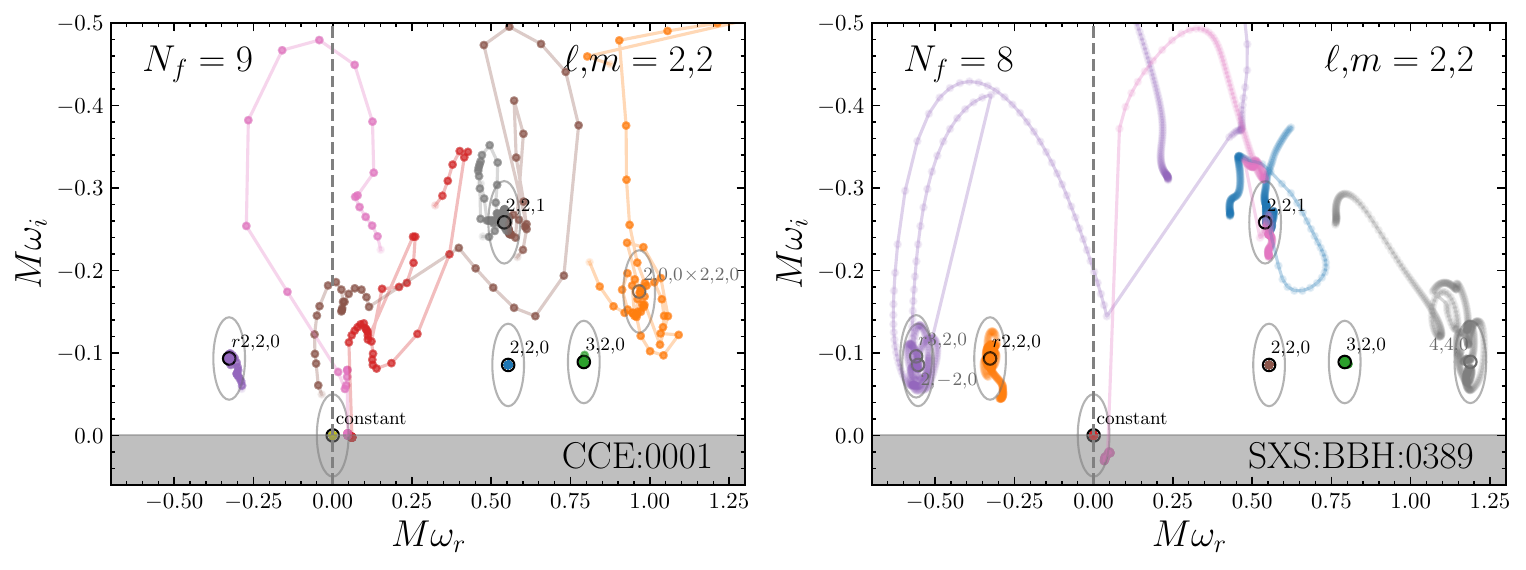}
	\includegraphics[width=0.99\textwidth]{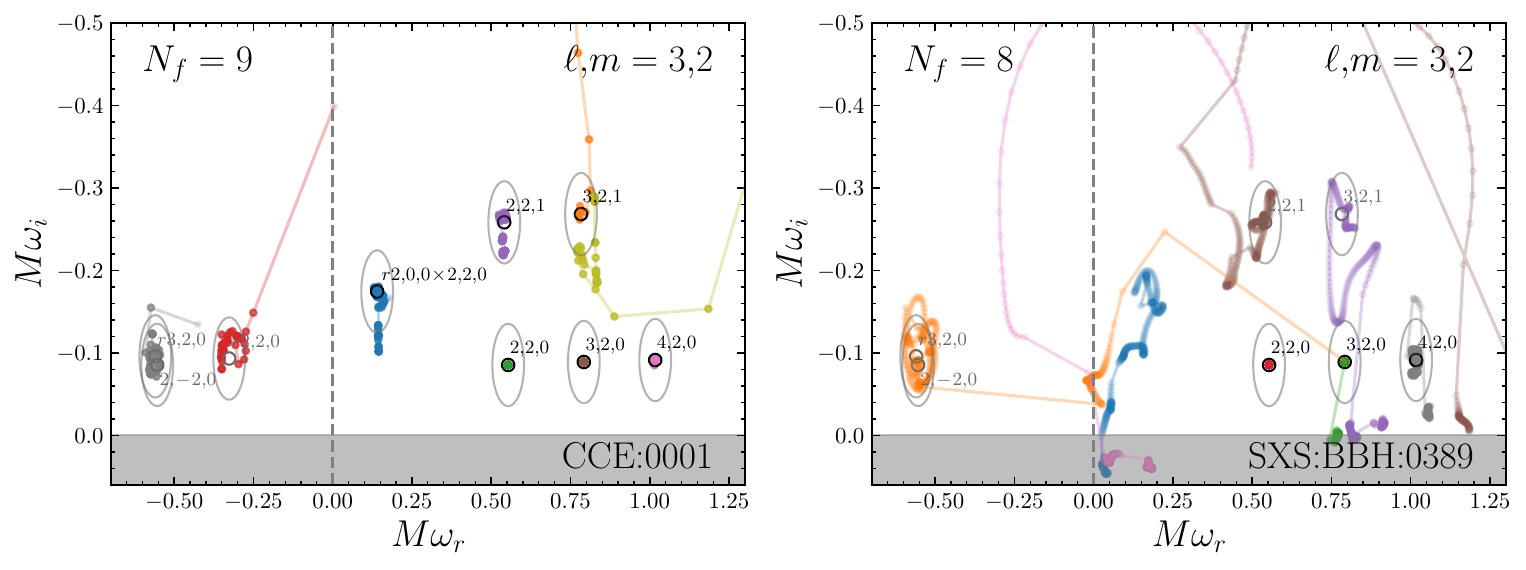}
	\includegraphics[width=0.99\textwidth]{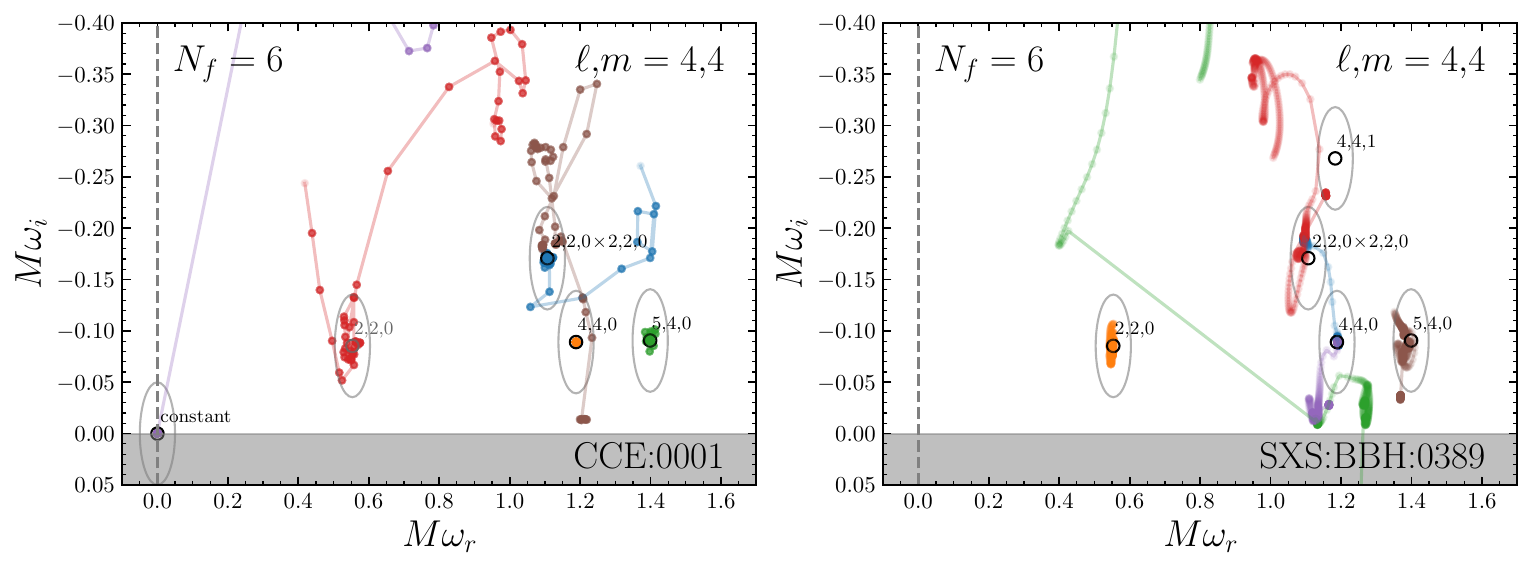}
	\caption{Frequency-agnostic fit results for the SXS:BBH\_ExtCCE:0001 waveform (left column, labeled as CCE:0001 for simplicity) and SXS:BBH:0389 (right column). These are both equal-mass, nonspinning binary simulations, so $q = 1, \chi_1 = \chi_2 = 0$.
		The rows corresponds to fits of the $\ell{,}m = 2{,}2, 3{,}2$ and $4{,}4$ multipoles of the waveform, respectively.
		For each panel, we show the best result among $N_f \in \{5, 6, \dots 10\}$, following the procedure outlined in Sec.~\ref{subsec:finalize}.
	}
	\label{fig:CCE_compare_omega}
\end{figure*}

Given a QNM extracted in its natural multipole with amplitude $A_{\ell{,}m{,}n | \ell{,}m}$, we can compute by Eq.~\eqref{eq:mixing_coeff} its theoretical amplitude $A^{\rm t}_{\ell{,}m{,}n|\ell^\prime{,}m}$ when mixed into the $\ell{,}m^\prime$ mode, and compare with the same quantity extracted through our procedure, $A^{\rm e}_{\ell{,}m{,}n|\ell^\prime{,}m}$, by defining
\begin{equation}
	\delta A_{\rm mix}    = A^{\rm e}_{\ell{,}m{,}n|\ell^\prime{,}m} - |A^{\rm t}_{\ell{,}m{,}n | \ell{,}m}|. \label{eq:delta_A_mix}
\end{equation}
This quantifies the relative difference between the theoretical mixing amplitudes and those extracted from the fit.
We define the relative error $\delta A / A$ by choosing the denominator to be $A = |A^{\rm t}_{\ell{,}m{,}n | \ell{,}m}|$.

We perform the test for the $2{,}2{,}0$ mode mixing into the $\ell{,}m = 3{,}2$ multipole, and for the $3{,}2{,}0$ mode mixing into the $\ell{,}m = 2{,}2$ multipole.
We use these mixing modes as examples because they are often dominant.
The results are shown in Fig.~\ref{fig:22_32_mixing} for the main set of 188 SXS simulations.
The difference between the theoretical and extracted mixing amplitudes shows broad consistency between the two. The exception are modes corresponding to remnants with small spin $\chi_{\rm rem}$, because then the recoil modes can interfere with the spherical-spheroidal mixing modes, as explained in Appendix~\ref{app:procedure}.

\section{Comparing the results with and without Cauchy Characteristic Extraction}
\label{app:CCE}

As discussed in the main text, the postmerger signal of a given multipole is contaminated by modes from other multipoles due to a combination of spherical-spheroidal mixing and gravitational recoil.
By applying CCE and by  appropriately fixing the BMS frame, the recoil modes can in principle be removed~\cite{MaganaZertuche:2021syq,Ma:2022wpv}. This can produce a cleaner ringdown signal and allow for better extraction of the modes, especially for slower spinning remnants.
The publicly available CCE waveform catalog currently contains 13 waveforms, and they do not cover the progenitor parameter space as extensively as the original SXS waveform catalog.
Nonetheless, we can compare the results between the SXS and CCE waveforms for selected cases to understand the improvements that may be possible by using the CCE waveforms.
For the CCE waveforms, we use the \texttt{scri} code~\cite{mike_boyle_2020_4041972,Boyle:2013nka,Boyle:2014ioa,Boyle:2015nqa} to map the waveform into the correct BMS frame for QNM extraction.
In principle, this removes the recoil modes and the constant shift in the waveform~\cite{MaganaZertuche:2021syq,Ma:2022wpv}.

While the waveforms we compare come from effectively the same progenitor settings, the sampling rate of the CCE waveforms is around an order of magnitude lower than the sampling rate of the SXS waveforms. Each data point for CCE is $\Delta t \approx 1 M$ apart, while that for SXS is $\Delta t \approx 0.1 M$.
This limits the scope of the comparisons done in this appendix.

As an example, we compare the CCE:0001 and SXS:BBH:0389 waveforms, which are both equal-mass simulations with nonspinning progenitors
(the CCE waveforms are usually labeled as SXS:BBH\_ExtCCE:XXXX; for simplicity, we will call them CCE:XXXX in this appendix).
We first test the frequency-agnostic fitting procedure by fitting both waveforms with $N_f = \{5, 6, \dots, 10 \}$, and selecting the case in which we find the highest number of potential modes (cf. Sec.~\ref{subsec:potential_modes}).

In Fig.~\ref{fig:CCE_compare_omega} we compare the results between the best instance among $N_f \in \{5, 6, \dots, 10 \}$ for both CCE:0001 and SXS:BBH:0389 for the $(\ell{,}m) \in \{(2{,}2), (3{,}2), (4{,}4)\}$ multipoles.
In the SXS results we observe traces of the recoil modes, including for example the $4{,}4{,}0$ and $2{,} \! - \! 2{,}0$ modes in the $2{,}2$ multipole.
As expected, in the CCE waveforms, the recoil modes are suppressed and we do not observe them.
An exception is the $2{,}2{,}0$ mode in the $4{,}4$ multipole, which is still identified as a potential mode in the CCE waveform.
However, this mode did not pass the amplitude flatness test (unlike the corresponding mode found in its SXS counterpart).

As the CCE waveforms are cleaner, in some cases we can identify more modes, especially overtones and quadratic modes.
For example, in the $\ell{,}m = 3{,}2$ multipole of the CCE waveform we observe the $2{,}2{,}1$ and $3{,}2{,}1$ overtones and even the retrograde $r2{,}0{,}0 \! \times \! 2{,}2{,}0$ quadratic mode, with all of them passing the flatness test. By contrast, in the SXS waveform only the $2{,}2{,}1$ and $3{,}2{,}1$ overtones are listed as potential modes, and both of them fail the flatness test.

\begin{figure*}[t!]
	\centering
	\includegraphics[width=0.49\textwidth]{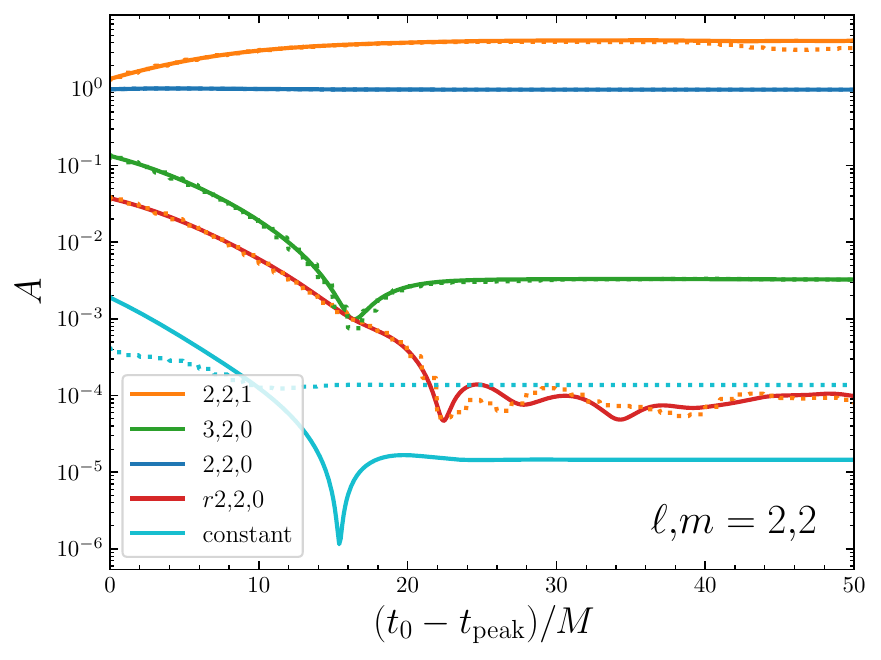}\\
	\includegraphics[width=0.49\textwidth]{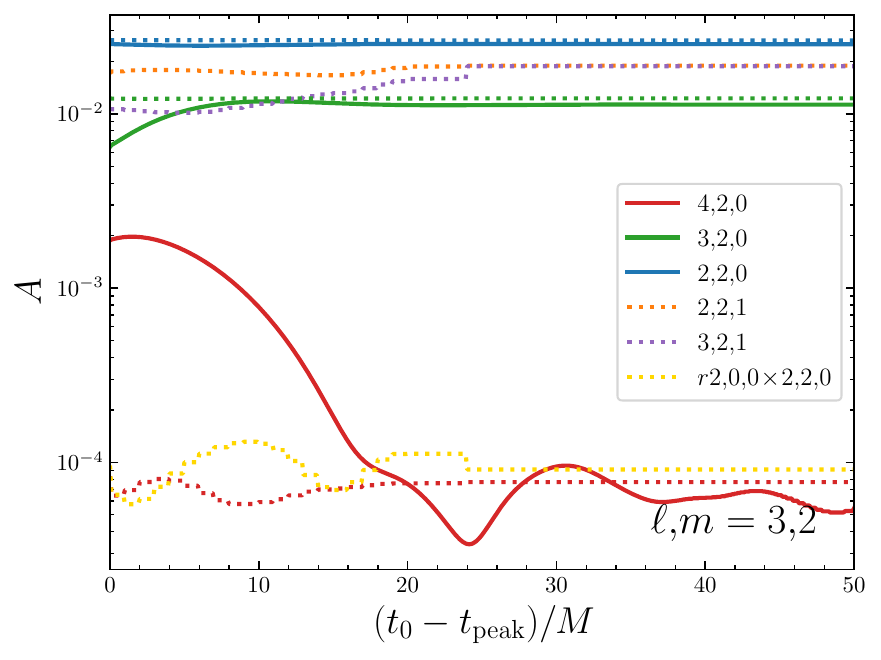}
	\includegraphics[width=0.49\textwidth]{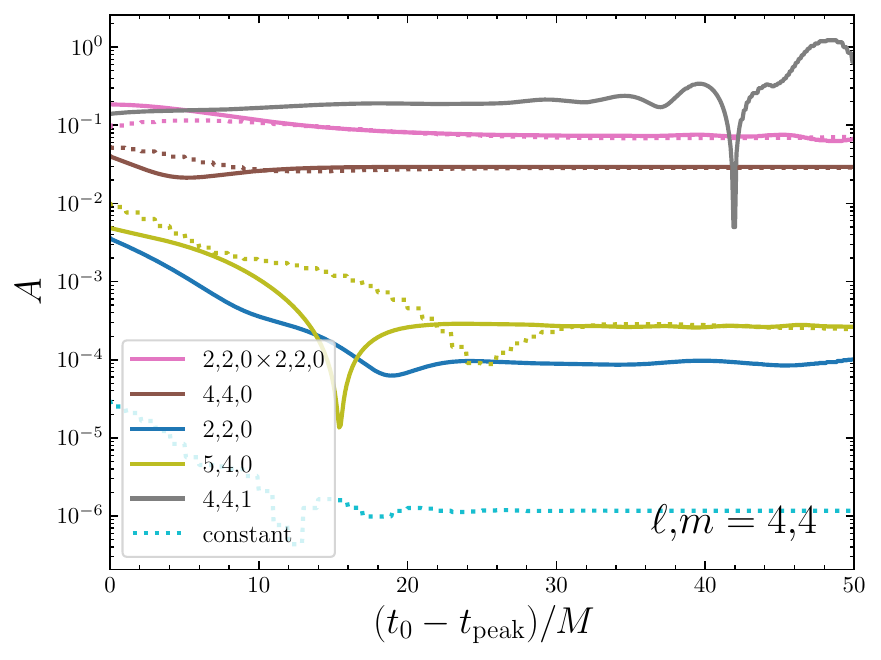}
	\caption{The fitted amplitude of the SXS:BBH\_ExtCCE:0001 waveform (dotted lines) and the SXS:BBH:0389 waveform (solid lines) for the $(\ell{,}m) \in \{(2{,}2), (3{,}2), (4{,}4)\}$ multipole, after discarding modes that failed the amplitude stability test as laid out in Sec.~\ref{subsec:stability_test}.
	}
	\label{fig:CCE_compare_amp}
\end{figure*}

Next, we compare the results of the fixed-frequency fitting analysis.
For both waveforms, following Sec.~\ref{subsec:stability_test}, we perform the stability test on the list of potential modes to arrive at a list of modes that we can consider ``robust.''
Then we can do a final fixed-frequency fit to extract the amplitude of different modes.
In Fig.~\ref{fig:CCE_compare_amp} we compare the amplitudes extracted from the CCE and SXS waveforms for the same three multipoles examined above.
While the list of modes found is different, the amplitudes of the modes that were found in both waveforms are similar.
The amplitudes of the most dominant modes -- in particular, modes with multipolar indices $\ell$ and $m$ that match the NR waveform multipole under consideration -- are nearly equal across the two waveforms, unless some important overtones are missed.
For example, as the $2{,}2{,}1$ and $3{,}2{,}1$ overtones failed the stability test and were discarded for the $\ell{,}m = 3{,}2$ multipole of the SXS waveform, the extracted amplitude of the $3{,}2{,}0$ mode is different from the amplitude extracted from the CCE waveform at the percent level.
Note that CCE alters the spurious constant in the waveforms, so a difference in the amplitude of the constant mode is expected.

The tests performed in this appendix show that using CCE waveforms is particularly important for robustly extracting QNMs.
However, the results obtained by fitting nonCCE waveforms are still useful, as long as we identify all the important modes present in the waveform.
A bias could be introduced if some modes are missed or misincluded, but the bias is typically small when compared to (e.g.) the fluctuation of the mode amplitude with respect to $t_0$, or to the error of the hyperfits presented in Sec.~\ref{sec:hyperfit}.

\section{Adjusting the strength of the stability condition}
\label{app:stability}

In the main text we use certaint threshold parameters, such as $p_{\rm ag}$ and $\epsilon_{\rm stable}$, for determining whether a mode is robustly present in the ringdown.
While the mode identification and stability conditions are qualitatively sound, these thresholds are largely arbitrary, and they were determined ``experimentally'' by trial and error.
In our experience, the threshold parameter that most sensitively affects the mode extraction procedure is $\epsilon_{\rm stable}$, the tolerance in the fluctuation of the amplitude and phase ($\Delta_{k, {\rm stable}}$).
Changing the value of $\epsilon_{\rm stable}$ affects the number of modes that we find. In this appendix we show how this, in turn, affects the amplitudes of the dominant modes that we would like to extract.

\begin{figure*}
	\centering
	\includegraphics[width=0.99\textwidth]{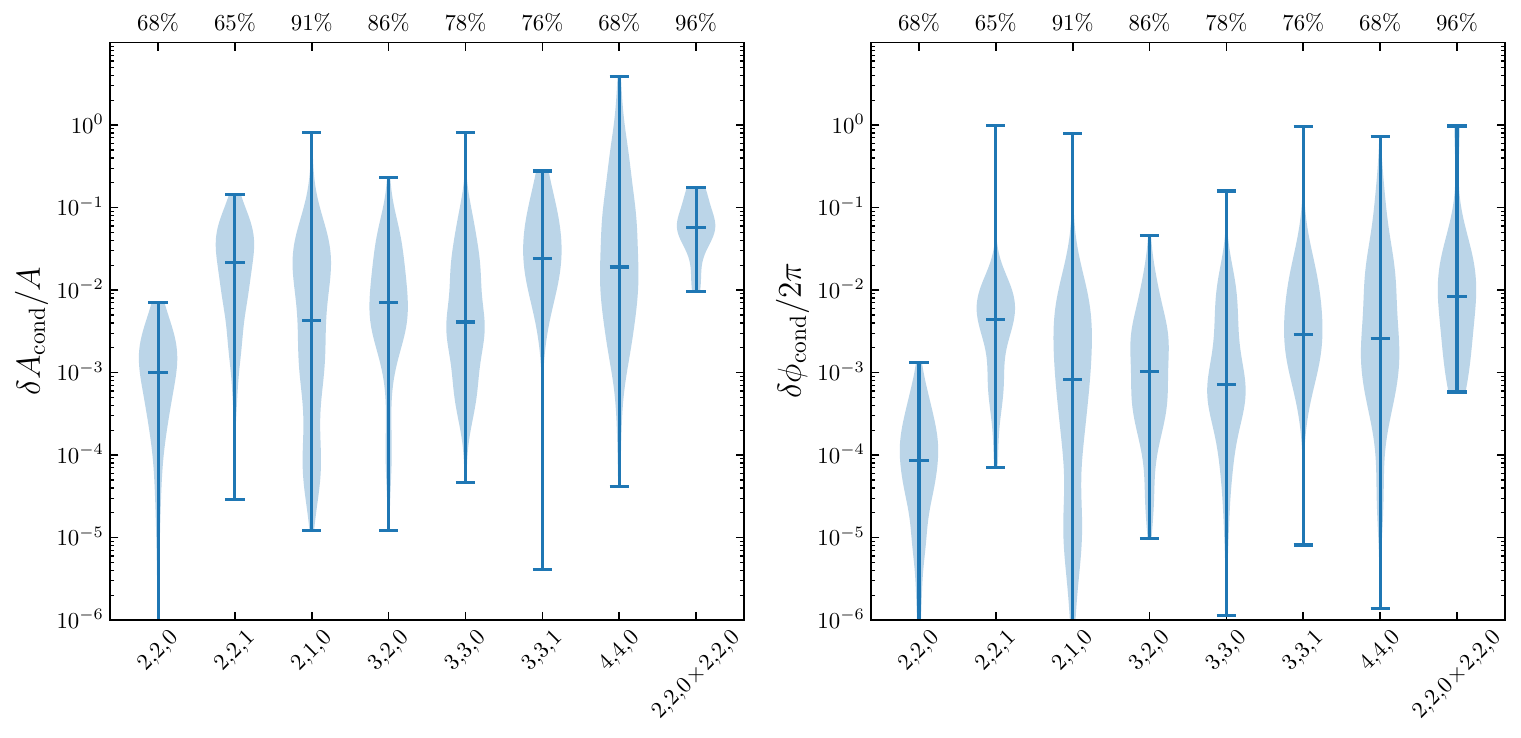}
	\includegraphics[width=0.99\textwidth]{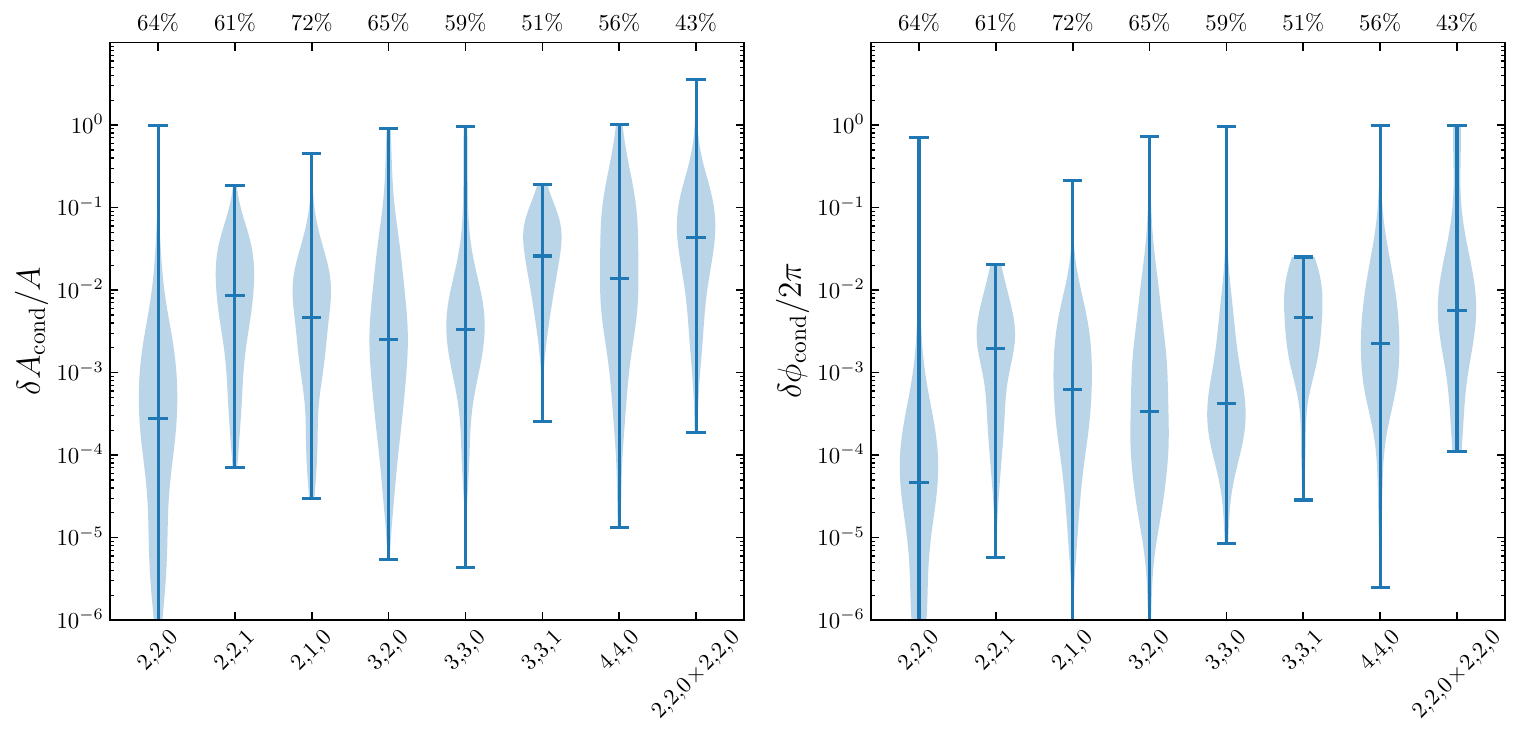}
	\caption{
		Violin plots of the relative difference in the extracted amplitude (left column) and phase (right column) for different modes. In the top row we compare the ``strong'' and ``normal'' conditions, while in the bottom row we compare the ``normal'' and ``weak'' conditions.
		The violins denote the distribution of the errors for all of the simulations considered.
		The numbers above the plotting panel show the percentage of instances for which the amplitude extracted for that particular mode has a nonzero change across the different conditions.
		The change in the condition only affects the number of other modes included when we are extracting the mode in question.
		Thus, if the two conditions we are comparing find the same list of modes, the amplitude of the extracted mode will be exactly the same, because the fit includes the same modes across the two different conditions.
	}
	\label{fig:strength_compare}
\end{figure*}

We will call our standard parameter choice in the main text ($\epsilon_{\rm stable} = 0.2$ and $A_{\rm tol} = 10^{-3}$) the ``normal'' condition.
Here we compare the ``normal'' results against ``strong'' and ``weak'' thresholds, such that $(\epsilon_{\rm stable}, A_{\rm tol}) = (0.1, 10^{-7})$ and $(0.4, 10^{-3})$, respectively.
We call the former ``strong'' because potential modes found by the frequency-agnostic fit are more easily discarded by the stability test. The opposite is true for the ``weak'' condition.
While $\epsilon_{\rm stable}$ affects mode extraction the most, we also change $A_{\rm tol}$ to $10^{-7}$ for the strong condition to verify that this parameter does not affect the extraction of the amplitudes and phases  the dominant modes.
As expected, using the strong (weak) condition we find fewer (more) modes.
In Fig.~\ref{fig:strength_compare} we show how this affects the extraction of amplitudes of different modes within their natural multipole.
We define the relative error due to using a different condition to be
\begin{align}
	\delta A_{\rm cond}    & = A_{\rm stronger} - A_{\rm weaker},       \\
	\delta \phi_{\rm cond} & = \phi_{\rm stronger} - \phi_{\rm weaker},
\end{align}
and we show the relative error $\delta A_{\rm cond} / A$ by choosing $A = A_{\rm weaker}$.
For example, in the top row of Fig.~\ref{fig:strength_compare} we use ``strong'' as the stronger condition and ``normal'' as the weaker condition.
The results include all the simulations used in the training/validation set of Sec.~\ref{sec:hyperfit}.
The relative errors $\delta A$ and $\delta \phi$ are mostly at the percent level, but they can be as large as $O(10\%)$ for relatively subdominant modes.
This is not too concerning: in real GW data, we would not expect the amplitudes of those modes to be measured better than that for current detectors.

\section{Outliers in the extracted amplitudes}
\label{sec:outliers}

The mode extraction algorithm outlined in the main text uses tolerance criteria to check whether the modes are present in the waveform. Choosing the best criteria is inherently difficult:
if the criteria are too stringent we will lose some modes; if they are too lax, we will identify spurious modes that should not be present.
In principle there is an infinite number of modes in the waveform, but most of them will be negligible.
Nonetheless, our goal is to extract all of the modes that we \textit{could} identify.
When we say that a mode is ``present'' in the waveform we mean that it is ``not negligible'' -- i.e., that the mode has a stable amplitude and phase, and missing it would cause a significant bias in the extraction of the other modes.

\begin{table}
	\label{table:outliers}
	\centering
	\begin{tblr}{
		colspec={ Q[c,m] Q[c,m]},
		vlines,
		hlines,
		hline{1,2,Z} = {1.25pt,solid},
		vline{1,Z} = {1.25pt,solid},
				vspan=even,
				row{1} = {3em},
				row{2-Z} = {1.7em},
			}

		Figure                                     & {Removed outlier            \\(SXS:BBH:XXXX)} \\
		Fig.~\ref{fig:quad_dep} top row            & \SetCell[r=2]{}$0156, 1432$ \\
		Fig.~\ref{fig:quad_ratio} left panel       &                             \\
		Fig.~\ref{fig:quad_dep} bottom row         & \SetCell[r=2]{} $0184$      \\
		Fig.~\ref{fig:quad_ratio} right panel      &                             \\
		Fig.~\ref{fig:overtone_ratio} right panel  & $0204$                      \\
		Fig.~\ref{fig:retrograde_ratio} left panel & $0207, 1428$                \\
	\end{tblr}
	\caption{Outlier data points removed in some figures in the main text.
		These simulations were removed either because they are significantly far from the trends observed in other simulations, or because their amplitudes or phases fluctuate significantly (i.e., their error bars are one or more orders of magnitude larger than the norm).}
\end{table}

In fact, as we are using the same criteria for all modes across all simulations, there is no guarantee that there exists a set of criteria that would identify all modes present in all simulations without finding spurious modes.
Therefore, as the misidentification of other modes (either missing modes or finding spurious modes) could affect the extraction of a given mode,
it is not surprising to see the extracted mode amplitude and phases being biased for a small fraction of simulations.
For example, the data points lying above the trend in the top left panel of Fig.~\ref{fig:quad_dep} (or in the left panel of Fig.~\ref{fig:quad_ratio}) could be biased for this reason.
In fact, by inspecting the mode extraction results for these simulations, we find that they often miss certain recoil modes that are identified in other simulations.

In some figures, we have removed one or two outliers that are significantly biased (e.g., one or more orders of magnitude away from the observed trends), or those with large fluctuations in the extracted amplitude and phase.
These simulations are listed in Table~\ref{table:outliers}.

\section*{Erratum I}

In a previous version of the paper, when we are constructing the hyperfit models in Sec.~\ref{sec:hyperfit}, we assumed that the leading dependence of the amplitude for the modes with an odd $m$ are $\sim \delta$ as $\delta \to 0$.
We thus divided the amplitudes by $\delta$ for those modes and fitted the result with a multivariate polynomial with the procedure explained in Sec.~\ref{sec:hyperfit}.
However, while the $\sim \delta$ dependence is exact for BBHs with equal spins in the progenitors, it is only an approximation for BBHs with unequal spins (i.e. with $\chi_- \neq 0$).
In the current version of the paper, we have introduced an additional term proportional to $\chi_-$ for some of the modes with odd $m$, so that the fits give the correct behavior as $\delta \to 0$, even when $\chi_- \neq 0$;
see Sec.~\ref{subsec:amplitude_adjustment} for the details.
We thank Costantino Pacilio, Swetha Bhagwat, Francesco Nobili and Davide Gerosa for pointing this out.

\section*{Erratum II}

\subsection{Corrections to Fig.~\ref{fig:overtone_ratio}}

In a previous version of the paper, when showing the ratios of excitation factors $|B_{2{,}2{,}1}/B_{2{,}2{,}0}|$ and $|B_{3{,}3{,}1}/B_{3{,}3{,}0}|$ in Fig.~\ref{fig:overtone_ratio} (purple dotted lines), we used the values $B_{\rm SN}$ computed in the Sasaki-Nakamura formalism presented in Ref.~\cite{Zhang:2013ksa}. 
However, our goal is to compare these ratios with the ones between the qausinormal-mode (QNM) amplitudes extracted from the waveform strain $h$, which is related to the $\psi_4$ scalar by $h = - \psi_4 / \omega^2$, where $\omega$ is the QNM frequency.
Therefore, we should use the excitation factors $B_{\rm T}$ computed in the Teukolsky formalism, where the perturbation wavefunction $\propto \psi_4$, instead of $B_{\rm SN}$.
We should also plot the ratios $|\omega_{2{,}2{,}0}^2 B_{2{,}2{,}1}/\omega_{2{,}2{,}1}^2 B_{2{,}2{,}0}|$ and $|\omega_{3{,}3{,}0}^2 B_{3{,}3{,}1}/\omega_{3{,}3{,}1}^2 B_{3{,}3{,}0}|$ so that they can be directly compared to the ratio of amplitudes extracted in $h$ instead of $\psi_4$.
Fig.~\ref{fig:overtone_ratio} has been corrected in this version of the paper.
This does not change any of the discussion or conclusions.
We thank Keefe Mitman and the other authors of Ref.~\cite{Mitman:2025hgy} for pointing out this error.

\subsection{Corrections to Eq.~\eqref{eq:overtone_fits}}

In a previous version of the paper, the symbols $\eta$ and $\chi_+$ were swapped erroneously in Eqs.~\eqref{eq:22_overtone_fit} and~\eqref{eq:33_overtone_fit}.
This has been corrected in this version of the paper.
The error is typographical and does not affect any of the figures or discussion in the paper.

\subsection{Note about the $2{,}1{,}1$ mode hyperfit model}

Upon further analysis, it was found that when modeling the ringdown of the $\ell{,}m = 2{,}1$ using QNMs generated according to the amplitude and phase of the hyperfit models in Eq.~\eqref{eq:A_2.1.1} and~\eqref{eq:phi_2.1.1}, including the $2{,}1{,}1$ overtone could decrease the match with NR simulations compared to the case when only the $2{,}1{,}0$ fundamental mode is used. 
This is because both the amplitude and the error of the $2{,}1{,}1$ fits are significantly higher than that of $2{,}1{,}0$, which could dominate the mismatch when comparing to NR simulations.
We also stress that all of the hyperfits should be used with caution when they are extrapolated to regions of parameter space where the mode is not found (i.e. passing all the criteria and tests of the full mode extraction algorithm) in simulations.
This is especially true for the $2{,}1{,}1$ mode, which was not found in simulations with high $\chi_+$.
A detailed analysis of the accuracy of the hyperfit models will be presented in future work~\cite{Crescimbeni_inprep}.

\bibliography{QNM_fitting}

\end{document}